\documentclass[10pt]{article}
\pdfoutput=1
\usepackage{jheppub,amsmath,amssymb}
\usepackage{enumerate}
\usepackage{bm}
\usepackage{bbm}
\usepackage{enumitem}
\usepackage[usenames,dvipsnames]{xcolor}
\usepackage{amsmath}
\usepackage{amsfonts}
\usepackage{amssymb}
\usepackage{graphicx}
\usepackage{hhline}
\usepackage{subfig}
\usepackage{hyperref}
\usepackage{color}
\usepackage{verbatim}
\usepackage{amsmath,amsthm,amssymb}
\usepackage{lscape}
\usepackage{float}
\usepackage{caption}
\usepackage{lscape}
\usepackage{breqn}
\usepackage{multicol}
\usepackage{graphics}
\usepackage{tikz,todonotes}
\usepackage[utf8]{inputenc}
\usepackage{autobreak}
\usetikzlibrary{arrows,positioning,decorations.markings,decorations.pathmorphing,calc}
\pgfdeclarelayer{edgelayer}
\pgfdeclarelayer{nodelayer}
\usetikzlibrary{decorations.pathreplacing}
\pgfsetlayers{edgelayer,nodelayer,main}
\tikzset{none/.style={draw=none}}
\tikzset{new edge style 2/.style={black}}
\tikzset{new style 0/.style={black}}
\tikzset{rednode/.style={draw=none, scale=0.3pt,fill=red,circle, draw}}
\tikzset{redline/.style={line width=0.3mm,red}}
\tikzset{greyE/.style={line width=0.1mm,gray}}
\usepackage[utf8]{inputenc}
\usetikzlibrary{arrows,positioning,decorations.markings,decorations.pathmorphing,calc}

\usepackage{enumitem}
\newlist{Results}{enumerate}{2}
\setlist[Results]{label={\color{blue}\arabic*.},leftmargin=*,ref={\color{blue}\  \arabic*}}

\definecolor{hyperref}{RGB}{026,028,087}

\newcommand{\beq}{\begin{equation}}
\newcommand{\eeq}{\end{equation}}
\newcommand{\bea}{\begin{eqnarray}}
\newcommand{\eea}{\end{eqnarray}}
\def\be{\begin{equation}}
\def\ee{\end{equation}}

\def\beq{\begin{equation}}
\def\eeq{\end{equation}}

\newcommand{\mpl}{M_{\rm Pl}}

\newcommand{\M}{M}
\newcommand{\K}{\mathcal K}
\newcommand{\U}{\mathcal U}

\renewcommand{\[}{\left[}
\renewcommand{\]}{\right]}
\renewcommand{\L}{\mathcal L}

\def\be{\begin{equation}}
\def\ee{\end{equation}}
\def\ba{\begin{eqnarray}}
\def\ea{\end{eqnarray}}

\def\nn{\nonumber}

\def\d{\mathrm{d}}
\DeclareMathOperator{\tr}{tr}

\usepackage[normalem]{ulem}

\def\ba{\begin{eqnarray}}
\def\ea{\end{eqnarray}}

\def\H{\mathcal{H}}

\def\L{\mathcal{L}}
\def\K{\mathcal{K}}

\def\stu{St\"uckelberg }
\newcommand{\Lic}{{Lichnerowicz }}

\def\d{\mathrm{d}}
\def\mn{_{\mu \nu}}
\def\ab{_{\alpha \beta}}
\def\mupn{^\mu_{\, \nu}}
\def\({\left(}
\def\){\right)}

\def\mpl{M_{\rm Pl}}
\def\p{\partial}
\def\ie{{\em i.e. }}

\begin{document}

\title{EFT of Interacting Spin-2 Fields}

\author[a]{Lasma Alberte,}
\author[a,b]{Claudia de Rham,}
\author[a]{Arshia Momeni,}
\author[a]{Justinas Rumbutis,}
\author[a,b]{Andrew J. Tolley}
\affiliation[a]{Theoretical Physics, Blackett Laboratory, Imperial College, London, SW7 2AZ, U.K.}
\affiliation[b]{CERCA, Department of Physics, Case Western Reserve University, 10900 Euclid Ave, Cleveland, OH 44106, USA}

\emailAdd{l.alberte@imperial.ac.uk}
\emailAdd{c.de-rham@imperial.ac.uk}
\emailAdd{arshia.momeni17@imperial.ac.uk}
\emailAdd{j.rumbutis18@imperial.ac.uk}
\emailAdd{a.tolley@imperial.ac.uk}

\abstract{We consider the effective field theory of multiple interacting massive spin--2 fields. We focus on the case where the interactions are chosen so that the cutoff is the highest possible, and highlight two distinct classes of theories. In the first class, the mass eigenstates only interact through potential operators that carry no derivatives in unitary gauge at leading order. In the second class, a specific kinetic mixing between the mass eigenstates is included non--linearly. Performing a decoupling and ADM analysis,  we point out the existence of a ghost present at a low scale for the first class of interactions. For the second class of interactions where kinetic mixing is included, we derive the full $\Lambda_3$--decoupling limit and confirm the absence of any ghosts. Nevertheless both formulations can be used to consistently describe an EFT of interacting massive spin--2 fields which, for a suitable technically natural tuning of the EFT, have the same strong coupling scale $\Lambda_3$. We identify the generic form of EFT corrections in each case. By using Galileon Duality transformations for the specific case of two massive spin-2 fields with suitable couplings, the decoupling limit theory is shown to be a bi-Galileon. }

\maketitle


\section{Introduction}
The search for all possible consistent field theories for various spins is a longstanding and ongoing research topic of fundamental importance. Although renormalizable theories only exist for $s
\le 1$, effective field theories (EFTs) can be constructed for higher spin values, notably for spin-2. As well as the formal interest in constructing such theories, there is significant phenomenological interest. For instance effectively spin-2 states may play a role in dark matter models \cite{Bernal:2018qlk,Marzola:2017lbt,Babichev:2016bxi} or if chosen sufficiently light in effective field theory descriptions of cosmic acceleration \cite{deRham:2014zqa}. Spin-2 particles appear as bound states of QCD and in some particle physics models of physics beyond the Standard Model \cite{Chivukula:2017fth}. Theories of multiple spin-2 fields have a wide range of applications in physics. For example, an infinite number of spin-2 fields appears in Kaluza-Klein (KK) tower of states from dimensional reduction of higher dimensional gravity (see, \emph{e.g.}, \cite{Overduin:1998pn, deRham:2014zqa} for a review) and massive spin--2 states arise in the spectrum of string theory. An important question from the effective field theory viewpoint is to understand the consistency of the mutual interactions of the multiple massive spin-2 fields appearing in the KK tower. For instance in recent work it is noted that scattering amplitudes of the KK modes grow much slower than anticipated from the individual components \cite{Chivukula:2019rij} as a consequence of the underlying higher dimensional symmetry which explains in part how the cutoff of the KK theory is higher than that typically expected for an interacting theory of massive spin-2 states. In the condensed matter context, EFTs for spin-2 states have been used to describe a gapped collective excitation in fractional quantum Hall effect \cite{Gromov:2017qeb}. \\

The action for free massless gauge fields of any spin is well known \cite{Fronsdal:1978rb, Fang:1978wz} and in recent years there has been significant progress in constructing interacting higher-spin theories albeit on anti-de Sitter spacetime \cite{Vasiliev:1990en,Vasiliev:1992av}. As is well known, in the case of massless spin-1 particles, adding interactions between different species of fields uniquely leads to the Yang--Mills theory as the low energy description. For a single massless self-interacting spin-2 field the only possible two-derivative action
is the Einstein--Hilbert action and it is known to be impossible to have interacting multiple species of massless spin-2 fields (gravitons) \cite{Boulanger:2000rq}.  Extensions to higher spin typically need infinite towers of spin states precluding a straightforward local low energy EFT description for the massless case. A review on massless higher-spin gauge theories with $s>2$ can be found in \cite{Vasiliev:1995dn,Rahman:2015pzl,Rahman:2013sta} with much progress being made also in the recent years.\\

Adding mass to the various fields have turned out to be a non-trivial problem in itself. In general no straightforward analogue of the (super-)Higgs mechanism is known\footnote{It is known that a weakly coupled Higgs mechanism for which the high energy behaviour of scattering amplitudes is improved to $\lim_{s \rightarrow \infty } A(s,t)<s^2$ at fixed $t$ would require an infinite number of spin states \cite{deRham:2018qqo,Caron-Huot:2016icg} and so is very different from the usual spin-1 Higgs. This is confirmed by explicit attempts using $s \le 1$ \cite{Bonifacio:2019mgk}. These arguments do not preclude a strongly coupled Higgs mechanism. Explicit UV completion in AdS has been proposed recently in \cite{Bachas:2017rch,Bachas:2018zmb,Bachas:2019rfq}. For related discussion on constraints on UV completion of spin-2 see \cite{deRham:2018dqm}}. It is however possible to construct \stu type EFTs that can be viewed as low energy EFTs of some unknown (partial) UV completion. In the case of spin-1 fields, the free theory is described by the Proca action, and the EFT interactions may be organized in several ways depending on the choice of additional symmetries for the \stu field (see for example the discussion in \cite{deRham:2018qqo}). Massive spin-1 generalizations to the classic Proca action that restrict to second order equations of motion have been studied, for example, in \cite{Heisenberg:2014rta,Jimenez:2016isa,ErrastiDiez:2019ttn,Jimenez:2019hpl}, but general EFTs allow for infinitely more interactions. A novel shift-symmetric extension to interacting massive spin-1 theories in de Sitter and anti-de Sitter spacetimes has recently been made in \cite{Bonifacio:2019hrj}. In the case of a single massive spin-2 which can be coupled to matter, the most general ghost-free (two derivative, five propagating degrees of freedom) nonlinear theory in four spacetime dimensions is the de Rham--Gabadadze--Tolley theory of massive gravity \cite{deRham:2010kj}. When viewed as an effective field theory, this is the theory with the highest energy cutoff, with perturbative unitarity broken at the scale $\Lambda_3 = (m^2 M)^{1/3}$. The spin-2 interaction scale $M$ here is associated with the Planck mass $\mpl$ via a dimensionless weak coupling constant $g_*\ll 1$ needed to satisfy `improved' (loop corrected) positivity bounds \cite{Bellazzini:2016xrt,deRham:2017imi,deRham:2017xox}. Specifically in four dimensions $\mpl^2=M^2/g_*^2$ and positivity bounds impose $g_* \lesssim m/\Lambda_3 \ll 1$ \cite{deRham:2017xox,Bellazzini:2017fep}. More general interactions are allowed, but lead to a lower cutoff scale (signalled by the presence of a massive ghost). Interestingly it has recently been found that the application of non-forward limit positivity bounds derived in \cite{deRham:2017zjm} assuming a standard UV completion impose some of the tunings necessary to raise the cutoff to this special case \cite{deRham:2018qqo}. \\

These `ghost-free' or `highest cutoff' effective field theories can be easily generalized to a theory of arbitrary number of interacting spin-2 fields in an arbitrary number of spacetime dimensions $D$, as proposed in \cite{Hinterbichler:2012cn}. The allowed interactions take a very simple form in the vierbein formalism where the interactions can be written as wedge products between the various vierbeins. Each interaction vertex can thus contain up to $D$ interacting vierbeins.  However, it is known that the theories presented in \cite{Hinterbichler:2012cn} are only ghost-free for pairwise couplings between the spin-2 fields.  The reason for this is (partially) the loss of equivalence between the vierbein and metric formulations of the interactions which occurs where there are cycles of interactions \cite{Scargill:2014wya}. The equivalence between the two formulations is usually recovered by the so-called \emph{symmetric vierbein condition} that arises dynamically once the additional Lorentz transformations (present in the vierbein language) are integrated out. In the case of either tri- (tetra-, etc.) metric interaction vertices as well as for cyclic trimetric interactions \cite{Hinterbichler:2012cn,Scargill:2014wya} if we begin in an {\emph{unconstrained}} vierbein formulation, for which all 16 components of each vierbein are independent,  the symmetric vierbein condition is lost signalling a potential pathology. In \cite{deRham:2015cha} it was shown that these specific unconstrained vierbein multi-gravity theories contain the Boulware-Deser (BD) ghost \cite{Boulware:1973my} otherwise absent in the theories introduced based on the ghost-free interactions~\cite{deRham:2010kj,Hassan:2011hr,deRham:2014zqa}.  A similar result is obtained from working in the metric (or {\emph{constrained} vierbein}) formulation as we shall see below.
These results preclude the existence of a `ghost-free' (at all scales) interacting theory of multiple massive spin-2 particles with (what we will refer to as) cycle interactions. We shall demonstrate this explicitly in what follows through both a standard ADM and decoupling limit (DL) analysis. However, when viewed in EFT terms, they remain consistent and can be interpreted as a particular realization of the {\emph{EFTs for multiple massive spin-2 particles}.  The behaviour of the strong coupling scale in various classes of  multi-gravity graphs (not cycles) was explored in \cite{Scargill:2015wxs}. \\

Spin-2 particles may also arise in non-gravitational theories, where we expect the symmetry that is spontaneously broken by the mass to simply be linearized diffeomorphisms, sometimes called spin-2 gauge invariance \cite{Wald:1986bj}. Interacting theories of single and multiple spin-2 fields with this breaking have been considered recently in \cite{Folkerts:2011ev,Hinterbichler:2013eza,Bonifacio:2016wcb,Bonifacio:2019pfg}. These theories are considerably easier to analyze due to the linearity of the underlying gauge symmetry, and so will be free from many of the issues to be discussed below. We refer to \cite{Bonifacio:2019pfg}, for a related discussion of interactions between multiple such spin-2 fields. Although sometimes referred to as pseudo-linear massive gravitons, such spin-2 states cannot couple to the stress energy by standard arguments \cite{Wald:1986bj} and so these are best thought of as non-gravitational theories. \\

In the present work we shall focus on the EFT of multiple interacting massive spin-2 fields, whose masses arising from breaking nonlinear diffeomorphism symmetry --- a case that so far has not been explored in detail using the EFT methodology --- with the interactions chosen such that the EFT has the highest possible cutoff. This latter assumption significantly reduces the allowed number of interactions, and will allow us in a forthcoming work to derive strong statements about the possibility of UV completion \cite{AlberteTA}. To be concrete we shall focus most attention on the case of two massive spin-2 fields, although many of the results will easily generalize. \\

The rest of this paper is organized as follows. In Section~\ref{sec:main} we first introduce the EFT of two interacting massive spin-2 fields that will be our main focus in the rest of this work. This will be constructed in the `metric' formulation, meaning that each spin-2 field is described by a 10 component tensor. However for reasons of calculational simplicity, it is helpful to write the metric in terms of a {\it constrained} symmetric vierbein which also has 10 components. Nevertheless, this formulation differs from the so-called {\it unconstrained} vierbein formulation where all 16 components of the vierbein are allowed to vary in the action.
We perform the ADM constraint analysis of this theory in subsection~\ref{adm} which will lead to the same conclusion as found in the analogous (but different)  unconstrained vierbein formulation that interactions between the two massive spin-2 particles lead to a ghost at some scale. By itself the ADM analysis is blind to the scale of the ghost, which should really be interpreted as the cutoff scale of the low energy effective theory.  To rectify this, in Section~\ref{sec:cycle} we perform the DL analysis for the theory with cycle of interactions. Doing this both clearly identifies the presence of the ghost and its associated energy scale, and hence determines the overall cutoff of the two massive spin-2 effective theory. In fact our DL analysis is equally valid for unconstrained and constrained vierbein formulations. We establish the strong coupling scale of our EFT of cycle interactions in Section~\ref{sec:strong} and determine the appropriate technically natural tuning, that puts the cutoff at the highest possible scale for two massive spin-2 particles, namely the $\Lambda_3=(m^2 M)^{1/3}$ scale, the same as that for individual self-interactions. For the cycle of interactions, the main conclusions of this analysis are:
\begin{Results}
\item The helicity-zero/helicity-two interactions occur at the scale $\Lambda_3$.\label{mainpoint1}
\item The cutoff of the theory is set be the helicity-one/helicity-zero interactions. Those occur at the scale $\Lambda_{7/2}$ unless (a) the cubic interactions vanish in which case the scale is $\Lambda_{10/3}$, or (b) generic quadratic mass mixings are included which lower the scale to $\Lambda_4$, or (c) the mixed interactions are suppressed by $m/\Lambda_3$, in which case, interactions arise at the scale  $\Lambda_3$. \label{mainpoint2}
\item The BD ghosts in the ADM analysis of the constrained and unconstrained cycle theories are clearly associated with the new mixed interactions which are not of a two-derivative nature, and give a non-zero contribution to the scattering amplitude. \label{mainpoint3}
\end{Results}
The decoupling limit of the theory with a line of interactions is then analyzed in Section~\ref{sec:line}, confirming the expected result that the cutoff scale remains as $\Lambda_3$, and this remains true for all helicity interactions. In Section~\ref{sec:biGalileon} we show that the decoupling limit of a subclass of two interacting spin-2 line theories is described by a bi-Galileon theory \cite{Padilla:2010de}, making significant use of the Galileon Duality transformations \cite{Fasiello:2013woa,Curtright:2012gx,deRham:2013hsa,deRham:2014lqa}. We generalize these arguments to multiple massive spin-2 fields in Section~\ref{sec:multi}. We conclude in Section~\ref{sec:conclusions}. Some technical results are reserved for Appendices~\ref{sec:metric}, \ref{sec:Leom}  and  \ref{App:biGalileondetails}.

\section{The EFT of Interacting Massive Spin-2 Fields}\label{sec:main}
In this Section we shall introduce the relevant EFT for interacting massive spin-2 fields where our goal is to construct the theory with the `highest cutoff' or strong coupling\footnote{See \cite{deRham:2014wfa} for a discussion on the distinction between those two scales. In principle the strong coupling scale may not be related to the cutoff of the theory or the onset of new physics to be included, but in this work we shall adopt a standard `weakly coupled' EFT picture where new physics enters at or below the strong coupling scale.} scale as proposed in \cite{Hinterbichler:2012cn} following the construction of \cite{deRham:2010kj}. We  focus on the case of two spin-2 fields to start with, and generalize to multiple interacting spin-2 fields in Section~\ref{sec:multi}.
As is well known, without any particular tunings, the generic theory for a single interacting massive spin-2 has a cutoff at the scale $\Lambda_5=(m^4 M)^{1/5}$, where $m$ is the mass of the (lightest) spin-2 field and $M$ the scale of nonlinearities (bearing in mind that the EFT only makes sense if $m \ll M$) \cite{ArkaniHamed:2002sp}. In what follows we make the implicit assumption that in considering two massive spin-2 particles, the hierarchy between the two masses $m_1$ and $m_2$ is negligible relative to the hierarchy between $m_1$ and $M$. This does not necessarily imply that the masses are similar, only that it is consistent to neglect corrections suppressed by $m_1/M$ or $m_2/M$ while keeping terms scaling as $m_1/m_2$. Stated differently, this allows us to consider the scaling or decoupling limit $M \rightarrow \infty$ where the ratio $m_1/m_2$ is kept fixed. For this reason we will generally denote the common mass scale via $m=m_2$.\\

It is clear that in order to raise the cutoff to a scale larger than $\Lambda_5$, one should tune the operators that enter the EFT Lagrangian. In particular the self-interactions should be tuned similarly as is done for the single spin-2 case leading to the ghost-free theory  \cite{deRham:2010ik,deRham:2010kj} with the strong coupling scale $\Lambda_3=(m^2M)^{1/3}$. As for the mixing between both  fields, it will become transparent when taking the decoupling limit in Sections~\ref{sec:cycle} and~\ref{sec:line} that keeping the same-type of  structure for the mixed interactions (as will be done in what follows) is what leads to the highest strong coupling scale. Indeed we will see that in the helicity-zero/helicity-two sector decoupling limit alone, this choice will be sufficient to ensure that the multi spin-2 system has a cutoff at the scale $\Lambda_3$. However, we shall see that in some cases choosing the double--epsilon interactions is not sufficient to maintain the $\Lambda_3$ strong coupling scale in the presence of the mixing between the two spin-2 fields due to the interactions in the helicity-zero/helicity-one sector. In what follows we will then show explicitly that by further tuning the interactions we can raise the cutoff of the effective field theory to a parametrically larger scale that will be determined in Section~\ref{sec:strong}.

\subsection{Cycle vs. Line of Interactions}
\label{sec:cycle_line}
\paragraph{Cycle of Interactions - Mass eigenstates:}  For free fields, the most natural way to describe two spin-2 massive particles $h_{\mu\nu}$ and $f_{\mu\nu}$ is simply to consider the sum of two independent (diagonalized) copies of the Fierz-Pauli action, coupled to their own sources $T_{1,2}^{\mu\nu}$
\ba
\label{eq:2FP}
\L_{\rm FP}=&-& h^{\mu \nu }\mathcal{E}^{\alpha\beta}\mn h\ab-\frac 12 m_1^2 \([h^2]-[h]^2\)+\frac{1}{ M_1}h\mn T_1^{\mu\nu}\\
&-&  f^{\mu \nu }\mathcal{E}^{\alpha\beta}\mn f\ab-\frac 12 m_2^2 \([f^2]-[f]^2\)+\frac{1}{ M_2}h\mn T_2^{\mu\nu}\,,\nn
\ea
where $\mathcal{E}$ is the standard \Lic operator defined as (linearized Einstein tensor)
\be\label{quadratic}
\mathcal E^{\alpha\beta}_{\mu\nu}h_{\alpha\beta}=-\frac{1}{2}\left[\Box h_{\mu\nu}-\partial_\alpha\partial_{\mu}h^\alpha_{\nu}-\partial_\alpha\partial_\nu h^\alpha_\mu+\partial_\mu\partial_\nu h-\eta_{\mu\nu}\left(\Box h-\partial_\alpha\partial_\beta h^{\alpha\beta}\right)\right]\,.
\ee
In this form $h$ and $f$ are seen to be mass eigenstates with no kinetic mixing between them. We recall that we are just dealing with (massive) spin-2 fields living on Minkowski, and all indices are naturally raised and lowered with respect to the Minkowski metric. Interactions (mixing) between the two interacting spin-2 fields can then be realized nonlinearly by considering purely potential interactions of the form $h^n f^\ell$, with $n+\ell>2$. Nonlinearly this corresponds to theories with cycles of interactions as in Fig.~\ref{Fig:cycle}. \\

\begin{figure}[h]
    \centering
\includegraphics[width=\textwidth]{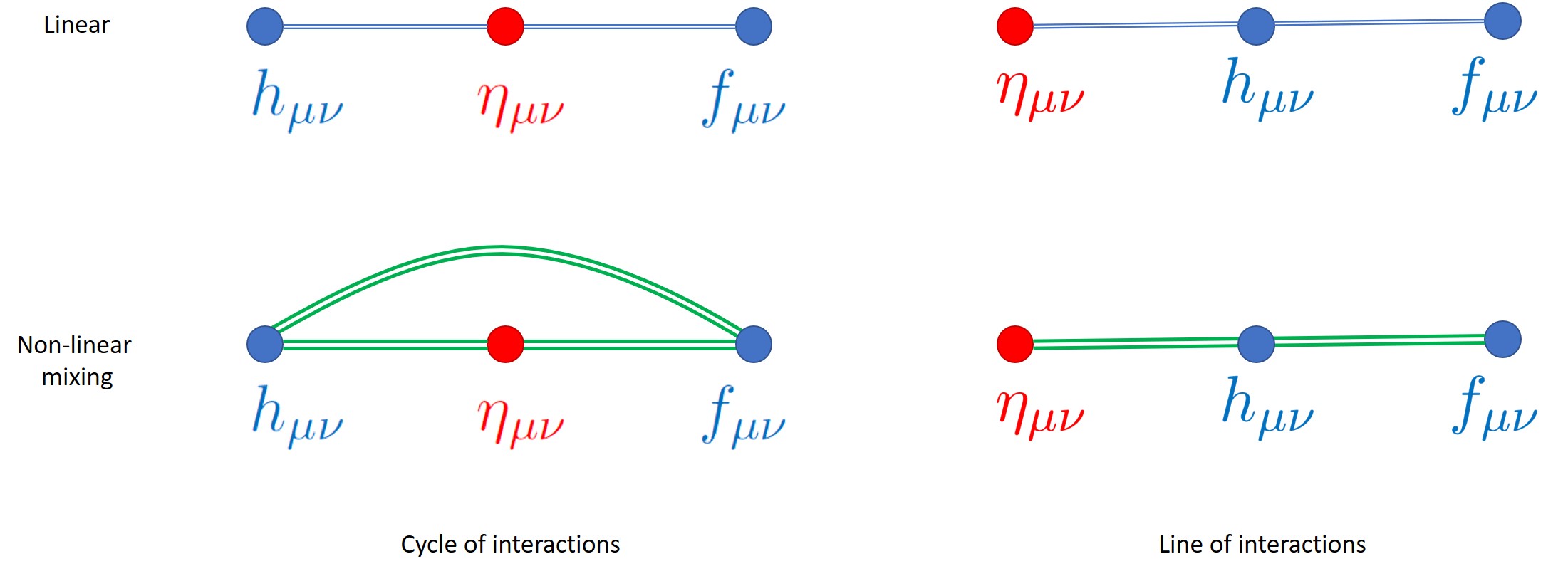}\\
\caption{Different interacting topologies \cite{Hinterbichler:2012cn}. For free fields both models are equivalent (after diagonalization), but the models differ nonlinearly. In terms of the mass eigenstates, the cycle of interactions [left] only includes potential interactions, while the line of interactions [right] also includes a specific class of kinetic mixing.}
\label{Fig:cycle}
\end{figure}\vspace{0.3cm}

\paragraph{Line of Interactions - Mass Mixings:} An alternative approach to that considered previously in Eq.~\eqref{eq:2FP}, is to have the  mass for the second spin-2 field $\tilde f\mn$ arising from a mixing with $\tilde h\mn$,
\ba
\label{eq:line}
\L_{\rm line}&=-& \tilde h^{\mu \nu }\mathcal{E}^{\alpha\beta}\mn \tilde h\ab-\frac 12 \tilde m_1^2 \([\tilde h^2]-[\tilde h]^2\)+\frac{1}{M_1}\tilde h\mn \tilde T_1^{\mu\nu}
\\
&-& \tilde f^{\mu \nu }\mathcal{E}^{\alpha\beta}\mn \tilde f\ab
-\frac{1}{4} \tilde m_2^2 \([(\tilde f-\tilde h)^2]-[\tilde f-\tilde h]^2\)
+\frac{1}{M_2}\tilde f\mn \tilde T_2^{\mu\nu}\nn\,,
\ea
leading (in this formulation) to a non-diagonal mass matrix of the form
\ba
\mathcal{M}_{ab}=\(
\begin{array}{ccc}
  \tilde m_1^2+\tilde m_2^2/2 & & -\tilde m_2^2/2 \\
  -\tilde m_2^2/2 & & \tilde m_2^2/2
\end{array}
\)\,.
\ea
For free fields, the theory can be diagonalized by performing an appropriate rotation in field space leading to a formulation which is equivalent to \eqref{eq:2FP}, with
\ba
\label{eq:m12}
m_{1,2}^2=\frac12 \(\tilde m_1^2+\tilde m_2^2\pm\sqrt{\tilde m_1^4+\tilde m_2^4}\)\,,
\ea
and the mass eigenmodes given by
$(h, f)=R_\theta\ (\tilde h, \tilde f)$, where
$R_\theta$ is the rotation matrix with angle $\theta$ given by
\ba
\label{eq:RotAngle1}
\cos^2 \theta=\frac12\(1+\frac{\tilde m_1^2}{\sqrt{\tilde m_1^4+\tilde m_2^4}}\)\,.
\ea
An essential difference with the previous construction of \eqref{eq:2FP}, is that accounting for this field space rotation inevitably leads to non-trivial kinetic term mixing at the nonlinear level. The explicit construction of the nonlinear theory is most clear in this non-diagonalized formalism, where it corresponds to a line of interactions (Fig.~\ref{Fig:cycle}), but the kinetic mixing is intrinsic throughout the construction and will be an essential element of the EFT, for instance in determining low energy scattering amplitudes~\cite{AlberteTA}. \\

The field space rotation also affects the coupling to external sources. In terms of the mass eigenstates $h\mn$ and $f\mn$, the coupling to matter is given by
\ba
\L_{\rm line}^{\rm sources}= h\mn \(\frac{\cos \theta }{M_1}\tilde T_1^{\mu\nu}+\frac{\sin \theta }{M_2}\tilde T_2^{\mu\nu}\)
+
 f\mn \(\frac{\cos \theta }{M_2}\tilde T_2^{\mu\nu}-\frac{\sin \theta }{M_1}\tilde T_1^{\mu\nu}\)\,.
\ea
At this stage one may be inclined to simply redefine the external sources, for instance, $M_1^{-1}T_1=M_1^{-1}\cos \theta \,\tilde T_1+M_2^{-1}\sin \theta \,\tilde T_2$ so that $h\mn$ would only couple to $T_1$ (and similarly for $T_2$ and $f\mn$). However such a procedure would be impossible as soon as one would consider dynamical fields in $\tilde T_{1,2}$.
\\

\paragraph{Diffeomorphisms versus Spin-2 Gauge Invariance:}

In constructing an interacting theory, we need to distinguish two cases. A free massless spin-2 field admits a copy of spin-2 gauge invariance (\emph{i.e.} linearized diffeomorphisms)
\be
h_{\mu\nu} \rightarrow h_{\mu\nu} + \partial_{\mu} \xi_{\nu}+ \partial_{\nu} \xi_{\mu} \, .
\ee
This symmetry is spontaneously broken by the mass term already at the linear level. Nonlinearly, there are only two nonlinear completions of the symmetry, full nonlinear diffeomorphisms, or the identical same spin-2 gauge invariance \cite{Wald:1986bj}. Thus in the interacting theory, it is natural to imagine that the local symmetry which is broken by the mass term is either (a) Diffeomorphism symmetry, or (b) Spin-2 gauge symmetry. The construction of the nonlinear theory will be different in each case, in particular how we choose to introduce \stu fields to describe the EFT in the broken state is different, and hence how we choose to organize the EFT expansion is different. The case of spin-2 gauge invariance is significantly simpler, and we refer to \cite{Bonifacio:2019pfg} for a recent discussion of interactions between multiple such spin-2 particles. We note in particular, that for this case the distinction between what we refer to as cycle and line interactions is minor, since the linear transformation which diagonalize the quadratic action can easily be performed within the full nonlinear action, and simply mixes allowed interactions into each other.  In what follows, we focus exclusively on the more involved case where the symmetry broken is diffeomorphism symmetry, \emph{i.e.} where the spin-2 particles may appropriately be called gravitons.

\paragraph{Spin-2 fields as Perturbations of Metric-like Objects:}
Any two massive spin-2 fields $h_{\mu\nu}$, $f_{\mu\nu}$, which arise from breaking a diffeomorphism symmetry, may be considered as the perturbations of `would-be' metrics $g^{(1)}$ and $g^{(2)}$. The particular relation we will find useful is $g^{(1)}_{\mu\nu}= (\eta_{\mu\nu}+h_{\mu\nu}/{M_1})^2$, and $g^{(2)}_{\mu\nu} = (\eta_{\mu\nu}+{f_{\mu\nu}}/{M_2})^2$ where $(A_{\mu\nu})^2 \equiv A_{\mu \alpha} \eta^{\alpha \beta} A_{\beta \nu}$. For instance, within the context of multi-gravity, the interacting theory we are considering here would arise as the decoupling limit $M_3\to \infty$ of tri-gravity where both metrics $g^{(1)}\mn$ and $g^{(2)}\mn$ couple to one another and couple to a third metric $g^{(3)}\mn=(\eta\mn+h^{(3)}\mn /M_3 )^2$. This was for instance one of the scenarios considered in \cite{deRham:2015cha}, corresponding to a `cycle' of interactions as defined in \cite{Hinterbichler:2012cn,Scargill:2014wya}, see Fig.~\ref{Fig:cycle} [left]. At the linear level $h$ and $f$ only couple to $\eta\mn$, but nonlinearly $h$ and $f$ directly couple to one another hence leading to a  cycle of interactions.\\

\subsection{Vierbein Formalism for Two Fields}
In this and the following subsections we present the highest cutoff EFTs separately for cycle and line interactions. Independently of whether we are dealing with a cycle or a line of interactions,
the EFT with the highest cutoff scale is that inspired by the single $\Lambda_3$ spin-2 field \cite{deRham:2010ik,deRham:2010kj} (we review it in Appendix~\ref{drgt rev}). This is immediately clear by performing the decoupling limit analysis in the helicity-zero/helicity-two sector as we do in Sections ~\ref{sec:hel02} and~\ref{sec:hel02_line}. In this sector, the key requirement is that the helicity-zero mode interactions have the {\it double--epsilon} structure which ensures that the helicity zero modes have second order equations of motion and that the cutoff scale is raised to $\Lambda_3$, at least for these interactions. In considering the interactions of $N$ massive spin-2 modes, we have $N$ helicity-zero modes, and each should interact via the same double--epsilon terms. Precisely how those interactions occur depends on whether we are dealing with a line or a cycle of interactions as we shall see.

For the purposes of this paper, it is most convenient to work in the vierbein formulation of $\Lambda_3$ theories \cite{Hinterbichler:2012cn} (for earlier work using vierbeins in massive gravity see \cite{Nibbelink:2006sz,Chamseddine:2011mu}).  In particular, the nonlinear action for two interacting spin-2 fields $h$ and $f$ can be written directly in the vierbein form with the respective vierbeins defined as\footnote{For future convenience we will normalize with a common interaction scale $M$.}
\be\label{identify0}
E^a_\mu\equiv\delta^a_\mu+\frac{h^a_\mu}{M}\,,\qquad F^a_\mu\equiv\delta^a_\mu+\frac{f^a_\mu}{M}\,,
\ee
 related to the `would-be' metrics $g^{(1)}$ and $g^{(2)}$ through $g_{\mu\nu}^{(1)}=E^a_\mu E^b_\nu\eta_{ab}$ and $g_{\mu\nu}^{(2)}=F^a_\mu F^b_\nu\eta_{ab}$. Since a vierbein contains 16 components in four dimensions, but a spin-2 needs only to be described by a 10 component tensor, the remaining 6 components for each vierbein must either be fixed by auxiliary equations, or by constraints. Depending on choice, this will lead us to different formulations. \\

Each of the vierbeins is equipped with its own Einstein--Hilbert term
\be\label{act_kinetic}
g_*^2S_{\text{kinetic}}=\frac{M_1^2}{8}\int \varepsilon _{abcd}\,E^a\wedge E^b\wedge R^{cd}[E]+\frac{M_2^2}{8}\int \varepsilon _{abcd}\,F^a\wedge F^b\wedge R^{cd}[F]\,,
\ee
and we have introduced the weak coupling parameter $g_*$, which distinguishes the Planck scales $M_{1,2}/g_*$ from the interaction scales $M_{1,2}$. Although for massive fields, this form of kinetic term is not protected by any symmetry, in practice corrections to it must be suppressed in order not to introduce ghosts/irrelevant operators at a lower scale \cite{deRham:2013tfa,deRham:2015rxa}. The allowed form of corrections that preserves the highest cutoff nature of the EFT is specified in Sections~\ref{sec:highdercycle} and \ref{HigherderLine}.
In addition, improved positivity bounds  \cite{Bellazzini:2016xrt,deRham:2017imi,deRham:2017xox} require that in order for the full effective field theory (with the interaction terms included) to admit a standard UV completion we need $g_* \ll 1$. In the context of the $\Lambda_3$ theory this condition is effectively \cite{deRham:2017xox,Bellazzini:2017fep}
\be
g_* \lesssim \frac{m}{\Lambda_3}\ll 1 \, .
\ee
Since the improved positivity bounds apply equally well in the case of two interacting spin-2 fields, we will assume the same scale of weak coupling throughout. The precise scale of the weak coupling is however not important in what follows.

If the symmetric vierbein conditions
\be\label{sym_vierb_gen2}
\eta E  =(\eta E )^T\,, \qquad \eta F = (\eta F )^T
\ee
equivalent to $E_{a\mu}=E_{\mu a}$ and $F_{a\mu}=F_{\mu a}$, are satisfied (or imposed) this implies that $h_{a\mu}=h_{\mu a}$ and $f_{a\mu}=f_{\mu a}$ thus providing an equivalent mapping between the vierbein and metric formulations. These particular conditions will follow automatically if we separately give masses to $E$ and $F$ with no interactions between them. That is, for two decoupled copies of massive gravity, the unconstrained vierbein and constrained vierbein formulations with conditions \eqref{sym_vierb_gen2} imposed are equivalent. We shall discuss the origin of these conditions in more detail below, see also \cite{Ondo:2013wka}. Henceforth, we shall use the term \emph{constrained vierbeins} whenever these (or related \eqref{sym_vierb_gen3}) symmetric vierbein conditions are imposed from the outset in our theory, and \emph{unconstrained vierbeins}  when all 16 components of the vierbein are taken to be independent. The constrained vierbein formulations can all be written explicitly in metric form and are discussed in detail in Appendices~\ref{sec:metric_cycle} and~\ref{sec:metric_line}. The constrained and unconstrained formulations of cycle of interactions are known not to be equivalent \cite{Hinterbichler:2012cn} and the existence of ghosts in the unconstrained vierbein was proven in \cite{deRham:2015cha}. We shall recover these results below, in particular the inequivalence is discussed in Appendix~\ref{sec:Leom}.

\subsubsection{Cycle of Interactions}
The most general non-derivative interactions between the two spin-2 fields suitable for describing the cycle of interactions and leading to the highest EFT cutoff reads \cite{deRham:2010ik,Hinterbichler:2012cn}:
\ba\label{act_nonder_general}
g_*^2 S_{\text{non-der}}&=& \frac{m^2\M^2}{4}\int \d^4x\ \sum_{n=0}^4\sum_{m=0}^{4-n} \kappa_{nm}\,\varepsilon\varepsilon I^{4-(n+m)}(E-I)^n(F-I)^m\,,
\ea
where $m$ and $M$ are some mass scales that depend on our choice of normalization. For the double--epsilon mass potentials we use the short-hand notations as, for example,
\be\label{Unm}
\begin{split}
\kappa_{21}\,\int \d^4 x\,\varepsilon\varepsilon I(E-I)^2(F-I)\equiv\kappa_{21}\,\int \varepsilon_{abcd}I^a\wedge(E-I)^b\wedge(E-I)^c\wedge(F-I)^d\,.
\end{split}
\ee
The cycle of interactions between the spin-2 fields $h_{\mu\nu}$ and $f_{\mu\nu}$ was defined in Section~\ref{sec:cycle_line} as the case when at linear level both fields are decoupled and start interacting at nonlinear level only.  Having the identification \eqref{identify0} in mind we see that this situation is described by the action \eqref{act_nonder_general} with $\kappa_{11}=0$. We find that it is more intuitive to keep the individual mass terms separate from the interactions between the dynamical vierbeins, $E^a$ and $F^a$. We therefore split the above action as:
\ba\label{act_nonder2}
g_*^2S_{\text{non-der}}&=&\frac{m_1^2M_1^2}{4}\int \d^4x\,\sum_{n=0}^4\kappa^{(1)}_{n}\,\mathcal U_{n}(I,E-I)+\frac{m_2^2M_2^2}{4}\int \d^4x\,\sum_{n=0}^4\kappa^{(2)}_{n}\,\mathcal U_{n}(I,F-I)\\
&+&\frac{m^2M^2}{4}\int \d^4x\,\sum_{n=1}^4\sum_{m=1}^{4-n}\kappa_{nm}\,\mathcal U_{nm}(I,E-I,F-I)\,,\nn
\ea
where $\kappa_{11}=0$ and we have introduced the notations
\be\label{Unml}
\begin{split}
&\mathcal U_{n}(I,E-I)\equiv \varepsilon\varepsilon I^{4-n}(E-I)^n\,,\qquad\\
&\mathcal U_{nm}(I,E-I,F-I)\equiv \varepsilon\varepsilon I^{4-(n+m)}(E-I)^n(F-I)^{m}\,.
\end{split}
\ee
The mass coefficients are in turn expressed as
\be\label{kappa single}
\kappa^{(1)}_n=\frac{m^2M^2}{m_1^2M_1^2}\kappa_{n0}\,,\qquad\kappa^{(2)}_n=\frac{m^2M^2}{m_2^2M_2^2}\kappa_{0n}\,,
\ee
and $m_i$ and $M_i$ are the corresponding masses and Planck masses of the two spin-2 fields. In the following we shall fix the mass scalings as
\be\label{scaling_mass}
m_2\equiv m\,,\quad \frac{m_1}{m_2}\equiv x\,;\qquad M_2\equiv\M\,,\quad\frac{M_1}{M_2}\equiv\gamma\,.
\ee
Let us emphasize again that written this way the action \eqref{act_nonder2} together with the Einstein--Hilbert terms \eqref{act_kinetic} can be used to describe a theory of two dynamical unconstrained vierbeins $E^a$ and $F^a$ coupled to a reference vierbein $I^a$. In the parlance of the standard vierbein formulation of theories of interacting spin-2 fields this is a cycle theory of three interacting vierbeins $E, F, I$. \\

A different theory, and the one which shall be our main focus, can be easily obtained from \eqref{act_nonder2} by imposing the symmetric vierbein conditions \eqref{sym_vierb_gen2} from the outset. These conditions together with the definition \eqref{identify0} allow us to make the identification with the `would-be' metric perturbations defined as
\begin{align}\label{metric_pert1}
g^{(1)}_{\mu\nu}\equiv\(\eta_{\mu\nu}+\frac{h_{\mu\nu}}{M}\)^2\,,\qquad g^{(2)}_{\mu\nu}\equiv\(\eta_{\mu\nu}+\frac{f_{\mu\nu}}{M}\)^2\,.
\end{align}
With this identification  the vierbein action \eqref{act_nonder2} together with the corresponding kinetic terms becomes a nonlinear action describing two massive spin-2 fields $h_{\mu\nu}$ and $f_{\mu\nu}$. The resulting action in the metric notation is given in Appendix~\ref{sec:metric_cycle}. For this cycle of interactions, there is no kinetic mixing in terms of the mass eigenstates. However it was shown in \cite{Hinterbichler:2012cn,deRham:2015cha} that the constrained and unconstrained formulations for this cycle are not equivalent and the unconstrained formulation carries a ghost. In what follows, we shall show that also the distinct constrained formulation carries a ghost, however it can still be used as a consistent EFT, and  with suitable technically natural rescalings the strong coupling scale remains as $\Lambda_3$.

\subsubsection{Line of Interactions}\label{sec:line_2}
Here we give the explicit action for the theory describing a line of interactions, as defined in Section~\ref{sec:cycle_line} and depicted on the right of Fig.~\ref{Fig:cycle}. In the metric language this means that only one of the dynamical metrics, $g^{(1)}$, couples to the reference metric $\eta$, while the second dynamical metric, $g^{(2)}$, only couples to $g^{(1)}$ both at linear and nonlinear level. This situation is most conveniently captured by the vierbein action written as \eqref{act_nonder} with only pairwise interactions between the three vierbeins, however, avoiding forming a cycle when mixing them. In doing so it is also more convenient for this analysis to keep working in terms of the mixed fields rather than the physical mass eigenstates which couple through both potential and kinetic interactions nonlinearly. \\

The theory we consider is described by constraining the coefficients in \eqref{act_nonder} as $\beta_{0m}=0$ and $\beta_{nm}=0$ for $n+m<4$ giving:
\be\label{act_nonder_line}
\begin{split}
g_*^2S_{\text{non-der}}=-\frac{m_1^2M_1^2}{2}\int \d^4x\,&\sum_{n=0}^4\frac{\tilde\beta^{(1)}_{n}}{n!(4-n)!}\,\mathcal U_{n}(I,E)\\
-\frac{m^2M^2}{2}\int \d^4x\,&\sum_{n=1}^4\frac{\tilde\beta^{(2)}_{n}}{n!(4-n)!}\,\mathcal U_{n}(E,F)\,,
\end{split}
\ee
with the coefficients $\tilde\beta^{(i)}_n$ related to $\beta_{nm}$'s through \eqref{rel_betas_1} and \eqref{rel_betas_2}.
Written in this form we recognize the first term as the standard mass term for a spin-2 field with vierbein $E$ with respect to a flat reference vierbein while the second term is the standard mass term for a spin-2 field with vierbein $F$ with respect to the vierbein $E$.\\

For later use, let us note that the action above can be rewritten in a different form by factoring out the determinant of the vierbein $E$ from the last three terms, as was done in \cite{Hinterbichler:2012cn}. Indeed, each particular interaction term between the vierbeins $E$ and $F$ can be expressed as, \emph{e.g.}
\be
\begin{split}
\varepsilon_{abcd}\,E^a\wedge F^b\wedge F^b\wedge F^c=&\ \d^4 x \left(\det E\right)\,\varepsilon_{\mu\nu\alpha\beta}\varepsilon^{\mu'\nu'\alpha'\beta'}\delta^\mu_{\mu'}\left(E^{-1}F\right)^\nu_{\nu'}\left(E^{-1}F\right)^\alpha_{\alpha'}\left(E^{-1}F\right)^\beta_{\beta'}\\
=&\ \d^4 x \left(\det E\right)\,\mathcal U_3\left(I, E^{-1}F\right)\,.
\end{split}
\ee
This leads to
\ba\label{act_nonder_line2}
g_*^2S_{\text{non-der}}=-\frac{m_1^2M_1^2}{2}\int \d^4x\,&&\sum_{n=0}^4\frac{\tilde\beta^{(1)}_{n}}{n!(4-n)!}\,\mathcal U_{n}(I,E)\\
-\frac{m^2M^2}{2}\int  \d^4x \left(\det E\right)\,&&\sum_{n=1}^4\frac{\tilde\beta^{(2)}_n}{n!(4-n)!}\,\mathcal U_n\left(I, E^{-1}F\right)\,.\nn
\ea
\paragraph{Symmetric Vierbeins:}
Let us emphasize  that  the vierbeins $E$ and $F$ can be viewed as constrained or unconstrained just as was considered for cycle of interactions. However, in this case, a slightly different version of the symmetric vierbein conditions is a direct consequence of the equations of motion, thus ensuring the equivalence between the unconstrained vierbein and metric formalisms. These conditions read:
\be\label{sym_vierb_gen3}
\eta E =(\eta E )^T\,,\qquad E^T \eta  F =F^T \eta E\,,
\ee
and imply that the vierbein $E$ satisfies the symmetric vierbein condition with respect to $I$ while $F$ obeys the symmetric condition with respect to $E$.
Crucially \eqref{sym_vierb_gen3} is not the same as \eqref{sym_vierb_gen2}. This means however that the previous identification, given in \eqref{identify0} for the cycle of interactions, of the two vierbeins with the metric perturbations defined as in \eqref{metric_pert1} is not suitable for this case anymore. Instead, we shall use
\be\label{def_pert_new}
g_{\mu\nu}^{(1)}\equiv (\eta_{\mu\nu}+\tilde h_{\mu\nu})^2\,,\qquad g_{\mu\nu}^{(2)}\equiv(g_{\mu\alpha}^{(1)}+\tilde f_{\mu\alpha})g^{\alpha\beta}_{(1)}(g_{\beta\nu}^{(1)}+\tilde f_{\beta\nu})\,.
\ee
Due to the symmetric vierbein condition \eqref{sym_vierb_gen3}, we can relate the vierbeins $E$ and $F$ to the metric perturbations $\tilde h_{\mu\nu}$ and $\tilde f_{\mu\nu}$ as:
\be\label{invEF}
E^a_\mu=\delta^a_\mu+\tilde h^a_\mu\,,\qquad \left(E^{-1}F\right)^\mu_\nu\equiv E_a^\mu F^a_\nu=\delta^\mu_\nu+\left(g^{-1}_{(1)}\tilde f\right)^\mu_\nu\,,
\ee
where we have introduced the inverse vierbein defined through $g^{\mu\nu}_{(1)}=E_a^\mu E_b^\nu\eta^{ab}$. This makes it straightforward to express the potential \eqref{act_nonder_line2} in the metric formalism. We give the result in Appendix~\ref{sec:metric_line}.\\

As mentioned previously, the theory considered in \eqref{act_nonder_line2} is an alternative approach in building the EFT for two interacting massive spin-2 fields based on the nonlinear completion of \eqref{eq:line}. In particular, it is different from the cycle of interactions described by the action \eqref{act_nonder2}. The key difference with the previous approach is that the second spin-2 field $\tilde f\mn$ cannot directly couple to the reference Minkowski metric and only acquires a mass through its mixing with $\tilde h\mn$. As eluded in subsection~\ref{sec:cycle_line} (see Eqn.~\eqref{eq:line}), in this representation the two fields $\tilde h\mn$ and $\tilde f\mn$ mix already at the linear level and therefore $\tilde h\mn$ and $\tilde f\mn$ are not the diagonalized mass eigenstates. Similarly the quantities $m, m_{1}$ here do not represent the physical masses and $M,M_{1}$ do not present the physical coupling scales. To determine the physical scales one should first diagonalize the free fields. This will imply further mixing between both fields nonlinearly, including kinetic term mixing. We shall refer to the construction \eqref{act_nonder_line2} as the `line of interactions' for massive spin-2 field or massive spin-2 field interacting theory that includes specific type of nonlinear kinetic mixing between the mass eigenstates.

\subsection{ADM Constraint Analysis}\label{adm}
As the first consistency check of the theories \eqref{act_nonder2} and \eqref{act_nonder_line2} we discuss the ADM phase space constraint analysis of these systems. In fact, the ADM analysis for both cycle and line theories in the unconstrained vierbein form was already done in \cite{Hinterbichler:2012cn,deRham:2015cha}. It was found that the theory of cycle interactions between unconstrained vierbeins is inequivalent to the theory of constrained vierbeins and that the former one carries a ghost. The theory of line interactions between unconstrained vierbeins was found to be healthy and equivalent to its metric formulation. In the following we therefore focus on the yet unexplored case of cycle of interactions between constrained vierbeins.\\

We present the analysis for the action written in the metric form given in \eqref{action} with two of the quartic interaction terms absent, \emph{i.e.} with $\kappa_{31}=\kappa_{13}=0$. In the absence of interactions between the two metrics, it is known that the ghost-free mass terms  ensure the existence of two Hamiltonian constraints---one for each metric (this was proven in \cite{deRham:2010kj} for specific dimensions and for special cases in four dimensions, generalizing the proof for generic cases in four-dimensions was performed in \cite{Hassan:2011hr}). In a Lorentz invariant theory this is sufficient to claim the absence of the BD ghosts. Thus, in a theory of two non-interacting massive spin-2 fields there are in total $2\times 5=10$ degrees of freedom. By allowing the interactions between the fields there is a potential danger of reintroducing either one or both of the BD ghosts leading to at most $2\times6=12$ degrees of freedom. To check whether this happens in our theory we count the number of degrees of freedom using the ADM language as often used in the context of massive gravity \cite{Creminelli:2005qk,deRham:2010kj,deRham:2011rn,Hassan:2011hr}.
The existence of a constraint would manifest itself as the vanishing of the determinant of the Hessian defined as \cite{deRham:2011rn}:
\begin{equation}
L_{\mu\nu}=\frac{\partial^2\H }{\partial N^{\mu}\partial N^{\nu}},
\end{equation}
where $\H$ is the Hamiltonian and $ N^{\mu}=\{N,M,N^{i},M^{i}\}$ are two sets of the canonical ADM variables of the lapses and shifts. The Hessian is thus an $8\times 8$ matrix. For a vanishing determinant there must be one or more zero eigenvalues signalling the presence of a constraint. The number of constraints coincides with the degree of the zero eigenvalue. The total number of degrees of freedom in this theory is then equal to $2\times6-(\#\text{ of constraints})$. \\

Another equivalent way to determine whether the system contains constraints is to first integrate over both sets of shifts (being auxiliary variables this is always possible), and then inspect whether the lapses are Lagrange multipliers generating two primary second class constraints (\ie whether the resulting Hamiltonian is linear in the lapses or possibly a combination of them). We do so perturbatively and use the following ADM decomposition for both metric perturbations \cite{Arnowitt:1962hi}:
\ba
g^{(1)}_{\mu\nu}\d x^\mu \d x^\nu &=&-(1+\epsilon \delta N)^2\d t^2
+(\delta_{ij}+\epsilon \gamma_{ij}) (\d x^i+\epsilon  \delta N^i\d t)(\d x^j+\epsilon  \delta N^j\d t)=(\eta_{\mu\nu}+h_{\mu\nu})^2  \d x^{\mu}\d x^{\nu}\,,\nn \\
g^{(2)}_{\mu\nu}\d x^\mu \d x^\nu  &=&-(1+\epsilon \delta M)^2\d t^2+(\delta_{ij}+\epsilon    \sigma_{ij})(\d x^i+\epsilon \delta  M^i\d t)(\d x^j+\epsilon \delta M^j\d t)=(\eta_{\mu\nu}+f_{\mu\nu})^2 \d x^{\mu}\d x^{\nu}\,,\nn
\ea
where $\epsilon$ has been introduced as a bookkeeping parameter, counting the order in perturbation theory.  The total Hamiltonian for the  theory can be schematically written as (ignoring overall $g_*^2$ factor)
\be
\begin{split}
 \H=\H_{\text{GR}}(h)+\H_{\text{GR}}(f)&-\frac{x^{2}\gamma^{2}m^{2}M^{2}}{4}\,\sum_{n=0}^4\kappa^{(1)}_n\,\U_n\left[\eta^{-1}h\right]-\frac{m^{2}M^{2}}{4}\,\sum_{n=0}^4\kappa^{(2)}_n\,\U_n\left[\eta^{-1}f\right]\\
&-\frac{m^{2} M^{2}}{4}\,\left(\kappa_{21}\,\L_{hhf}+\kappa_{12}\,\L_{hff}+\kappa_{22}\,\L_{hhff}\right),
\end{split}
\ee
where $\H_{\text{GR}}(h,f)$ are the Hamiltonians derived from Einstein-Hilbert action. They are linear in their respective lapses and shifts due to the diffeomorphism invariance,
\ba
\H_{\text{GR}}(h) &=& \epsilon^2 \delta N \H_0 (\gamma, p_\gamma) +\epsilon^2 \delta N_i R_h^i (\gamma, p_\gamma)\,,\\
\H_{\text{GR}}(f) &=& \epsilon^2 \delta M \H_0 (\sigma, p_\sigma) +\epsilon^2 \delta M_i R_f^i (\sigma, p_\sigma)\,,
\ea
where $\mathcal H_0$ and $R^i$ are non-trivial functions of the three-dimensional metric perturbations and their associated conjugate momenta as well as their spatial derivatives. They each start at first order in perturbations. Expressing $h\mn$ in terms of  the ADM perturbations
$\{\delta N, \delta N_i, \gamma_{ij}\}$ and similarly for $f\mn$ in terms of $\{\delta M, \delta M_i, \sigma_{ij}\}$, we can now solve for all the shifts $\delta N_i$ and $\delta M_i$ perturbatively in $\epsilon$ and substitute these auxiliary variables back in the Hamiltonian,
\ba
\H_{\rm red}(\gamma, p_\gamma, \sigma, p_\sigma, \delta N, \delta M)=\int \mathcal{D} \delta N_i \int \mathcal{D} \delta M_i\
\H (\gamma, p_\gamma, \sigma, p_\sigma, \delta N, \delta M, \delta N_i, \delta M_i)\,.
\ea
Up to quintic order in $\epsilon$, the resulting Hamiltonian remains linear in both lapses (with no mixing between both lapses). It therefore follows that both lapses generate one primary second-class constraint each\footnote{In the case of a single field, for a Lorentz invariant and parity preserving system, the existence of a primary second class constraint automatically implies the existence of a secondary constraint removing a whole physical degree of freedom since there cannot be half-propagating number of degrees of freedom, \cite{deRham:2014zqa}. In this two-field system, one may \emph{a priori} worry that both half-degrees of freedom could combine to give a full degree of freedom. However, since a secondary constraint exists related to each of the lapses in the case of the one-field system, one is guaranteed that the primary constraint does not commute with the Hamiltonian. Adding a mixing term between each metric cannot change this result and the existence of secondary constraints is therefore automatic.}, and up to that order in perturbation theory the constraints are satisfied. In particular, it implies that the trivial vacuum solution $\langle h\mn \rangle = \langle f\mn \rangle =0 $ is ghost-free (since the constraints are satisfied at second order in $\epsilon$).

However, upon following the same procedure to higher order we see the following operator entering the reduced Hamiltonian,
\ba
\H_{\rm red} \widetilde{\supset} - \frac{(1+x)\epsilon^6}{32 m^2 x^3}\(\delta M \, R_h^i - x \delta N \, R_f^i\)\(\delta M \, R_h^j - x \delta N \, R_f^j\)\(\zeta_i^k \zeta_{kj}-2 \zeta_{ij}\zeta^k_k+\zeta^k_k \zeta^\ell_\ell \delta_{ij}\)\,,
\ea
where we have defined the following tensor $\zeta_{ij}=(\kappa_{21} \gamma_{ij}+\kappa_{12} \sigma_{ij})/2$. Hence as soon as the cubic couplings between both spin-2 fields are introduced, \emph{i.e.} $\kappa_{21},\kappa_{12}\ne 0$, the resulting reduced Hamiltonian is quadratic in both lapses, spoiling both\footnote{Indeed one can easily see from the way the lapses enter, quadratically in the vector $\delta M \, R_h^i - x \delta N \, R_f^i$, that one could never simply redefine one of the lapses so as to keep the Hamiltonian linear in the second lapse. Hence both constraints are ruined.} second-class constraints.
Nonlinearly the theory therefore carries not one but two BD ghosts, and we refer to Section \ref{sec:cycle} and \ref{sec:strong} to determine the scale at which those ghosts would come in. However we emphasize here that so long as we treat the theory as an EFT, and the mass of the ghosts as the cutoff, there is no issue working with this EFT. The existence of ghostly operators is simply signalling the failure of the EFT at the cutoff scale \cite{deRham:2014fha,deRham:2014naa}.
In principle we could set $\kappa_{12}=\kappa_{21}=0$ and just have both fields interacting through the quartic interactions, however this interaction also carries a ghost. The above ADM analysis can also easily be extended to include $\kappa_{31},\kappa_{13}$ with similar conclusions.

\section{Decoupling Limit for the Cycle of Interactions}\label{sec:cycle}
In this Section we work out the decoupling limit Lagrangian of two interacting spin-2 fields described by the action for cycle interactions given by the non-derivative interactions \eqref{act_nonder2} together with the Einstein--Hilbert kinetic terms \eqref{act_kinetic}. We start by presenting the general framework and some  technical points that will be equally useful also for the decoupling limit of line interactions in Section~\ref{sec:line}. We generalize our conclusions to arbitrary number of interacting spin-2 fields in Section~\ref{sec:multi}. \\

We take a double scaling limit
\be\label{DL}
m\to 0\,,\quad \M\to\infty\,,
\ee
while maintaining the lowest strong coupling scale that arises fixed. This will automatically zoom in on the dominant interactions in the theory. At the moment it is as yet unclear, which scale this would be so we will make no assumptions, and for every family of interactions we focus on the operators that arise at the lowest possible energy scale. We will use the standard notation
\ba
\label{eq:Lambdan}
\Lambda_n\equiv \(m^{n-1}M\)^{1/n}\,,
\ea
in particular, $\Lambda_3=(m^2M)^{1/3}$. We derive the full nonlinear decoupling limit for both the helicity-0/helicity-2 and helicity-0/helicity-1 sectors of interactions by using the vierbein \stu language presented below. This, in addition to the standard diffeomorphism St\"uckelberg fields, also includes Lorentz \stu fields that can either be constrained by the symmetric vierbein conditions (``constrained'' vierbein) or treated as auxiliary fields (``unconstrained" vierbein).  We shall state clearly the difference between the two approaches when appropriate. \\

For the case of cycle interactions we find that the helicity-0/helicity-1 interactions break perturbative unitarity at a scale lower than $\Lambda_3$ in general. These new interactions are responsible for the two BD ghosts in the two spin-2 system that are absent when interactions are switched off between the two spin-2 fields.
We discuss the new scale and how to make sense of a $\Lambda_3$ EFT  in Section~\ref{sec:strong}.

\subsection*{St\"uckelberg Fields}
Beginning with the non-derivative cycle interactions given in the vierbein form in Eq.~\eqref{act_nonder2}, together with the kinetic terms \eqref{act_kinetic}, we now restore the local Lorentz invariance and the diffeomorphism invariance of the action. For this we follow closely the formalism of \cite{Ondo:2013wka} and  introduce two sets of St\"uckelberg fields $\phi^a$ and $\psi^a$ for diffeomorphisms and two sets of fields $\Lambda^a\,_b$ and $\Gamma^a\,_b$ for the Lorentz transformations. In the action this amounts to replacing
\be\label{diff_stuck}
E^a_\mu\to\tilde E^a_\mu=\Lambda^a\,_bE^b_c(\phi)\partial_\mu\phi^c\,,\qquad F^a_\mu\to\tilde F^a_\mu=\Gamma^a\,_bF^b_c(\psi)\partial_\mu\psi^c\,.
\ee
In order to compute the decoupling limit interactions to all orders, we then use the following decomposition for the two helicity-2 modes and the two sets of St\"uckelberg fields:
\be\label{stuck}
\begin{split}
&E^a_\mu=\delta^a_\mu+\frac{h^a_\mu}{\M}\,,\qquad F^a_\mu=\delta^a_\mu+\frac{f^a_\mu}{\M}\,,\\
&\Lambda^a\,_b=e^{\hat\omega^a\,_b}=\delta^a_b+\hat\omega^a\,_b+\frac{1}{2}\hat\omega^a\,_c\,\hat\omega^c\,_b+\dots\,,\qquad\Gamma^a\,_b=e^{\hat\sigma^a\,_b}\,,\\
&\hat\omega^a\,_b=\frac{\omega^a\,_b}{\Lambda_2^2}\,,\qquad\hat\sigma^a\,_b=\frac{\sigma^a\,_b}{\Lambda_2^2}\,,\\
&\phi^a=x^a+\frac{A^a}{\Lambda_2^2}+\frac{\partial^a\pi}{\Lambda_{3}^3}\,,\qquad\psi^a=x^a+\frac{B^a}{\Lambda_2^2}+\frac{\partial^a\chi}{\Lambda_{3}^3}\,, \\
&\hat \Pi^\mu_\nu\equiv \frac{\partial^\mu\partial_\nu\pi}{\Lambda_3^3} \, , \qquad \hat{\mathbb X}^\mu_\nu\equiv \frac{\partial^\mu\partial_\nu\chi}{\Lambda_3^3} \, .
\end{split}
\ee
After inserting this decomposition in $\tilde E^a_\mu =\Lambda^a\,_bE^b_c(\phi)\partial_\mu\phi^c$, we have
\be
\begin{split}
\tilde E^a\,_\mu=&\, \delta^a_\mu+\hat \Pi^a_\mu+\frac{1}{\Lambda_2^2}\left(\partial_\mu A^a+\omega^a\,_\mu+\omega^a\,_b\hat \Pi^b_\mu\right)+\frac{1}{M}\left(h^a_\mu(\phi)+h^a_\nu(\phi)\hat \Pi^\nu_\mu\right)\\
+&\frac{1}{\Lambda_2^4}\left(\omega^a\,_b\partial_\mu A^b+\frac{1}{2}\omega^a\,_b\,\omega^b\,_\mu+\frac{1}{2}\omega^a\,_b\,\omega^b\,_c\,\hat \Pi^c_\mu\right)+\mathcal O\left(\frac{1}{M\Lambda_2^2},\frac{1}{\Lambda_2^6}\right)\,.
\end{split}
\ee
We have neglected terms further suppressed by powers of $\Lambda_2$ as we already know from the standard double--epsilon structure of the helicity-2/helicity-0 interactions that the strong coupling scale is at most $\Lambda_3$. Hence in the decoupling limit we will necessarily have $\Lambda_2\to \infty$.

Schematically, the above decomposition of the vierbein takes the form (we drop the tilde in the following):
\be\label{E1}
\begin{split}
E = &\, I +\hat \Pi+\frac{1}{\Lambda_2^2}\left(\partial A+\omega+\omega\cdot\hat \Pi\right)+\frac{1}{M}\left(h(\phi)+h(\phi)\cdot\hat \Pi\right)\\
+& \frac{1}{\Lambda_2^4}\left(\omega\cdot\partial A+\frac{1}{2}\omega\cdot\omega+\frac{1}{2}\omega\cdot\omega\cdot\hat \Pi\right)+\mathcal O\left(\frac{1}{\M\Lambda_2^2}\right)\,.
\end{split}
\ee
Similarly, we decompose also the other vierbein according to \eqref{stuck}. This gives
\be\label{F1}
\begin{split}
F = &\, I +\hat {\mathbb X}+\frac{1}{\Lambda_2^2}\left(\partial B+\sigma+\sigma\cdot\hat {\mathbb X}\right)+\frac{1}{M}\left(f(\psi)+f(\psi)\cdot\hat {\mathbb X}\right)\\
+&\frac{1}{\Lambda_2^4}\left(\sigma\cdot\partial B+\frac{1}{2}\sigma\cdot\sigma+\frac{1}{2}\sigma\cdot\sigma\cdot\hat {\mathbb X}\right)+\mathcal O\left(\frac{1}{\M\Lambda_2^2}\right)\,.
\end{split}
\ee

\subsection*{Constrained Vierbein}
In the vierbein decomposition \eqref{stuck}, the symmetric vierbein conditions \eqref{sym_vierb_gen2}  become constraints on the Lorentz St\"uckelberg fields $\omega$ and $\sigma$~\cite{Ondo:2013wka}. Indeed, \eqref{sym_vierb_gen2} written in index notations becomes:
\be\label{sym_vierb_1}
\begin{split}
&\partial_a A_b-\partial_bA_a=2\omega_{ab}-\left(\omega_{da}\hat \Pi^d_b-\omega_{db}\hat \Pi^d_a\right)\,,\\
&\partial_a B_b-\partial_bB_a=2\sigma_{ab}-\left(\sigma_{da}\hat {\mathbb X}^d_b-\sigma_{db}\hat {\mathbb X}^d_a\right)\,.
\end{split}
\ee
By introducing the short-hand notation $(\hat \Pi\cdot\omega)_{\[ab\]}\equiv\hat \Pi^c_a\omega_{cb}-\hat \Pi^c_b\omega_{ca}$ and defining $F_{ab}\equiv\partial_a A_b-\partial_b A_a$ and $J_{ab}\equiv\partial_aB_b-\partial_bB_a$ these can be written as
\be\label{sym_vierb_2}
2\omega_{ab}=F_{ab}-(\hat \Pi\cdot\omega)_{\[ab\]}\,,\qquad 2\sigma_{ab}=J_{ab}-(\hat {\mathbb X}\cdot\sigma)_{\[ab\]}\,.
\ee

We wish to emphasize that when imposing these conditions we are implicitly making the choice to work in the ``metric" language (even though we may not necessarily be dealing with a gravitational theory). In this case, using the vierbein inspired decomposition \eqref{stuck} is merely a convenient choice of variables for what are otherwise 10 component constrained symmetric vierbeins. However, the non-derivative interactions as given in Eq. \eqref{act_nonder2} (and, similarly, in Eq.~\eqref{act_nonder_line} for the line of interactions) are written in the unconstrained vierbein form with each of the vierbeins $E^a_\mu$ and $F^a_\mu$ having $16$ \emph{a priori} arbitrary components that can be treated as auxiliary fields. Only when the symmetric vierbein conditions \eqref{sym_vierb_gen2} (or \eqref{sym_vierb_gen3} for the line interactions) are imposed do we recover the metric formulation of the interactions. As we shall see below, most of our analysis and main results are in fact independent on whether we work with the constrained or unconstrained vierbeins.

Finally, let us remark that the symmetric vierbein condition \eqref{sym_vierb_gen2} also implies that the following combination is symmetric:
\be\label{scalar_sym_vierb}
h_{a\mu}(\phi)+h_{a\nu}(\phi)\hat \Pi^\nu_\mu=h_{\mu a}(\phi)+h_{\mu\nu}(\phi)\hat \Pi^\nu_a\,.
\ee
We shall see in subsection~\ref{sec:hel02} that this condition is automatically imposed by the decoupling limit theory.

\subsection{Helicity-two/Helicity-zero Sector}\label{sec:hel02}
In order to study the decoupling limit interactions between the helicity-0 and helicity-2 modes it is sufficient to truncate the decompositions \eqref{E1} and \eqref{F1} of the two vierbeins as
\be\label{EF_scalar}
\begin{split}
E = &I +\frac{\Pi}{\Lambda_3^3}+\frac{1}{M}\left(h\left[x+\frac{\partial\pi}{\Lambda_3^3}\right]+\frac{h\left[x+\frac{\partial\pi}{\Lambda_3^3}\right]\cdot\Pi}{\Lambda_3^3}\right)\,,\\
F = &I +\frac{ {\mathbb X}}{\Lambda_3^3}+\frac{1}{M}\left(f\left[x+\frac{\partial\chi}{\Lambda_3^3}\right]+\frac{f\left[x+\frac{\partial\chi}{\Lambda_3^3}\right]\cdot  {\mathbb X}}{\Lambda_3^3}\right)\,,
\end{split}
\ee
where the square brackets denote the argument of $h$ and $f$. This decomposition captures all interactions that survive in the decoupling limit in this sector. Note that we have reintroduced the $\Lambda_3^3$ scale by redefining $\Pi=\hat \Pi \Lambda_3^3$ and $\mathbb X=\hat{\mathbb X} \Lambda_3^3$, and use the shorthand $(h\cdot\Pi)^a_\mu\equiv h^a_\nu\Pi^\nu_\mu$.

\subsubsection*{Self--Interactions}
In the absence of interactions between the two vierbeins, the action \eqref{act_nonder2} reduces to two decoupled massive spin-2 fields. Let us therefore start by reviewing the decoupling limit of a single massive graviton described by the action
\ba
\label{drgt_vierb}
\begin{split}
g_*^2 S=\frac{\gamma^2M^2}{8}\int &\varepsilon _{abcd}\,E^a\wedge E^b\wedge R^{cd}[E] \\
+\frac{m^2\M^2}{4}\gamma^2x^2\int \Bigg[ & \varepsilon_{abcd}I^a\wedge I^b\wedge(E^c-I^c)\wedge(E^d-I^d)\\
+\kappa_3^{(1)}&\varepsilon_{abcd}I^a\wedge(E^b-I^b)\wedge(E^c-I^c)\wedge(E^d-I^d) \\
+\kappa_4^{(1)}&\varepsilon_{abcd}(E^a-I^a)\wedge(E^b-I^b)\wedge(E^c-I^c)\wedge(E^d-I^d)
\Bigg]\,.
\end{split}
\ea
It is known that in the decoupling limit the action above contains seemingly dangerous operators for the scalar field $\pi$ that arise as:
\be\label{tot_der}
\frac{4g_*^2}{\gamma^2 x^2}\L_{\text{DL}}\supset\frac{1}{m^2}\varepsilon\varepsilon II\Pi\Pi
+\frac{\kappa^{(1)}_3}{\Lambda_{5}^5}\varepsilon\varepsilon I\Pi\Pi\Pi+\frac{\kappa^{(1)}_4}{\Lambda_{4}^8}\varepsilon\varepsilon\Pi\Pi\Pi\Pi\,,
\ee
where $\Lambda_{5}^5\equiv m^4\M$, $\Lambda_4^4\equiv m^3 M$ and we have also introduced the shorthand notations
\be
\varepsilon\varepsilon I\Pi\Pi\Pi\equiv\varepsilon_{\mu\nu\alpha\beta}\varepsilon^{\mu'\nu'\alpha'\beta'}\delta^\mu_{\mu'}\Pi^\nu_{\nu'}\Pi^{\alpha}_{\alpha'}\Pi^\beta_{\beta'}\,.
\ee
One can easily check that these contributions are actually total derivatives and there are therefore no physical interactions neither at the scale $\Lambda_5$ nor $\Lambda_4$. The first interactions occur at the scale $\Lambda_3$ and are given by \cite{deRham:2010ik,deRham:2010kj,deRham:2011qq}
\ba
g_*^2 \L_{\text{DL}}=-\gamma^2h^{\mu\nu}\mathcal E^{\alpha\beta}_{\mu\nu}h_{\alpha\beta}
+\frac{\gamma^2x^2}{4}\varepsilon\varepsilon\left(h\[..\]+\frac{h\[..\]\cdot\Pi}{\Lambda_3^3}\right)
\(2\Pi+\frac{3\kappa_3}{\Lambda_3^3}\Pi^2+\frac{4\kappa_4}{\Lambda^6_3}\Pi^3
\)\,,
\ea
where the \Lic operator is defined in \eqref{quadratic}. The notation $h\[..\]$ reminds us that the field $h_{\mu\nu}$ is evaluated at $h\left[x+\frac{\partial\pi}{\Lambda_3^3}\right]$.

\subsubsection*{Mixed--Interactions}\label{sec:hel02int}
In the presence of interactions between the two spin-2 fields there are new interactions mixing the two helicity-0 modes. Similarly, naively the leading order ones seem to appear at the scales $\Lambda_5$ and $\Lambda_{4}$:
\be\label{tot_der2}
\begin{split}
g_*^2 \L_{\text{DL}}\supset\frac{1}{4}\left(\kappa_{21}\frac{1}{\Lambda_{5}^5}\varepsilon\varepsilon I\mathbb X\Pi\Pi+\kappa_{12}\frac{1}{\Lambda_{5}^5}\varepsilon\varepsilon I\mathbb X\mathbb X\Pi
+\kappa_{22}\frac{1}{\Lambda_{4}^8}\varepsilon\varepsilon \mathbb X\mathbb X\Pi\Pi +\kappa_{31}\frac{1}{\Lambda_{4}^8}\varepsilon\varepsilon \mathbb X \Pi \Pi\Pi +\kappa_{13}\frac{1}{\Lambda_{4}^8}\varepsilon\varepsilon \mathbb X\mathbb X\mathbb X \Pi\right)
\end{split}
\ee
where, as before, $\Pi_{\mu\nu}\equiv \partial_\mu\partial_\nu\pi$ and $\mathbb X_{\mu\nu}\equiv\partial_\mu\partial_\nu\chi$. One can easily verify that all these interactions are total derivatives, \emph{e.g.}
\be
\varepsilon\varepsilon I\mathbb X\Pi\Pi\equiv\varepsilon_{\mu\nu\alpha\beta}\varepsilon^{\mu'\nu'\alpha'\beta}\mathbb X^\mu_{\mu'}\Pi^\nu_{\nu'}\Pi^{\alpha}_{\alpha'}
=\partial^\mu\left(\varepsilon_{\mu\nu\alpha\beta}\varepsilon^{\mu'\nu'\alpha'\beta}\partial_{\mu'}\chi
\Pi^\nu_{\nu'}\Pi^{\alpha}_{\alpha'} \right)\,.
\ee
As a result, the total decoupling limit for the helicity-zero and helicity-two interactions only has the following $\Lambda_3$ interactions:
\ba
\begin{split}\label{totalDL20A}
g_*^2 \L_{\text{DL}}=&-\gamma^2h^{\mu\nu}\mathcal E^{\alpha\beta}_{\mu\nu}h_{\alpha\beta}
+\frac{1}{4}\left(h^{a}_{\mu}\[..\]+\frac{h^{a\nu}\[..\]\Pi_{\nu\mu}}{\Lambda_3^3}\right)X_{a}^{\mu}[\Pi,\mathbb X]\\
&-f^{\mu\nu}\mathcal E^{\alpha\beta}_{\mu\nu}f_{\alpha\beta}
+\frac{1}{4}\left(f^{a}_{\mu}\[..\]+\frac{f^{a\nu}\[..\]\mathbb X_{\nu\mu}}{\Lambda_3^3}\right)Y_{a}^{\mu}[\Pi,\mathbb X]\, ,
\end{split}
\ea
where $X^\mu_a$ can be split as
\ba
X_{a}^{\mu}[\Pi,\mathbb X]&\equiv&\,(X^{(1)})^\mu_a+(X^{(2)})^\mu_a+(X^{(3)})^\mu_a \, ,
\ea
with the different $X^{(n)}$'s defined as:
\be
\begin{split}
&(X^{(1)})^\mu_a=2\gamma^2x^2\left(\varepsilon\varepsilon II\Pi\right)_{a}^{\mu}\,\\
&(X^{(2)})^\mu_a=\gamma^2x^2\frac{3\kappa_3^{(1)}}{\Lambda_3^3}\left(\varepsilon\varepsilon I\Pi\Pi\right)_{a}^{\mu}+\frac{2\kappa_{21}}{\Lambda_3^3}\left(\varepsilon\varepsilon I\mathbb X\Pi\right)_{a}^{\mu}+\frac{\kappa_{12}}{\Lambda_3^3}\left(\varepsilon\varepsilon I\mathbb X\mathbb X\right)_{a}^{\mu}\,,\\
&(X^{(3)})^\mu_a=\gamma^2x^2\frac{4\kappa_4^{(1)}}{\Lambda_3^6}\left(\varepsilon\varepsilon \Pi\Pi\Pi\right)_{a}^{\mu} +\frac{2\kappa_{22}}{\Lambda_{3}^6}\left(\varepsilon\varepsilon\mathbb X\mathbb X\Pi\right)_{a}^{\mu}+\frac{3 \kappa_{31}}{\Lambda_{3}^6}\left(\varepsilon\varepsilon\mathbb X\Pi \Pi\right)_{a}^{\mu}+\frac{ \kappa_{13}}{\Lambda_{3}^6}\left(\varepsilon\varepsilon\mathbb X\mathbb X\mathbb X \right)_{a}^{\mu}\,.
\end{split}
\ee
We also have similarly introduced
\be
Y_{a}^{\mu}[\Pi,\mathbb X]\equiv X_{a}^{\mu}\left[\Pi\leftrightarrow\mathbb X\,;\,\kappa_n^{(1)}\leftrightarrow\kappa_n^{(2)}\,;\,\kappa_{nm}\text{ with }n\leftrightarrow m\,;\,\gamma=x=1\right]\, ,
\ee
and the corresponding splitting in $Y^{(n)}$'s. We note that it is the combination
\be\label{new_h}
h^{a}_{\mu}\left[x+\frac{\partial\pi}{\Lambda_3^3}\right]+\frac{h^{a\nu}\left[x+\frac{\partial\pi}{\Lambda_3^3}\right]\Pi_{\nu\mu}}{\Lambda_3^3} \, ,
\ee
that mixes with $X^\mu_a$ and not $h_{\mu\nu}(x)$ as in the standard derivation of the decoupling limit (see, \emph{e.g.}~\cite{deRham:2010ik}). \\

There are two subtleties that need to be clarified regarding this. First, the above combination is only symmetric if the symmetric vierbein condition \eqref{scalar_sym_vierb} is imposed and one might wonder whether the symmetric vierbein condition is anyhow necessary for these interactions to be healthy. However, due to the fact that $X_{\mu a}$ is itself symmetric these interactions automatically pick up the symmetric part of \eqref{new_h}. 

Second, the field $h$ is seemingly non-local due to the fact that it is evaluated at $x+\frac{\partial\pi}{\Lambda_3^3}$. This can be undone by the Galileon duality transformation \cite{Fasiello:2013woa,Curtright:2012gx,deRham:2013hsa}. This amounts to redefining the coordinate as $x'=x+\partial\pi(x)/\Lambda_3^3$ and introducing the dual Galileon field $\pi'$ through the inverse transformation $x=x'+\partial'\pi'(x')/\Lambda_3^3$ (and similarly for $\chi$), as was pointed out in \cite{Fasiello:2013woa,Curtright:2012gx,deRham:2013hsa}. The relevant steps are closely analogous to those performed in Section~\ref{sec:biGalileon}. However there is a crucial difference between the line and cycle theories. On comparing our result for the cycle theories \eqref{totalDL20A} with that for the line theories \eqref{DL_sc_1} and \eqref{DL_sc_2}, we see that in the latter case $f$ only couples directly to $\chi$, whereas in \eqref{totalDL20A} both helicity-2 modes couple to both helicity-0 modes. One consequence of this is that in attempting to repeat the procedure outlined in Section~\ref{sec:biGalileon}, it is necessary to perform two separate Galileon duality transformations to render both $h$ and $f$ as functions of the same $x$. The transformation on $f$ taking the form $x'=x+\partial\chi(x)/\Lambda_3^3$. On doing so the $\chi$ and $\pi$ interactions generated from mixing with $h$ will be non-locally related to those that arise from $f$ by virtue of effectively being in two different duality frames. Thus even when we can demix the helicity-2 mode fluctuations, the resulting effective two scalar Lagrangian will be non-local, \emph{i.e.} it will contain an infinite number of derivative interactions. This is in stark contrast to the line theories where the resulting Lagrangian is that of a (local) bi-Galileon as discussed in Section~\ref{sec:biGalileon}. Nevertheless, since the scale of non-locality is the $\Lambda_3$ scale, there remains no inconsistency in viewing these interactions perturbatively.
Hence, putting the helicity-1 modes aside, this cycle theory decoupling limit, as given in \eqref{totalDL20A}, is exact and has no higher order contributions, and by itself is a consistent EFT for helicity-2 field and helicity-0 modes with a strong coupling scale $\Lambda_{3}$ which is the highest scale possible in a Lorentz invariant theory of (hard) massive gravity. The real difficulties arise in fact in the helicity-1 sector, which we turn to now.

\subsection{Helicity-one/Helicity-zero Sector}\label{sec:hel1}
In order to derive the decoupling limit interactions involving the helicity-1 fields we follow the formalism developed in \cite{Ondo:2013wka} (see also \cite{Gabadadze:2013ria}). We shall work in terms of the vierbeins defined above in \eqref{E1} and \eqref{F1}. As will become clear below, most of our analysis and main results are in fact independent on whether we work with the constrained or unconstrained vierbeins. In this subsection we focus on the constrained vierbein theory. We discuss the unconstrained case later in Section~\ref{sec:unconstrained}.

For interactions between the helicity-0 and helicity-1 fields, the relevant St\"uckelberg expansion of the vierbeins simplifies to:
\ba\label{E2}
E &=& I +\hat{\Pi}+\frac{1}{\Lambda_2^2}\left(\partial A+\omega+\omega\cdot\hat{\Pi}\right)
+\frac{1}{\Lambda_2^4}\left(\omega\cdot\partial A+\frac{1}{2}\omega\cdot\omega+\frac{1}{2}\omega\cdot\omega\cdot\hat{\Pi}\right)+\mathcal O\left(\frac{h}{M}\right)\,,\\
\label{F2}
F &=& I +\hat{\mathbb X}+\frac{1}{\Lambda_2^2}\left(\partial B+\sigma+\sigma\cdot\hat{\mathbb X}\right)
+\frac{1}{\Lambda_2^4}\left(\sigma\cdot\partial B+\frac{1}{2}\sigma\cdot\sigma+\frac{1}{2}\sigma\cdot\sigma\cdot\hat{\mathbb X}\right)+\mathcal O\left(\frac{f}{\M}\right)\,,
\ea
where we have neglected the subleading $h^a_\mu(\phi)/\M$ and $f^a_\mu(\psi)/M$ corrections. Indeed the helicity-2 modes are $1/\M$ suppressed and can be set to zero when discussing the Lorentz \stu fields and the helicity-1/helicity-0 decoupling limit.

\subsubsection*{Self--Interactions}
Using the above decompositions \eqref{E2}, \eqref{F2} for the self-interaction terms in \eqref{act_nonder2} gives the following:
\ba\label{DLvec1}
g_*^2 \L_{\text{DL, self}}&=&\frac{\gamma^2x^2}{4} \,
 \varepsilon\varepsilon \Bigg\{
\left[I^2+3\kappa_3^{(1)} I \hat{\Pi} + 6\kappa_4^{(1)}\hat{\Pi} \hat{\Pi} \right](\partial A+\omega+\omega\cdot\hat{\Pi})^2\\
&&\phantom{\frac{\gamma^2x^2}{4} \,
 \varepsilon\varepsilon }
+ \left[2I^2+3\kappa_3^{(1)} I \hat{\Pi} + 4\kappa_4^{(1)}\hat{\Pi} \hat{\Pi} \right]\hat{\Pi}  \left(\omega\cdot A+\frac{1}{2}\omega\cdot\omega+\frac{1}{2}\omega\cdot\omega\cdot\hat{\Pi}\right)
\Bigg\}\nn \\
&+&\ \left(1\leftrightarrow 2,\hat{\Pi}\leftrightarrow\hat{\mathbb X},A\leftrightarrow B,\omega\leftrightarrow\sigma\,,\gamma=x=1\right)\,.\nn
\ea
A few remarks are in order. First, let us point out that only terms at most quadratic in $\omega$ (or $\sigma$) appear in the decoupling limit action. This is due to the fact that any higher powers of $\omega$ are suppressed by powers of $\Lambda_2$ and we already know from the helicity-2/helicity-0 interactions that in the decoupling limit one will necessarily have $\Lambda_2\to \infty$. Second, we observe that the leading interactions of the Lorentz \stu fields arise, in fact, from the quadratic mass term as follows
\ba\label{U20}
m^2\M^2\int \varepsilon_{abcd} I^a\wedge I^b\wedge (E^c-I^c)\wedge (E^d-I^d)
=2m \M \int  \d^4x\,\varepsilon\varepsilon I^2\hat{\Pi}\left(\partial A+\omega+\omega\cdot\hat{\Pi}\right) +\mathcal O(1)\,.\qquad
\ea
It is straightforward to see that these potentially dangerous higher derivative terms  are actually harmless. Indeed, the first term is a total derivative while all the rest vanish because of the antisymmetry of $\omega$. Indeed, one can write
\begin{align}
&\varepsilon\varepsilon I^2\hat{\Pi}\omega=2\left(\[\hat{\Pi}\]\[\omega\]-\[\hat{\Pi}\omega\]\right)\,,\\
&\varepsilon\varepsilon I^2\hat{\Pi}\omega\cdot\hat{\Pi}=2\left(\[\hat{\Pi}\]\[\hat{\Pi}\omega\]-\[\hat{\Pi}^2\omega\]\right)\,, \label{badequation}
\end{align}
where as before we use the square brackets to denote the traces as $\[\hat{\Pi}\]=\hat{\Pi}^a_a\,,\[\hat{\Pi}\omega\]=\hat{\Pi}^a\,_b\,\omega^b\,_a,$ etc. All the contractions above vanish. The cubic and quartic mass terms give similar trace structures, all of which vanish. We show below that this holds only in the case when interactions between the vierbeins $E^a\,_\mu$ and $F^a\,_\mu$ are absent.\\

Finally, as we show in Appendix~\ref{sec:Leom}, in the absence of interactions between the two spin-2 fields, if we were in the unconstrained vierbein formalism and were to consider the Lorentz St\"uckelbergs $\omega^a\,_b$ (and $\sigma^a\,_b$) as auxiliary fields, their equations of motion would be independent on the self-interaction parameters $\kappa_{3,4}^{(i)}$ and would be equivalent to the symmetric vierbein conditions \eqref{sym_vierb_1}. Indeed, the equation of motion for $\omega$ derived from the quadratic mass term and using the variation law
\be
\frac{\delta\omega^{ab}}{\delta\omega^{cd}}=\frac{1}{2}\delta^{ab}_{cd}\equiv\frac{1}{2}(\delta^a_c\delta^b_d-\delta^a_d\delta^b_c)
\ee
reads
\be
\begin{split}
\left(\delta^c_a-\[\hat{\Pi}\]\delta^c_a+2\hat{\Pi}^c_a\right)\bigg(2\omega_{cb}-F_{cb}+(\hat{\Pi}\cdot\omega)_{\[cb\]}\bigg)-(a\leftrightarrow b)=0\,.
\end{split}
\ee
This demonstrates the equivalence between the constrained and unconstrained vierbein formalism in the absence of interactions between the two spin-2 fields.

\subsubsection*{Mixed--Interactions}\label{sec:dl_vierb}
The mixed interaction terms give the following leading order interactions
\be\label{DLvec2}
\begin{split}
g_*^2 \L_{\text{DL, int}}=\frac{m \M}{4}  \varepsilon\varepsilon \bigg[& \kappa_{21} I\left(2\hat{\Pi}\hat{\mathbb X}\omega+\hat{\Pi}^2\sigma+2\hat{\Pi}\hat{\mathbb X}\,\omega\cdot\hat{\Pi}+\hat{\Pi}^2\sigma\cdot\hat{\mathbb X}\right) \\
&+ \kappa_{12} I\left(2\hat{\Pi}\hat{\mathbb X}\sigma+\hat{\mathbb X}^2\omega+2\hat{\Pi}\hat{\mathbb X}\,\sigma\cdot\hat{\mathbb X}+\hat{\mathbb X}^2\omega\cdot\hat{\Pi}\right)\\
&+ 2\kappa_{22} \left(\hat{\Pi}\hat{\mathbb X}^2\omega+\hat{\Pi}^2\hat{\mathbb X}\sigma+\hat{\Pi}\hat{\mathbb X}^2\,\omega\cdot\hat{\Pi}+\hat{\Pi}^2\hat{\mathbb X}\,\sigma\cdot\hat{\mathbb X}\right)\\
&+\kappa_{31} \left(\hat{\Pi}^3\sigma+3\hat{\Pi}^2\hat{\mathbb X}\omega+\hat{\Pi}^3\sigma\cdot\hat{\mathbb X}+3\hat{\Pi}^2\hat{\mathbb X}\,\omega\cdot\hat{\Pi}\right)\\
&+\kappa_{13} \left(\hat{\mathbb X}^3\omega+3\hat{\mathbb X}^2\hat{\Pi}\sigma+\hat{\mathbb X}^3\omega\cdot\hat{\Pi}+3\hat{\mathbb X}^2\hat{\Pi}\,\sigma\cdot\hat{\mathbb X}\right)\bigg]\,.
\end{split}
\ee
We have neglected the total derivative interactions in the above expression. Together with the self-interaction terms the total decoupling limit action for the vector fields and Lorentz \stu fields is given by
\be
\L_{\text{DL,vec}}=\L_{\text{DL,self}}+\L_{\text{DL,int}}\,.
\ee
It is important to emphasize that there are no higher order terms contributing to the decoupling limit action for the Lorentz \stu fields at this energy scale. Moreover, due to the prefactor $mM$, the terms coming from the interactions \eqref{DLvec2} arise at a higher scale (we determine the scale in Section~\ref{sec:strong}) than the terms coming from the mass terms \eqref{DLvec1}. We shall therefore focus our attention on the contribution arising from the mixed terms \eqref{DLvec2}.

Let us start by analyzing the first two terms appearing on each of the lines of \eqref{DLvec2}. These are all similar in structure, so that one of the Lorentz \stu fields, $\omega$ or $\sigma$, is contracted with some combination of the helicity-0 fields $\hat{\Pi}$ and $\hat{\mathbb X}$ directly through the epsilon structure:
\be
\varepsilon\varepsilon I \hat{\Pi}\hat{\mathbb X}\omega = \varepsilon_{abcd}\varepsilon^{\mu\nu\alpha\beta}\,\delta^a_\mu\hat{\Pi}^b_\nu\hat{\mathbb X}^c_\alpha\omega^d\,_\beta\,.
\ee
By using the fact that the double--epsilon structure is invariant under the transpose together with $(\omega^T)^a\,_b=-\omega^a\,_b$, we see that all interactions of this type vanish
\be
\varepsilon\varepsilon I \hat{\Pi}\hat{\mathbb X}\omega=\varepsilon\varepsilon (I \hat{\Pi}\hat{\mathbb X}\omega)^T=-\varepsilon\varepsilon I \hat{\Pi}\hat{\mathbb X}\omega=0\,.
\ee
For the same reason also $\varepsilon\varepsilon I\hat{\Pi}^2\sigma =\varepsilon\varepsilon I\hat{\mathbb X}^2\omega=\varepsilon\varepsilon \hat{\Pi}\hat{\mathbb X}^2\omega=\varepsilon\varepsilon \hat{\Pi}^2\hat{\mathbb X}\sigma= 0$. As a result, the surviving interactions are
\be\label{DLvec3}
\begin{split}
g_*^2 \L_{\text{DL, int}}=\frac{m \M}{4} \varepsilon\varepsilon \bigg[& \kappa_{21} I\left(2\hat{\Pi}\hat{\mathbb X}\,\omega\cdot\hat{\Pi}+\hat{\Pi}^2\sigma\cdot\hat{\mathbb X}\right) \\
&+\kappa_{12} I\left(2\hat{\Pi}\hat{\mathbb X}\,\sigma\cdot\hat{\mathbb X}+\hat{\mathbb X}^2\omega\cdot\hat{\Pi}\right)\\
&+ 2\kappa_{22} \left(\hat{\Pi}\hat{\mathbb X}^2\,\omega\cdot\hat{\Pi}+\hat{\Pi}^2\hat{\mathbb X}\,\sigma\cdot\hat{\mathbb X}\right)\\
&+\kappa_{31} \left(\hat{\Pi}^3\sigma\cdot\hat{\mathbb X}+3\hat{\Pi}^2\hat{\mathbb X}\,\omega\cdot\hat{\Pi}\right)\\
&+\kappa_{13} \left(\hat{\mathbb X}^3\omega\cdot\hat{\Pi}+3\hat{\mathbb X}^2\hat{\Pi}\,\sigma\cdot\hat{\mathbb X}\right)\bigg]\,.
\end{split}
\ee
These are the leading interactions arising in the decoupling limit. Supplemented with the symmetric vierbein conditions \eqref{sym_vierb_2} the expressions above give the full decoupling limit helicity-1 interactions in a closed form. In what follows we shall determine the physical scale of those interactions and determine at which scale the ghosts enter in this cycle of interactions. We point out that while the existence of at least one ghost was pointed out in \cite{deRham:2015cha}, the precise scale at which the (what turns out to be two) ghosts enter was not established. In what follows we shall see that the ghosts in the cycle of interactions typically enter below the scale $\Lambda_3$ unless a specific (stable) tuning is considered.

\section{Strong Coupling Scale for Cycle of Interactions}\label{sec:strong}
We now turn to the
 cycle of interactions \eqref{act_nonder2} coming from the quartic helicity-1 and helicity-0 interactions presented in \eqref{DLvec3}. These give, for example,
\be
m \M \varepsilon\varepsilon I\hat{\Pi}\hat{\mathbb X}(\omega\cdot\hat{\Pi})\to\frac{1}{\Lambda_{7/2}^7}\,\varepsilon\varepsilon I \partial^2\pi\,\partial^2\chi\,(\omega\cdot\partial^2\pi)\,,
\ee
with $\Lambda_{7/2}^{7/2}=m^{5/2}\M$. This scale, $\Lambda_{7/2}$, is lower than $\Lambda_3$ and is the true strong coupling scale at which the leading derivative interactions involving the helicity-1 fields arise if the cubic couplings $\kappa_{21},\kappa_{12}$ are assumed to be of order unity. This is a new result and has no analogue in the case of a single or two decoupled massive spin-2 theories. It is only because there are two helicity-0 modes entering, namely $\chi$ and $\pi$, that this term is not simply zero as in the case of \eqref{badequation}.

\subsection{Physical Interaction}

To establish whether this interaction introduces higher order derivatives in the equations of motion, it is sufficient to work perturbatively and use the leading order symmetric vierbein condition \eqref{sym_vierb_2}. This gives for the above interaction:
\be\label{bad_op}
\begin{split}
\varepsilon\varepsilon I \partial^2\pi\,\partial^2\chi\,(\omega\cdot\partial^2\pi)&=\frac{1}{2}\varepsilon_{\mu abc}\varepsilon^{\mu\alpha\beta\gamma} \partial^a\partial_\alpha\pi\,\partial^b\partial_\beta\chi\,(F\cdot\partial^2\pi)^c\,_\gamma\\
&=\frac{1}{2}\varepsilon_{\mu abc}\varepsilon^{\mu\alpha\beta\gamma} \partial^a\partial_\alpha\pi\,\partial^b\partial_\beta\chi\,(\partial^cA^d-\partial^d A^c)\partial_d\partial_\gamma\pi\,.
\end{split}
\ee
Such an interaction term in the Lagrangian will introduce higher order derivatives in the equations of motion.
One can see it, for example, by varying the above term with respect to the vector field $A^\nu$:
\be
\frac{\delta S}{\delta A^\nu}\supset-\frac{1}{2}\varepsilon_{\mu abc}\varepsilon^{\mu\alpha\beta\gamma} \partial^a\partial_\alpha\pi\,\partial^b\partial_\beta\chi\partial^c\partial_\gamma\partial_\nu\pi+\frac{1}{2}\varepsilon_{\mu ab\nu}\varepsilon^{\mu\alpha\beta\gamma} \partial^d\left(\partial^a\partial_\alpha\pi\,\partial^b\partial_\beta\chi\,\partial_d\partial_\gamma\pi\right)\,.
\ee
We note however that although troublesome, those do not necessarily imply the existence of ghost(s) at that order. Indeed when multiple fields are involved, higher derivatives are not necessarily linked with an Ostrogradsky instability \cite{deRham:2011rn} (see also \cite{deRham:2016wji}).

One can in fact prove that the interaction \eqref{bad_op} is not simply a redundant one, removable by a field redefinition (as was the case of the lower scale interactions found in the helicity decomposition of a single massive spin-2 field, \cite{deRham:2011qq}).  Indeed,  computing the contribution to the scattering process $A\pi\to\pi\,\chi$ allowed by the interaction vertex \eqref{bad_op} for a generic choice of polarization for the massless vector field, $   \epsilon^\mu=(0,a,b,0)$, and  scattering angle $\theta$,
we find for the non-vanishing scattering amplitude:

\begin{equation}
    \mathcal{A}\sim 2 g_*^2 a \frac{\textbf{p}^7}{\Lambda_{7/2}^7} \sin (2 \theta )\,.
\end{equation}
This result shows that this vertex clearly gives a non-vanishing contribution at the scale $\Lambda_{7/2}$ thus confirming that the interaction above is physical and could not be removed by a field redefinition. As a result, we conclude that $\Lambda_{7/2}\equiv(m^{5/2}\M)^{2/7}$ is the actual strong coupling scale in the theory of two interacting massive spin-2 fields given by the action \eqref{act_nonder2} provided $\kappa_{21},\kappa_{12}$ are of order unity and at least one is non-vanishing (one can easily check that any higher order operator of the form $\varepsilon\varepsilon F \hat{\Pi}^k \hat{\mathbb X}^\ell$, with $k+\ell>2$ would necessarily enter at a higher energy scale.)

\subsection{Ghost in the DL and ADM}

Since this interaction is physical and since it involves higher than second order equations of motion, the absence of ghosts at that scale is not guaranteed (and indeed from the ADM analysis we had found the existence of a ghost in the full theory).
Let us recall however that the ADM analysis of Section~\ref{adm} showed that the primary constraint removing the BD ghost only disappeared at the sixth order in perturbations. A natural question that arises is how is this compatible with the fact that we see higher derivative interactions appearing already at quartic order. The reason for this becomes clear when considering the dynamics of the time component of the vector field, $A^0$. The quadratic decoupling limit action for the helicity-one modes, \eqref{DLvec1}, after imposing the symmetric vierbein constraint \eqref{sym_vierb_2} gives the standard Maxwell kinetic term for the vector field, meaning that there appear no time derivatives of $A^0$ at quadratic level. However, the time derivatives of $A^0$ do arise in the dangerous interaction \eqref{bad_op}.  Indeed, due to its double--epsilon structure it contains the time-derivatives of $A^0$ as
\be
\varepsilon_{\mu0ij}\varepsilon^{\mu 0i'j'}\,\dot\pi\,\partial^i\partial_{i'}\chi\,\partial^j\dot A^0\,\partial_{j'}\dot\pi\,.
\ee
To see how this affects the order at which the primary constraint removing the BD ghost disappears it is instructive to consider a purely schematic analogous example of a perturbative Lagrangian
\be
\mathcal L[x,z] = \frac{1}{2}\epsilon^2 \dot x^2+\epsilon^4\dot x^3\dot z\,.
\ee
By looking at the Hessian with respect to the time derivatives defined as
\be
L_{ij}\equiv\frac{\partial^2 \mathcal L}{\partial \dot x^i\partial\dot x^j}=
\begin{pmatrix}
\epsilon^2&3\dot x^2\epsilon^4\\
3\dot x^2\epsilon^4&0
\end{pmatrix}\,,
\ee
we see that the two eigenvalues are $\lambda_1 = \epsilon^2+9\dot x^4\epsilon^4$ and $\lambda_2 = -9\dot x^4\epsilon^6$. Hence we see that the vanishing eigenvalue, $\lambda_2$, responsible for the primary constraint becomes non-zero at sixth order in perturbations although the interaction responsible for it appeared at the quartic order in the action. This is exactly what is happening also in our case and thus reconciles the findings of this section with the ADM analysis.

\subsection{Correct Scaling of the EFT}

\label{CorrectScaling}

There are a few specific configurations, worth mentioning, in which the strong coupling scale can be changed. We discuss these in the following. First, let us point out that the interactions that appear at the $\Lambda_{7/2}$ energy scale are the ones quartic in fields in \eqref{DLvec3} and arise due to the cubic operators $\kappa_{21}\varepsilon\varepsilon Ihhf$, $\kappa_{12}\varepsilon\varepsilon Ihff$. Setting these two coefficients to zero the relevant operators would arise from the quartic interaction, $\kappa_{22}\varepsilon\varepsilon hhff$, (and also from $\kappa_{31}, \kappa_{13}$) and lead to a higher strong coupling scale. Indeed, one has
\be
\kappa_{22} \,m\M\,\varepsilon\varepsilon\hat{\Pi}\hat{\mathbb X}^2\,\omega\cdot\hat{\Pi}\to\frac{\kappa_{22}}{(m^{7/3}\M)^3}\varepsilon\varepsilon\,\partial^2\pi\,(\partial^2\chi)^2(\omega\cdot\partial^2\pi)\,,
\ee
corresponding to a strong coupling scale $\Lambda_{10/3}\equiv\left(m^{7/3}\M\right)^{3/10}>\Lambda_{7/2}$.  \\

There is however a far more interesting possibility that is consistent and technically natural for raising the strong coupling scale. Instead of making the assumption that the mixed cubic and quartic interactions determined by $\kappa_{nm}$'s are of order unity, we rescale them in the following manner
\be\label{scaling}
\kappa_{nm}\to\frac{m}{\Lambda_{3}}\kappa_{nm}\,,\qquad \kappa_{nm}=\{\kappa_{12},\kappa_{21},\kappa_{22},\kappa_{13},\kappa_{31}\}\,.
\ee
After restoring the original fields in the decomposition \eqref{stuck}, this gives for the problematic interaction terms
\be
\begin{split}
\label{DLmixedint}
g_*^2 S_{\text{DL, int}}^{(4)}=\frac{1}{4}\int \d^4x&\bigg[\frac{\kappa_{21}}{\Lambda_3^7}\varepsilon\varepsilon I\left(2\partial^2\pi\,\partial^2\chi\,\omega\cdot\partial^2\pi+(\partial^2\pi)^2\sigma\cdot\partial^2\chi\right)\bigg.\\
&+\frac{\kappa_{12}}{\Lambda_3^7}\varepsilon\varepsilon I\left(2\partial^2\pi\,\partial^2\chi\,\sigma\cdot\partial^2\chi+(\partial^2\chi)^2\omega\cdot\partial^2\pi\right)\,\\
&+\bigg.\frac{2\kappa_{22}}{\Lambda_3^{10}}\varepsilon\varepsilon \left(\partial^2\pi\,(\partial^2\chi)^2\,\omega\cdot\partial^2\pi+(\partial^2\pi)^2\,\partial^2\chi\,\sigma\cdot\partial^2\chi\right) \\
&+\bigg.\frac{\kappa_{31}}{ \Lambda_3^{10}}\varepsilon\varepsilon \left(\,(\partial^2\pi)^3\,\sigma\cdot\partial^2\chi+3 (\partial^2\pi)^2\,\partial^2\chi\,\omega\cdot\partial^2 \pi\right) \\
&+\bigg.\frac{\kappa_{13}}{ \Lambda_3^{10}}\varepsilon\varepsilon \left(\,(\partial^2\chi)^3\,\omega\cdot\partial^2\pi+3 (\partial^2\chi)^2\,\partial^2\pi\,\sigma\cdot\partial^2\chi\right)\bigg]
\,.
\end{split}
\ee
We see that, indeed, the dangerous interactions are now shifted to $\Lambda_3$ thus restoring the validity of the $\Lambda_3$ decoupling limit. However, by doing so we have also rescaled the mixed helicity-0/helicity-2 interactions presented in Sec.~\ref{sec:hel02int}. As a result, effectively we are looking at the limit when $\kappa_{nm} \ll \kappa_n^{(i)}\sim \mathcal O(1)$ in the original action \eqref{act_nonder2}. Moreover, there are now additional $\Lambda_3$ interactions contributing to the helicity-1/helicity-0 decoupling limit arising from the self-interaction terms $\kappa_n^{(i)}$ in \eqref{act_nonder2}. These are of the same structure as derived for self-interactions below in \eqref{DL_vec_1}. When compared to the interactions in \eqref{DLmixedint}, we see that at each order in fields the decoupling limit is dominated by the self-interactions. For instance, comparing the quartic operators in \eqref{DLmixedint} and  \eqref{DL_vec_1} we see that these are operators of mass dimensions $11$ and $10$ respectively.

That this tuning is technically natural is simply because once it has been made, there are no other interactions that arise at a lower energy scale, and nothing to arise at any loop order that will push the cutoff scale down. Note that  this statement is manifest in the \stu formalism. In unitary gauge, one may worry about dangerous $1/m^2$ and $1/m^4$ terms that arise in the unitary gauge propagator for a massive spin-2 particle leading to a less straightforward EFT power counting (see \cite{deRham:2017xox} for more details on this power counting in the single spin-2 case), however the stability of the tuning is of course valid in any formulation, and the \stu one makes it manifest.
These arguments do not rely on the non-renormalization theorem that arises for the double--epsilon interactions by virtue of the Galileon form of the decoupling limit which applies to the self-interactions \cite{Luty:2003vm,deRham:2012ew}. As we have seen the leading mixed interactions described in \eqref{DLmixedint} are not of the standard Galileon form for which those non-renormalization theorems apply.

Where the non-renormalization theorems \cite{Luty:2003vm,deRham:2012ew} do come into play is when we choose to focus on the $\Lambda_{7/2}$ theory, \emph{i.e.} we do not make the rescaling given in \eqref{scaling}. In this case it remains technically natural to keep the self interactions $\kappa^{(i)}_n$ at the same scale, with their associated $\Lambda_3$ interactions. These $\Lambda_3$ interactions are not driven down to $\Lambda_{7/2}$ at any order in loops by virtue of the non-renormalization theorems.

\subsection{Quadratic Mixing}\label{Quadraticmixing}

Let us also remark that even the quadratic interaction term (though absent in our model \eqref{act_nonder2}) contains interactions that appear at an energy scale lower than $\Lambda_3$. In fact, the quadratic interaction term gives
\be
\begin{split}
&m^2\M^2\int \varepsilon_{abcd} I^a\wedge I^b\wedge (E^c-I^c)\wedge (F^d-I^d)\\
&= m \M \,\varepsilon\varepsilon I^2\left[\hat{\mathbb X}\left(\partial A+\omega+\omega\cdot\hat{\Pi}\right)+\hat{\Pi}\left(\partial B+\sigma+\sigma\cdot\hat{\mathbb X}\right)\right]+\mathcal O(1)\,.
\end{split}
\ee
It is straightforward to see that the interactions involving the vector fields $A,B$ are total derivatives while there are terms
\begin{align}
&\varepsilon\varepsilon I^2\hat{\mathbb X}\ \omega\cdot\hat{\Pi}=2\left(\[\hat{\mathbb X}\]\[\hat{\Pi}\omega\]-\[\hat{\Pi}\hat{\mathbb X}\omega\]\right)=-2\[\hat{\Pi}\hat{\mathbb X}\omega\]\neq0\,,\\
&\varepsilon\varepsilon I^2\hat{\Pi}\sigma\cdot\hat{\mathbb X}=2\left(\[\hat{\Pi}\]\[\hat{\mathbb X}\sigma\]-\[\hat{\Pi}\hat{\mathbb X}\sigma\]\right)=-2\[\hat{\Pi}\hat{\mathbb X}\sigma\]\neq0\,.
\end{align}
By restoring the $\Lambda_{3}^3$ hidden in our definition of $\hat{\Pi}\,,\hat{\mathbb X}$ this gives non-vanishing interactions at the scale
\be
m \M \varepsilon\varepsilon\,I^2\hat{\mathbb X}\omega\cdot\hat{\Pi}\to\frac{1}{\Lambda_{4}^4}\varepsilon\varepsilon \,\partial^2\chi\,(\omega\cdot\partial^2\pi)\,.
\ee
This is the lowest of the scales appearing in the discussion above, so that we have $\Lambda_4<\Lambda_{7/2}<\Lambda_{10/3}<\Lambda_3$ (the scale $\Lambda_n$ with arbitrary $n$ was defined in \eqref{eq:Lambdan}). As in the previous case, choosing the coefficient of this operator to be $m/\Lambda_3$ suppressed rather than order unity rescales this to a $\Lambda_3$ interaction:
\be
\frac{m}{\Lambda_3}\frac{1}{\Lambda_{4}^4}\varepsilon\varepsilon \,\partial^2\chi\,(\omega\cdot\partial^2\pi) =\frac{1}{\Lambda_{3}^4}\varepsilon\varepsilon \,\partial^2\chi\,(\omega\cdot\partial^2\pi) \,.
\ee
However, this quadratic term mixing between the vierbeins $E$ and $F$ gives a coupling with $h_{\mu\nu}$ and $f_{\mu\nu}$ at a linear level already and is thus not part of cycle interactions in the sense described in Section~\ref{sec:cycle_line}. We therefore do not consider it any further here. \\

\paragraph{Summary:} To summarize the conclusions of the analysis of this and the previous sections, the cycle theories do in general have ghosts from the perspective of an ADM analysis, but are consistent as EFTs with a cutoff scale that is at most $\Lambda_3$, which can be determined from the analysis of the helicity-2/helicity-0 sector alone. Interactions in the helicity-1/helicity-0 sector can occur at a parametrically lower scale, ranging from $\Lambda_4$ to $\Lambda_{7/2}$ to $\Lambda_{10/3}$, but with a suitable rescaling of the mixed interactions between the two spin-2 states, these interactions may in turn be put at the $\Lambda_3$ scale.

\subsection{Decoupling Limit for Unconstrained Vierbein Theory}\label{sec:unconstrained}

Although our principle focus has been the metric-like (constrained vierbein) formulation of the EFT for the interaction of multiple spin-2 fields, the above analysis gives us for free the decoupling limit also for the unconstrained vierbein formulation of the theory. As we have discussed the difference between the two is that in the latter, the Lorentz \stu fields are determined by their equations of motion, and as a consequence of the interactions between the two spin-2 fields, the usual symmetric vierbein constraint is lost \cite{Hinterbichler:2012cn,deRham:2015cha}. The new condition is discussed explicitly in Appendix \ref{sec:Leom} for the full theory of cycle interactions of two spin-2 fields. Since the above decoupling limit analysis was performed effectively in vierbein notation, the resulting DL actions are the same with the only difference being that $\omega$ and $\sigma$ are viewed as independent fields and are not fixed by the equations (\ref{sym_vierb_2}). Thus we may immediately draw all of the same conclusions as above (\emph{i.e.} the points \ref{mainpoint1}--\ref{mainpoint3} from the Introduction) about the unconstrained vierbein theory of cycle interactions. We stress once again that, while these theories are different nonlinearly, from the decoupling limit perspective they are extremely similar.

\subsection{Higher Derivative EFT corrections for Cycle Theories}\label{sec:highdercycle}

Up to now we have defined the cycle theories to be those given by the action with kinetic terms \eqref{act_kinetic} and non-derivative interactions \eqref{act_nonder_general}}. However, these are just the leading terms in an EFT expansion which contains an infinite number of higher derivative operators. The generic form of the higher derivative corrections for multiple massive spin-2 fields follows closely the discussion for the single massive spin-2 case (see for instance \cite{deRham:2017xox}).

\subsubsection{$\Lambda_3$ theory}

Provided we make the scaling given in Eq.~\eqref{scaling}, we have established that the lowest interaction scale from the leading interactions is $\Lambda_3$.
At leading order in the decoupling limit $M \rightarrow \infty$, $m \rightarrow 0$ we have
\ba
&& E_{a\mu} = \eta_{a\mu} + \frac{\partial_a \partial_{\mu} \pi}{\Lambda_3^3} + \dots\\
&& F_{a\mu} = \eta_{a\mu} + \frac{\partial_a \partial_{\mu} \chi}{\Lambda_3^3} + \dots \, .
\ea
Thus any function of these two vierbeins will, at leading order in the decoupling limit, be a dimensionless function of operators suppressed by the $\Lambda_3$ scale. Further derivatives of these functions should also be suppressed by the scale $\Lambda_3$. By contrast, the Riemann curvature constructed from each of these vierbeins will necessarily come suppressed by an additional power of $M$, since the above vierbeins are at leading order equivalent to Minkowski spacetime in a non-standard coordinate system, \emph{i.e.} if the tensors are set to zero, we have $E^a = \Lambda^a{}_{b} d \phi^b$, which is just a generic parameterization of Minkowski for which $R[E]=0$. Hence it is the next order $1/M$ corrections to $E$ that determine the leading corrections to the Riemann tensor, and so {\it very} schematically
\be
R[E]_{abcd} \sim \sum \frac{1}{M} \partial \partial h  \(\frac{\partial \partial \pi}{\Lambda_3^3}\)^m  \(\frac{m\partial A}{\Lambda_3^3}\)^n \(\frac{h}{M}\)^p \, ,
\ee
and similarly for $R[F]_{abcd}$, where $R[E]_{abcd}$/$R[F]_{efgh}$ denote the standard components of the Riemann tensor associated with the vierbein's $E_{a\mu}$/$F_{b \mu}$. \\

In the general helicity-2/helicity-0 decoupling limit, we have seen that there are interactions between the helicity-2 modes and scalars of the form $\Lambda_3^3h \hat \Pi \hat \Pi \hat \Pi \sim \frac{1}{\Lambda_3^6 }h \Pi \Pi \Pi$ etc. It is known from the case of single spin-2 massive gravity that these interactions are not removable with a local field redefinition. They will give rise to an $h \pi \pi \pi $ four point vertex of the form (for a general discussion of interactions in the DL theory see \cite{deRham:2012ew})
\be
\frac{1}{\Lambda_3^6} \epsilon_{abcd} \epsilon_{ABCD }{{\bf e}}^{aA}(k_1) k_2^b k_2^B k_3^c k_3^C k_4^d k_4^D = \frac{1}{\Lambda_3^6} \epsilon_{abcd} \epsilon_{ABCD }{{\bf e}}^{aA}(k_1) k_2^b k_2^B k_3^c k_3^C k_1^d k_1^D \, ,
\ee
where ${{\bf e}}^{aA}(k_1)$ is the helicity-2 polarization and we have made use of momentum conservation $k_4=-(k_1+k_2+k_3)$. In general this interaction vertex is non-zero for off-shell $\pi$, even when the helicity-2 mode is on-shell $k_1^2=0$, and will lead at three-loop level to a contribution to the $hhhh$ vertex (\emph{i.e.} the $2$--$2$ graviton scattering amplitude) which will require counter-terms of the very schematic form (multiplied by appropriate powers of $g_*$)
\be\label{counterterm}
\frac{1}{\Lambda_3^{24}} \partial^{24}  h^4 \, .
\ee
Crucially, these counterterms come in at the scale $\Lambda_3$, and are not $M$ suppressed. In other words, if in the theory there are non-trivial interactions between the helicity-2 modes and the helicity-0 modes already at the $\Lambda_3$ scale, then it is necessary in the EFT to allow for all possible such interactions at the $\Lambda_3$ scale, if only to act as counter-terms in loop diagrams. It is however not possible to build counter-terms of the form \eqref{counterterm} directly out of $E_{a \mu}$ without further qualification, since $h$ enters $E$ with an additional $M$ suppression. If we multiply by $M$ we will introduce helicity-0 interactions at the scale $m$, since $M (E-I) \sim \partial \partial \pi/m^2$. The solution is that such counterterms will arise from operators build out of the Riemann tensor, for which the dangerous helicity-0 interactions drop out by virtue of diffeomorphism invariance. Now since $\partial \partial h \sim \Lambda_3^3$,  $\partial \partial f\sim \Lambda_3^3$ when
\be
M R[E]_{abcd}  \quad \text{or} \quad M R[F]_{abcd}\sim \Lambda_3^3 \, ,
\ee
the EFT Lagrangian must contain combinations of the Riemann curvature in the combination ${M R[E]_{abcd}}/{\Lambda_3^3}$ and ${M R[F]_{efgh}}/{\Lambda_3^3}$ with any additional derivatives suppressed by $\Lambda_3$. For instance an operator of the form \eqref{counterterm} will arise from an interaction
\be
\Lambda_3^4  \(M \frac{\partial_a \partial_b \partial_c \partial_d R_{efgh}[E]}{\Lambda_3^7} \)^4 \, .
\ee
Putting this together, unlike the leading interactions, whose form relies on special properties of total derivative combinations, the generic higher derivative corrections will arise at their naive interaction scale. This is when (in unitary gauge)
\be\label{scaling1}
g_*^2 {S}_{\text{higher-der}} = \int \d^4 x  \Lambda_3^4 \, {\cal F}\[E^a_{\mu} , F^b_{\mu}, \frac{\partial_{\rho}}{\Lambda_3} , \frac{M R[E]_{abcd}}{\Lambda_3^3}, \frac{M R[F]_{efgh}}{\Lambda_3^3}\] \, ,
\ee
where ${\cal F}$ denotes the superposition of all Lorentz scalar combinations of the arguments, with dimensionless order unity coefficients. \\

In fact, the situation is typically better than that. When the coefficients of the self- and mixed-interactions are chosen so that the $h \Pi \Pi \Pi$/$f {\mathbb X} {\mathbb X}{\mathbb X}$/$h \Pi {\mathbb X}{\mathbb X}$ etc. terms vanish (\emph{i.e.} when $X^{(3)}=Y^{(3)}=0$), then it is possible to diagonalize/demix the interactions between the helicity-2 modes and the helicity-0 modes as we will do in Section~\ref{sec:biGalileon}, Eq.~\eqref{demix} for the case of line interactions. In this case there are no pure $\Lambda_3$ interactions containing helicity-2 modes. Thus if we focus on only those terms in the EFT that are needed to renormalize loops of light fields, any contribution from helicity-2 states will come suppressed by additional powers of the spin-2 interaction scale $M$. In this case the more appropriate power counting is
\be\label{scaling2}
g_*^2 {S}_{\text{higher-der}} = \int \d^4 x  \Lambda_3^4 \, {\cal F}\[E^a_{\mu} , F^b_{\mu}, \frac{\partial_{\rho}}{\Lambda_3} , \frac{R[E]_{abcd}}{\Lambda_3^2}, \frac{ R[F]_{efgh}}{\Lambda_3^2}\] \, ,
\ee
so that at leading order in the $M \rightarrow \infty$ limit with fixed $\Lambda_3$ the helicity-2 states do not enter in any of the higher derivative EFT corrections. This is the scaling considered for example in \cite{deRham:2017xox}. \\

As an illustrative example, the types of non-minimal kinetic terms considered in \cite{deRham:2013tfa,deRham:2015rxa} are allowed in the EFT context, however they come in at a parametrically smaller scale. For instance, with the worse case choice \eqref{scaling1}, we are allowed the kinetic term
\be
g_*^2 \Delta{S}_{\rm kinetic} \sim M \Lambda_3 \int \epsilon_{abcd} I^a \wedge E^b \wedge R^{cd}[E]+ \alpha M \Lambda_3 \int \epsilon_{abcd} I^a \wedge I^b \wedge R^{cd}[E]
\ee
but see that it is suppressed by $\frac{\Lambda_3}{M}$ relative to the leading kinetic term. In this way, the ghost implied by the analysis \cite{deRham:2013tfa,deRham:2015rxa} is moved to the cutoff scale of the EFT rendering it harmless. With the scaling \eqref{scaling1} this situation is only improved. \\

As a second example, we note that the double--epsilon structure of the leading non-derivative interactions is itself not stable under loop corrections. However, the corrections that are expected to arise to this structure are suppressed \cite{deRham:2012ew,deRham:2013qqa}. For either choice of scaling \eqref{scaling1} or \eqref{scaling2}, the non-double--epsilon mass terms (\emph{i.e.} those which are not of the double--epsilon form) will arise at the scale (highly schematically)
\be
g_*^2 \Delta{S}_{\rm mass}\sim  \Lambda_3^4  \int   \sum_{nm} \left[ b_{nm}(E-I)^2 E^n F^m+c_{nm}(F-I)^2 E^n F^m+d_{nm}(E-F)^2 E^n F^m \right]
\ee
which is suppressed by
\be
\frac{\Lambda_3^4}{m^2 M^2} \sim \frac{\Lambda_3}{M} \sim \(\frac{m}{M}\)^{2/3} \, ,
\ee
relative to the leading mass terms.

\subsubsection{$\Lambda_{7/2}$ theory}

If we do not make the scaling implied by Eq.~\eqref{scaling}, then the leading interactions of the cycle theories already imply a cutoff $\Lambda_{7/2}$ from the helicity-1/helicity-0 interactions (unless we include quadratic mixing as in Section~\ref{Quadraticmixing} or tune the cubic interactions to zero as discussed in Section~\ref{CorrectScaling}). As we have explained, these interactions are the origin of the Boulware-Deser ghosts seen in the ADM analysis. However, as already noted, when viewed as an EFT there is no problem with this theory provided we accept that $\Lambda_{7/2}$ is indeed the cutoff. One consequence of this, is that all higher derivative corrections are expected to arise now at the scale $\Lambda_{7/2}$.
In other words, in the worst case scenario, these EFT corrections will organize in the schematic form
\be\label{scaling3}
g_*^2 {S}_{\text{higher-der}} = \int \d^4 x  \Lambda_{7/2}^4 \, {\cal F}\[\frac{\Lambda_3^3(E^a_{\mu}-I^a_{\mu})}{\Lambda_{7/2}^3} , \frac{\Lambda_3^3(F^a_{\mu}-I^a_{\mu})}{\Lambda_{7/2}^3}, \frac{\partial_{\rho}}{\Lambda_{7/2}} , \frac{M R[E]_{abcd}}{\Lambda_{7/2}^3}, \frac{M R[F]_{efgh}}{\Lambda_{7/2}^3} \] \, ,
\ee
and in the more optimistic scenario in which the helicity-2 modes do not enter at leading order in the decoupling limit
\be\label{scaling4}
g_*^2 {S}_{\text{higher-der}} = \int \d^4 x  \Lambda_{7/2}^4 \, {\cal F}\[\frac{\Lambda_3^3(E^a_{\mu}-I^a_{\mu})}{\Lambda_{7/2}^3} , \frac{\Lambda_3^3(F^a_{\mu}-I^a_{\mu})}{\Lambda_{7/2}^3}, \frac{\partial_{\rho}}{\Lambda_{7/2}} , \frac{ R[E]_{abcd}}{\Lambda_{7/2}^2}, \frac{ R[F]_{efgh}}{\Lambda_{7/2}^2} \] \, .
\ee

\section{Decoupling Limit for the Line of Interactions}\label{sec:line}

The derivation of the decoupling limit action for the theory with a line of interactions (or mixed nonlinear kinetic terms in the mass eigenstates) turns out to be much simpler. We shall work with the action \eqref{act_nonder_line} for the line interactions and follow the same formalism that was used for deriving the decoupling limit for the cycle of interactions in Section~\ref{sec:cycle}. For line interactions we find that only the standard interactions known from the case of a single massive spin-2 field and  leading to strong coupling at $\Lambda_3$ scale are present, with no ghost. This is consistent with the ADM arguments of \cite{Hinterbichler:2012cn} and previous discussions \cite{Fasiello:2013woa,Noller:2015eda}. \\

For the sake of clarity we expand the terms on the second line of the action \eqref{act_nonder_line} here:
\be\label{act_vierb_line2}
\begin{split}
g_*^2S_{\text{non-der}}=-\frac{m_1^2M_1^2}{2}\int \d^4x\,&\sum_{n=0}^4\frac{\tilde\beta^{(1)}_{n}}{n!(4-n)!}\,\mathcal U_{n}(I,E)\\
=-\frac{m^2M^2}{2}\int \varepsilon_{abcd}&\left[\frac{\tilde\beta_{1}^{(2)}}{3!}E^a\wedge E^b\wedge E^c\wedge F^d\right.\\
&\left.+\frac{\tilde\beta_{2}^{(2)}}{2!2!}E^a\wedge E^b\wedge F^c\wedge F^d\right.\\
&\left.+\frac{\tilde\beta_{3}^{(2)}}{3!}E^a\wedge F^b\wedge F^c\wedge F^d\right.\\
&\left.+\frac{\tilde\beta^{(2)}_4}{4!}F^a\wedge F^b\wedge F^c\wedge F^d\right]\,.
\end{split}
\ee
Now, let us introduce the two sets of St\"uckelberg fields for both the local Lorentz transformations and spacetime diffeomorphisms as we did before in \eqref{diff_stuck}:
\be
E^a_\mu\to\tilde E^a_\mu=\Lambda^a\,_bE^b_c(\phi)\partial_\mu\phi^c\,,\qquad F^a_\mu\to\tilde F^a_\mu=\Gamma^a\,_bF^b_c(\psi)\partial_\mu\psi^c\,.
\ee
Since the Einstein--Hilbert terms are invariant under these transformations, the St\"uckelberg fields only appear in the non-derivative terms presented above.
With these fields in place the total action is again invariant under spacetime diffeomorphisms and under one overall local Lorentz transformation transforming all the Lorentz indices in the same fashion. There is in fact a more convenient way of writing the Lorentz transformations $\Gamma^a\,_b$ by splitting it as a product of two sequential transformations:
\be
\Gamma^a\,_b\equiv \Lambda^a\,_c\tilde\Gamma^c\,_b\,.
\ee
The interaction terms then schematically become
\be
\begin{split}
g_*^2S_{\text{non-der}}=-\frac{m_1^2M_1^2}{2}\int \d^4x\,&\sum_{n=0}^4\frac{\tilde\beta^{(1)}_{n}}{n!(4-n)!}\,\mathcal U_{n}(I,\Lambda E\partial\phi)\\
-\frac{m^2M^2}{2}\int \varepsilon_{abcd}&\left[\frac{\tilde\beta_{1}^{(2)}}{3!}\Big(\Lambda E\partial\phi\Big)^a\wedge\Big(\Lambda E\partial\phi\Big)^b\wedge \Big(\Lambda E\partial\phi\Big)^c\wedge \left(\Lambda\tilde\Gamma F\partial\psi\right)^d\right.\\
&\left.+\frac{\tilde\beta_{2}^{(2)}}{2!2!}\Big(\Lambda E\partial\phi\Big)^a\wedge \Big(\Lambda E\partial\phi\Big)^b\wedge \left(\Lambda\tilde\Gamma F\partial\psi\right)^c\wedge \left(\Lambda\tilde\Gamma F\partial\psi\right)^d\right.\\
&\left.+\frac{\tilde\beta_{3}^{(2)}}{3!}\Big(\Lambda E\partial\phi\Big)^a\wedge \left(\Lambda\tilde\Gamma F\partial\psi\right)^b\wedge \left(\Lambda\tilde\Gamma F\partial\psi\right)^c\wedge \left(\Lambda\tilde\Gamma F\partial\psi\right)^d\right.\\
&\left.+\frac{\tilde\beta^{(2)}_4}{4!}\left(\Lambda\tilde\Gamma F\partial\psi\right)^a\wedge \left(\Lambda\tilde\Gamma F\partial\psi\right)^b\wedge \left(\Lambda\tilde\Gamma F\partial\psi\right)^c\wedge \left(\Lambda\tilde\Gamma F\partial\psi\right)^d\right]\,,
\end{split}
\ee
making it apparent that the Lorentz transformation $\Lambda$ drops out from the last set of terms, thus decoupling the Lorentz St\"uckelbergs $\Lambda$ and $\tilde\Gamma$.

One can simplify the terms under the second integral in the above action even further by performing an inverse diffeomorphism to $\phi^a$. To demonstrate it clearly, let us write the $\tilde\beta^{(2)}_2$ term explicitly (we drop the normalization factors):
\be
\begin{split}
S_{\tilde \beta_2^{(2)}}=&\int \varepsilon_{abcd}\,\Big( E\partial\phi\Big)^a\wedge \Big(E\partial\phi\Big)^b\wedge \left(\tilde\Gamma F\partial\psi\right)^c\wedge \left(\tilde\Gamma F\partial\psi\right)^d\\
=&\int \varepsilon_{abcd}\,\Big( E(\phi)\partial\phi\Big)^a_\mu \d x^\mu\wedge \Big(E(\phi)\partial\phi\Big)^b_\nu \d x^\nu\wedge \left(\tilde\Gamma F(\psi)\partial\psi\right)^c_\alpha \d x^\alpha\wedge \left(\tilde\Gamma F(\psi)\partial\psi\right)^d_\beta \d x^\beta\\
=&\int \varepsilon_{abcd}\,E^a_{a'}(\phi)\frac{\partial\phi^{a'}}{\partial\phi^\mu} \d \phi^\mu\wedge E^b_{b'}(\phi)\frac{\partial\phi^{b'}}{\partial\phi^\nu} \d \phi^\nu\wedge {\tilde\Gamma}^c_{c'} F^{c'}_{c''}(\psi)\frac{\partial\psi^{c''}}{\partial\phi^\alpha} \d \phi^\alpha\wedge {\tilde\Gamma}^d_{d'} F^{d'}_{d''}(\psi)\frac{\partial\psi^{d''}}{\partial\phi^\beta} \d \phi^\beta
\end{split}
\ee
where in the last equality we have simply rewritten it in a different set of coordinates $x^\mu\equiv\phi^\mu$. Expanding the differentials $\d \phi^\mu=\frac{\partial\phi^\mu}{\partial x^{\mu'}}\d x^{\mu'}$ and recognising that $E^a_\mu(\phi)\frac{\partial\phi^\mu}{\partial x^{\mu'}}=\tilde E^a_{\mu'}(x)$ this simplifies to
\be
S_{\tilde \beta_2^{(2)}}=\int \varepsilon_{abcd}\,\tilde E^a_{\mu}(x) \d x^\mu\wedge \tilde E^b_{\nu}(x) \d x^\nu\wedge {\tilde\Gamma}^c_{c'} F^{c'}_{c''}(\psi)\partial_\alpha\psi^{c''}\d x^\alpha\wedge {\tilde\Gamma}^d_{d'} F^{d'}_{d''}(\psi)\partial_\beta\psi^{d''} \d x^\beta\,.
\ee
We also reorganize the terms on the first line of the action \eqref{act_vierb_line2} in a similar fashion by performing the inverse diffeomorphism to $\phi^a$ there as well. Taking now the $\tilde\beta^{(1)}_2$ term as an example this becomes
\be
\begin{split}
S_{\tilde \beta_2^{(1)}}=&\int \varepsilon_{abcd}\, I^a\wedge I^b\wedge \Big( \Lambda E\partial\phi\Big)^c\wedge \Big(\Lambda E\partial\phi\Big)^d\\
=&\int \varepsilon_{abcd}\,\delta^a_\mu \d x^\mu\wedge\delta^b_\nu \d x^\nu \wedge \Big( \Lambda E(\phi)\partial\phi\Big)^c_\alpha \d x^\alpha\wedge \Big(\Lambda E(\phi)\partial\phi\Big)^d_\beta \d x^\beta\\
=&\int \varepsilon_{abcd}\,\delta^a_\mu \d \phi^\mu\wedge\delta^b_\nu \d \phi^\nu \wedge\Lambda^c_{c'} E^{c'}_{c''}(\phi)\frac{\partial\phi^{c''}}{\partial\phi^\alpha} \d \phi^\alpha\wedge\Lambda^d_{d'} E^{d'}_{d''}(\phi)\frac{\partial\phi^{d''}}{\partial\phi^\beta} \d \phi^\beta\,,\\
\end{split}
\ee
leading to
\be
S_{\tilde \beta_2^{(1)}}=\int \varepsilon_{abcd}\,\partial_\mu\phi^a \d x^\mu\wedge \partial_\nu\phi^b \d x^\nu\wedge {\Lambda}^c_{c'} \tilde E^{c'}_\alpha(x) \d x^\alpha\wedge {\Lambda}^d_{d'} \tilde E^{d'}_\beta(x) \d x^\beta\,,
\ee
which makes it clear that there is only one set of diffeomorphism St\"uckelberg fields necessary to restore the diffeomorphism invariance of each of the integrals in the action \eqref{act_vierb_line2}. Hence, the St\"uckelberg trick needed to restore the diffeomorphism invariance in its most convenient form is given by
\be\label{stuck_line}
I^a_\mu\to \tilde I^a_\mu=\partial_\mu\phi^a\,,\qquad E^a_\mu\to\tilde E^a_\mu=\Lambda^a\,_bE^b_\mu(x)\,,\qquad F^a_\mu\to\tilde F^a_\mu=\Lambda^a\,_b\Gamma^b\,_cF^c_d(\psi)\partial_\mu\psi^d\,.
\ee
The final action for line interactions written in the vierbein form then becomes:
\be\label{act_vierb_line3}
\begin{split}
g_*^2S_{\text{non-der}}=&-\frac{m_1^2M_1^2}{2}\int \d^4x\,\sum_{n=0}^4\frac{\tilde\beta^{(1)}_{n}}{n!(4-n)!}\,\mathcal U_{n}(\partial\phi,\Lambda E)\\
&-\frac{m^2M^2}{2}\int  \d^4x\sum_{n=1}^4\frac{\tilde\beta^{(2)}_n}{n!(4-n)!}\,\mathcal U_n(E,\,\Gamma F(\psi)\partial\psi) \,,
\end{split}
\ee
where we have already taken into account that the Lorentz St\"uckelberg fields $\Lambda^a\,_b$ drop out from the second term. Henceforth we work with the mass scaling \eqref{scaling_mass} for concreteness. Importantly, the two sets of diffeomorphism and Lorentz St\"uckelberg fields $(\Lambda, \phi^a)$ and $(\Gamma,\psi^a)$ are only coupled to each other through the helicity-2 perturbations. We shall see below that this significantly simplifies the decoupling limit analysis.

\subsection{Helicity-two/Helicity-zero Sector}\label{sec:hel02_line}
The decomposition of the St\"uckelberg fields introduced in \eqref{stuck_line} suitable for the analysis of the helicity-0/helicity-2 sector decoupling limit interactions is:
\be\label{EF_scalar_line}
\begin{split}
\tilde I&=I+\frac{\Pi}{\Lambda_3^3}\,,\qquad E = I +\frac{h(x)}{M}\,,\\
F &= I +\frac{\mathbb X}{\Lambda_3^3}+\frac{1}{M}\left(f\left[x+\frac{\partial\chi}{\Lambda_3^3}\right]+\frac{f\left[x+\frac{\partial\chi}{\Lambda_3^3}\right]\cdot\mathbb X}{\Lambda_3^3}\right)\,,
\end{split}
\ee
where we neglect both Lorentz and vector St\"uckelberg fields.

The decoupling limit of the first line of non-derivative interactions in \eqref{act_vierb_line3} coincides with that of a single massive graviton. We have analyzed such terms in Section~\ref{sec:hel02} already. As presented there the naively dangerous higher derivative self-interactions \eqref{tot_der} that arise at the scales $\Lambda_4$ and $\Lambda_5$ are in fact total derivatives and therefore cancel. Instead the leading order interactions arise at the $\Lambda_3$ scale and take the standard form \cite{deRham:2010ik}:
\ba\label{DL_sc_1}
\begin{split}
g_*^2 \L_{\text{DL, 1}}&=-\gamma^2h^{\mu\nu}\mathcal E^{\alpha\beta}_{\mu\nu}h_{\alpha\beta}
+\frac{1}{4}\gamma^2 x^2h^{\mu\nu}\tilde X^{(1)}_{\mu\nu}[\Pi]\,,
\end{split}
\ea
with
\ba
\tilde X^{(i)}_{\mu\nu}[\Pi]&\equiv&\,\tilde\kappa_2^{(i)}\left(\varepsilon\varepsilon II\Pi\right)_{\mu\nu}
+\frac{\tilde\kappa_3^{(i)}}{\Lambda_3^3}\left(\varepsilon\varepsilon I\Pi\Pi\right)_{\mu\nu}
+\frac{\tilde\kappa_4^{(i)}}{\Lambda_3^6}\left(\varepsilon\varepsilon \Pi\Pi\Pi\right)_{\mu\nu} \,,
\ea
where $i = 1,2$ and
\be
\tilde\kappa_2^{(1)}=-\tilde\beta_1^{(1)}-2\tilde\beta_2^{(1)}-\tilde\beta^{(1)}_3\,,\quad\tilde\kappa_3^{(1)}=-\tilde\beta_1^{(1)}-\tilde\beta_2^{(1)}\,,\quad\tilde\kappa_4^{(1)}=-\frac{1}{3}\tilde\beta_1^{(1)}\,.
\ee

On the second line of \eqref{act_vierb_line3} there are also the mixed total derivative interactions appearing at scales $\Lambda_4$ and $\Lambda_5$ as in \eqref{tot_der2} that we disregard. The leading physical interactions in the helicity-0/helicity-2 decoupling limit coming from these terms are:
\ba\label{DL_sc_2}
\begin{split}
g_*^2 \L_{\text{DL,2}}&=-f^{\mu\nu}\mathcal E^{\alpha\beta}_{\mu\nu}f_{\alpha\beta}+\frac{1}{4}h^{\mu\nu}\tilde X_{\mu\nu}^{(2)}[\mathbb X]
+\frac{1}{4}\left(f^{a}_{\mu}[x+\p \chi/\Lambda_3^3]+\frac{f^{a\nu}[x+\p \chi/\Lambda_3^3]\mathbb X_{\nu\mu}}{\Lambda_3^3}\right)\tilde Y_{a}^{\mu}[\mathbb X]\,
\end{split}
\ea
where in the last term alone $f$ is evaluated at $x+\p \chi/\Lambda_3^3$\,, and
with
\be
\tilde\kappa_2^{(2)}=-\tilde\beta_1^{(2)}-2\tilde\beta_2^{(2)}-\tilde\beta^{(2)}_3\,,\quad\tilde\kappa_3^{(2)}=-\tilde\beta_2^{(2)}-\tilde\beta_3^{(2)}\,,\quad\tilde\kappa_4^{(2)}=-\frac{1}{3}\tilde\beta_3^{(2)}\,.
\ee
The $\tilde Y^\mu_a$ are given by
\ba
\label{tildeY}
\tilde Y_{a}^{\mu}[\mathbb X]&\equiv&\,2\kappa_2^{(2)}\left(\varepsilon\varepsilon II\mathbb X\right)_{a}^{\mu}
+\frac{3\kappa_3^{(2)}}{\Lambda_3^3}\left(\varepsilon\varepsilon I\mathbb X\mathbb X\right)_{a}^{\mu}
+\frac{4\kappa_4^{(2)}}{\Lambda_3^6}\left(\varepsilon\varepsilon \mathbb X\mathbb X\mathbb X\right)_{a}^{\mu}\, ,
\ea
where $\kappa_n^{(2)}=\mathbb C_{nm}\tilde\beta_n^{(2)}$ and the matrix $\mathbb C$ is defined in \eqref{kappa_beta}. The following combination amounts to an overall cosmological constant 
\be
\tilde \beta^{(i)}_0+4\tilde \beta^{(i)}_1+6\tilde \beta^{(i)}_2+4\tilde \beta^{(i)}_3+\tilde \beta^{(i)}_4\,,\qquad i=1,2\,.
\ee
and can be set to any value, \emph{e.g.} $\tilde \beta^{(i)}_0+4\tilde \beta^{(i)}_1+6\tilde \beta^{(i)}_2+4\tilde \beta^{(i)}_3+\tilde \beta^{(i)}_4=0$  (since we do not have a massless graviton here). In practice this means that say $\tilde \beta^{(i)}_4$ are not independent. The condition for the absence of tadpoles for both fields imposes
\ba
&\gamma^2 x^2\left(\tilde \beta^{(1)}_1+3\tilde \beta^{(1)}_2+3\tilde \beta^{(1)}_3+\tilde \beta^{(1)}_4\right)+3\tilde \beta^{(2)}_1+3\tilde \beta^{(2)}_2+\tilde \beta^{(2)}_3=0\,,\\
&\tilde \beta^{(2)}_1+3\tilde \beta^{(2)}_2+3\tilde \beta^{(2)}_3+\tilde \beta^{(2)}_4=0\,.
\ea
This condition is slightly different from the no-tadpole condition in standard massive gravity due to the non-trivial mixing between both spin-2 fields.

The total decoupling limit action for the helicity-0 and helicity-2 interactions for a theory with line interactions is then given by the sum of \eqref{DL_sc_1} and \eqref{DL_sc_2}. These interactions become strongly coupled at $\Lambda_3$. Let us also note that there is no direct coupling between the two helicity-0 modes $\pi$ and $\chi$. They only mix through their coupling to $h^{\mu\nu}$. Finally, we remark that, as in the case of cycle interactions, the last term in \eqref{DL_sc_2} appears to be non-local, since we are evaluating $f_{\mu\nu}$ as $f_{\mu\nu}\left[x+\frac{\partial\chi}{\Lambda_3^3}\right]$. This can be corrected by performing a Galileon duality transformation \cite{Fasiello:2013woa,Curtright:2012gx,deRham:2013hsa,deRham:2014lqa}. Defining $x'=x+\partial\chi(x)/\Lambda_3^3$ and the dual Galileon field $\chi'$ by $x=x'+\partial'\chi'(x')/\Lambda_3^3$, and then relabelling the dummy label $x' \rightarrow x$ we find only $f_{\mu\nu}(x)$ enters. We shall do this explicitly in Section~\ref{sec:biGalileon}, demonstrating that the resulting action is actually local. We can then apply the standard arguments known from the single spin-2 case leading to the conclusion that due to the double--epsilon structure of our non-derivative interactions, there are no dangerous higher derivative interactions contained in the helicity-0/helicity-2 decoupling limit of this theory.

\subsection{Helicity-one/Helicity-zero Sector}\label{sec:hel1_line}
In this subsection we work out the decoupling limit of the helicity-0 and helicity-1 interactions arising from the line-type non-derivative interactions \eqref{act_vierb_line3} between the two spin-2 fields. The relevant interactions can be captured by using the St\"uckelberg decomposition \eqref{stuck_line}, after integrating out the Lorentz St\"uckelberg fields. As already mentioned in the general analysis at the beginning of this section the two sets of Lorentz St\"uckelberg fields, $\Lambda^a\,_b$ and $\Gamma^a\,_b$, are decoupled in the case of line interactions. This means that the decoupling limit analysis for the mixed helicity-1 sector can be done separately for each of the two terms in \eqref{act_vierb_line3}, as in the case of two non-interacting massive spin-2 fields.  Hence, this means that, further decomposing the fields as \eqref{stuck}, and varying with respect to $\omega$ and $\sigma$, each of the fields will obey their respective symmetric vierbein conditions as in \eqref{sym_vierb_gen3} and that there are no new mixed helicity-1 interactions leading to the lowering of the strong coupling scale. For the sake of completeness we give the final result here, but refer the reader to the standard decoupling limit analysis \cite{Ondo:2013wka} of the helicity-1 sector for more details.

For the first term in \eqref{act_vierb_line3} we use the decomposition \eqref{stuck} as
\be\label{E3}
\begin{split}
E\Lambda=I+\frac{\omega}{\Lambda_2^2}+\frac{1}{2}\frac{\omega\cdot\omega}{\Lambda_2^4}+\mathcal O\left(\frac{1}{\M\Lambda_2^2}\right)\,,\qquad\partial\phi=I +\hat{\Pi}+\frac{\partial A}{\Lambda_2^2}\,,
\end{split}
\ee
where as for the cycle interactions we truncate the expansion by dropping higher orders of $\Lambda_2^2$. We also disregard the helicity-2 modes $h^a_\mu/M$ due to their additional $M^{-1}$ suppression. This leads to the decoupling limit interactions
\be\label{DL_vec_1}
\begin{split}
g_*^2 \L_{\text{DL,1}}=-\frac{\gamma^2 x^2}{2}\varepsilon\varepsilon&\left[\frac{\tilde\beta_1^{(1)}}{3!}\left(I+\hat{\Pi}\right)^2\left(\frac{1}{2}(I+\hat{\Pi})\,\omega\cdot\omega+3\,\partial A\,\omega\right)\right.\\
&+\frac{\tilde\beta^{(1)}_2}{2!2!}\left(I+\hat{\Pi}\right)\left[4I\partial A\,\omega+(I+\hat{\Pi})\left(I\,\omega\cdot\omega+\omega^2\right)\right]\\
&\left.+\frac{\tilde\beta^{(1)}_3}{2}I\left[I\partial A\,\omega+(I+\hat{\Pi})\left(\frac{1}{2}I\,\omega\cdot\omega+\omega^2\right)\right]\right]\,.
\end{split}
\ee
We note that this decoupling limit action has $\Lambda_3$ as the strong coupling scale. All terms appearing at a lower scale have cancelled either as total derivatives or because of the antisymmetric properties of~$\omega$. Let us also remark that it is known that the equations of motion for the Lorentz St\"uckelberg fields is independent on the coefficients $\tilde\beta_n^{(1)}$ (see \cite{Ondo:2013wka} and Appendix~\ref{sec:Leom}).
Hence varying any of the $\tilde\beta_n^{(1)}$ terms in the above action with respect to $\omega$ gives the standard symmetric vierbein condition allowing to express the Lorentz St\"uckelberg fields as before:
\be\label{con1}
2\omega_{ab}=\partial_{[a}A_{b]}-(\hat{\Pi}\cdot\omega)_{\[ab\]}\,.
\ee
For the second term in \eqref{act_vierb_line3} we use $E=I$ and
\be\label{F3}
\begin{split}
\Gamma F(\psi)\partial\psi = &I +\hat{\mathbb X}+\frac{1}{\Lambda_2^2}\left(\partial B+\sigma+\sigma\cdot\hat{\mathbb X}\right)\\
&+\frac{1}{\Lambda_2^4}\left(\sigma\cdot\partial B+\frac{1}{2}\sigma\cdot\sigma+\frac{1}{2}\sigma\cdot\sigma\cdot\hat{\mathbb X}\right)+\mathcal O\left(\frac{f}{\M}\right)\,.
\end{split}
\ee
In the decoupling limit this then  gives
\ba\label{DL_vec_2}
g_*^2 \L_{\text{DL,2}}&=&-\frac{1}{2}\varepsilon\varepsilon\Bigg\{\frac{\tilde\beta_1^{(2)}}{3!}I^3\left(\sigma\cdot\partial B+\frac{1}{2}\sigma\cdot\sigma+\frac{1}{2}\sigma\cdot\sigma\cdot\hat{\mathbb X}\right)\\
\nn
&+&\frac{\tilde\beta_2^{(2)}}{2!2!}I^2\left[(\partial\beta+\sigma+\sigma\cdot\hat{\mathbb X})^2+2(I+\hat{\mathbb X})\left(\sigma\cdot\partial B+\frac{1}{2}\sigma\cdot\sigma+\frac{1}{2}\sigma\cdot\sigma\cdot\hat{\mathbb X}\right)\right]\\
\nn
&+&\frac{\tilde\beta_3^{(2)}}{3!}I\left[3(I+\hat{\mathbb X})(\partial\beta+\sigma+\sigma\cdot\hat{\mathbb X})^2+3(I+\hat{\mathbb X})^2\left(\sigma\cdot\partial B+\frac{1}{2}\sigma\cdot\sigma+\frac{1}{2}\sigma\cdot\sigma\cdot\hat{\mathbb X}\right)\right]\\
\nn
&+& \frac{\tilde\beta_4^{(2)}}{4!}\left[6(I+\hat{\mathbb X})^2(\partial\beta+\sigma+\sigma\cdot\hat{\mathbb X})^2+4(I+\hat{\mathbb X})^3\left(\sigma\cdot\partial B+\frac{1}{2}\sigma\cdot\sigma+\frac{1}{2}\sigma\cdot\sigma\cdot\hat{\mathbb X}\right)\right]\Bigg\}\,,
\ea
which all enter at the scale $\Lambda_3$. The dangerous interactions present in the cycle case do not arise for the line of interactions. The reason for this is that only mixing between $
\sigma$ and $\hat{\mathbb X}$ (and not with $\hat{\Pi}$ occurs). Thus, similarly as in the case of the self interactions in the mass terms \eqref{badequation}, these terms vanish due to the antisymmetry of $\sigma$. Varying with respect to $\sigma$ gives the symmetric vierbein condition for $\sigma^a\,_b$:
\be\label{con2}
\sigma_{ab}=\partial_{[a}B_{b]}-(\hat{\mathbb X}\cdot\sigma)_{\[ab\]}\,.
\ee

The total helicity-1/helicity-0 decoupling limit is given in a closed form by the sum of \eqref{DL_vec_1} and \eqref{DL_vec_2}, together with the symmetric vierbein conditions \eqref{con1} and \eqref{con2}. Its strong coupling scale is $\Lambda_3$, as would be the case for two decoupled massive spin-2 fields. After integrating out the Lorentz St\"uckelberg fields (or, equivalently, imposing the symmetric vierbein conditions) this describes the interactions in the helicity-0/helicity-1 theory of line of interactions in a closed form. These coincide with the interactions in ghost-free massive gravity \cite{Ondo:2013wka}.

\subsection{Higher Derivative EFT corrections for Line Theories}\label{HigherderLine}
The higher derivative corrections that arise in the EFT for line theories will take the same form as those for the $\Lambda_3$ cycle theories in Sec.~\ref{sec:highdercycle}. The reason being is that the overall scaling of each of the arguments in the decoupling limit remains the same, and the higher derivative terms do not rely on any special cancellations, \emph{e.g.} through terms being total derivatives. Hence we either have
\be
g_*^2 {S}_{\text{higher-der}} = \int \d^4 x  \Lambda_3^4 \, {\cal F}\[E_{a\mu} , F_{b \mu}, \frac{\partial_{\rho}}{\Lambda_3} , \frac{M R[E]_{abcd}}{\Lambda_3^3}, \frac{M R[F]_{efgh}}{\Lambda_3^3}\] \, ,
\ee
assuming the helicity-two modes interact with other modes at the same $\Lambda_3$ scale or
\be
g_*^2 {S}_{\text{higher-der}} = \int \d^4 x  \Lambda_3^4 \, {\cal F}\[E_{a\mu} , F_{b \mu}, \frac{\partial_{\rho}}{\Lambda_3} , \frac{R[E]_{abcd}}{\Lambda_3^2}, \frac{ R[F]_{efgh}}{\Lambda_3^2}\] \, ,
\ee
assuming the helicity-two interactions are additionally suppressed, as is natural in models which for we may diagonalize/demix the helicity-2 interactions at leading order in the decoupling limit (as in Eq.~\eqref{demix}).

Technically speaking, when these higher derivative terms are added to the action, the unconstrained and constrained formulations are no longer equivalent. This is not a significant problem, and the same is true already in GR where different higher derivative Riemann operators are inequivalent in the first order and second order formulation. It is simply necessary to make a choice from the outset. Since the Lorentz \stu fields have no physical meaning, and it is enough to work with two 10 component symmetric tensors to describe two spin-2 fields, it is arguably better to work with the constrained formulation, as we did in the case of the cycle theories.

\section{Bi-Galileon Theory from the Line of Interactions}\label{sec:biGalileon}

Here we shall rewrite the decoupling limit action for the helicity-0/helicity-2 interactions derived in Section~\ref{sec:hel02_line} in the form of a bi-Galileon theory \cite{Padilla:2010de}. We shall follow closely a similar derivation derived in \cite{Fasiello:2013woa} in the context of bigravity theories which is generalized to multi-gravity in \cite{Noller:2015eda}. As a first step we rewrite the action \eqref{DL_sc_2} in the form:
\ba\label{DL_sc_3}
\begin{split}
g_*^2 S_{\text{DL,2}}=&-\int \d^4x\,f^{\mu\nu}\mathcal E^{\alpha\beta}_{\mu\nu}f_{\alpha\beta}+\frac{1}{4}\int \d^4x\,h^{\mu\nu}\tilde X_{\mu\nu}^{(2)}[\mathbb X]\\
&+\frac{1}{4}\int \d^4x\,f^{a}_{\nu}\left[x+\frac{\partial\chi}{\Lambda_3^3}\right]\left(\delta^{\nu}_\mu+\frac{\mathbb X^{\nu}_{\mu}}{\Lambda_3^3}\right)\tilde Y_{a}^{\mu}[\mathbb X]\,,
\end{split}
\ea
where the squared brackets indicate that $f_\mu^a$ is evaluated at $x+\partial\chi/\Lambda_3^3$. We also rewrite $\tilde Y_a^\mu$ defined in Eq.~\eqref{tildeY} as:
\be
\tilde Y_a^\mu[\mathbb X]=-2\Lambda_3^3\,\varepsilon^{\mu\dots}\varepsilon_{a\dots}\sum_{n=1}^4\frac{\tilde \beta^{(2)}_n}{(n-1)!(4-n)!}I^{4-n}\left(I+\frac{\mathbb X}{\Lambda_3^3}\right)^{n-1}\,,
\ee
where we use the notations such that, \emph{e.g.}
\be
\varepsilon^{\mu\dots}\varepsilon_{a\dots}I\left(I+\frac{\mathbb X}{\Lambda_3^3}\right)^2\equiv\varepsilon^{\mu\nu\alpha\beta}\varepsilon_{abcd}\,\delta^b_\nu\left(I+\frac{\mathbb X}{\Lambda_3^3}\right)^c_\alpha\left(I+\frac{\mathbb X}{\Lambda_3^3}\right)^d_\beta\,.
\ee

In order to write the term on the second line of \eqref{DL_sc_3} in a local form it is helpful to first recognize that it is a wedge product of combinations of the following one-forms
\ba
&& v_1^a = f^{a}_{\nu}\left[x+\frac{\partial\chi}{\Lambda_3^3}\right]\left(\delta^{\nu}_\mu+\frac{\mathbb X^{\nu}_{\mu}(x)}{\Lambda_3^3}\right) \d x^{\mu} \, , \\
&& v_2^a =\delta ^a_{\mu} \d x^{\mu} \, ,\\
&& v_3^a = \( \delta^{a}_{\mu} +\frac{\mathbb X^{a}_{\mu}(x)}{\Lambda_3^3} \) \d x^{\mu} \, .
\ea
We shall now perform a Galileon-duality transformation, which is best thought of in the present context as a field dependent diffeomorphism \cite{Fasiello:2013woa,Curtright:2012gx,deRham:2013hsa,deRham:2014lqa}.
We introduce new coordinates
\be\label{defZ1}
{x'}^a=x^a+\frac{\partial ^a\chi(x)}{\Lambda_3^3} \,,
\ee
and the dual Galileon field via
\be\label{defZ2}
x^a={x'}^a+\frac{{\partial'}^a\chi'(x')}{\Lambda_3^3}\,.
\ee
By differentiating both sides with respect to $x$ we have
\be\label{derZ}
\frac{\partial {x'}^a}{\partial x^b}=\left(I+\frac{\mathbb X(x)}{\Lambda^3_3}\right)^a{}_b\,.
\ee
or
\be
\frac{\partial x^a}{\partial {x'}^b}={\left(I+\frac{\mathbb X(x)}{\Lambda^3_3}\right)^{-1}}^a{}_b  =  \left(I+\frac{\mathbb X'(x')}{\Lambda^3_3}\right)^a{}_b    \,,
\ee
where we have defined $\mathbb X'\,^a_\mu(x') \equiv\partial'_\mu{\partial'}^a\chi'(x')$. In terms of the dual coordinates, the one-forms become
\ba
&& v_1^a = f^{a}_{\mu}[x'] \d {x'}^{\mu} \, ,\\
&& v_2^a =\( \delta^{a}_{\mu} +\frac{\mathbb {X'}^{a}_{\mu}(x')}{\Lambda_3^3} \) \d {x'}^{\mu} \, ,  \\
&& v_3^a =  \delta ^a_{\mu} \d {x'}^{\mu} \, .
\ea
Once these transformations have been done for every term on the second line of the action \eqref{DL_sc_3}, we now replace the dummy label $x'$ with $x$, leaving the interactions as a function of $ f^{a}_{\nu}[x]$ and the dual Galileon field $\chi'(x)$. The resulting local action is
\ba\label{DL_sc_4}
\begin{split}
g_*^2 S_{\text{DL,2}}=&-\int \d^4x\,f^{\mu\nu}\mathcal E^{\alpha\beta}_{\mu\nu}f_{\alpha\beta}+\frac{1}{4}\int \d^4x\,h^{\mu\nu}\tilde X_{\mu\nu}^{(2)}[\mathbb X]+\frac{1}{4}\int \d^4x\, f^{a}_{\mu}(x)\mathcal  Y_{a}^{\mu}[\mathbb {X'}]\,,
\end{split}
\ea
where we define
\be
\mathcal  Y_a^\mu[\mathbb {X'}]=-2\Lambda_3^3\,\varepsilon^{\mu\dots}\varepsilon_{a\dots}\sum_{n=1}^4\frac{\tilde \beta^{(2)}_n}{(n-1)!(4-n)!}\(I + \frac{\mathbb {X'}(x)}{\Lambda_3^3}\)^{4-n}I^{n-1}\,.
\ee
Combining the results here for the decoupling limit action \eqref{DL_sc_2} together with the other part of the action \eqref{DL_sc_1}, derived in Section~\ref{sec:hel02_line} we obtain the full helicity-0/helicity-2 decoupling limit theory for line interactions written in a local form. The result is a bi-Galileon theory of two helicity-2 fields $h_{\mu\nu}$, $f_{\mu\nu}$ and two Galileon fields $\pi$ and $\chi$ (or $\chi'$, related to $\chi$ through the duality transformations \eqref{defZ1} and \eqref{defZ2}). After some trivial restructuring the Lagrangian reads:
\ba\label{DL_total}
\begin{split}
g_*^2 \L_{\text{DL}}=&-\gamma^2h^{\mu\nu}\mathcal E^{\alpha\beta}_{\mu\nu}h_{\alpha\beta}-f^{\mu\nu}\mathcal E^{\alpha\beta}_{\mu\nu}f_{\alpha\beta}\\
&+\frac{1}{4}\gamma^2 x^2h^{\mu\nu}\tilde X^{(1)}_{\mu\nu}[\Pi]+\frac{1}{4}h^{\mu\nu}\tilde X_{\mu\nu}^{(2)}[\mathbb X]+\frac{1}{4} f^{\mu\nu}\tilde X^{(3)}_{\mu\nu}[\mathbb X']\,,
\end{split}
\ea
where similarly as in Section~\ref{sec:hel02_line} we define
\ba
\label{eq:BiGalX_1}
\tilde X^{(i)}_{\mu\nu}[\Pi]&\equiv&\,\tilde\kappa_2^{(i)}\left(\varepsilon\varepsilon II\Pi\right)_{\mu\nu}
+\frac{\tilde\kappa_3^{(i)}}{\Lambda_3^3}\left(\varepsilon\varepsilon I\Pi\Pi\right)_{\mu\nu}
+\frac{\tilde\kappa_4^{(i)}}{\Lambda_3^6}\left(\varepsilon\varepsilon \Pi\Pi\Pi\right)_{\mu\nu} \,,
\ea
but this time with $i=1,2,3$ and the various $\tilde \kappa^{(i)}_n$'s are defined in Appendix~\ref{App:biGalileondetails}, see Eqns.~(\ref{eq:exp1}--\ref{eq:exp2}).\\

In the special case when $\kappa^{(i)}_4=0$ for $i=1,2,3$ one can write the action \eqref{DL_total} in a manifestly bi-Galileon form \cite{Padilla:2010de}. This can be achieved by first decoupling/demixing the helicity-0 and helicity-2 modes by the following field redefinitions:
\be
\begin{split}\label{demix}
&h_{\mu\nu}\to h_{\mu\nu}+\frac{1}{4}x^2\tilde\kappa_2^{(1)}\pi\eta_{\mu\nu}-x^2\frac{\tilde\kappa^{(1)}_3}{4\Lambda_3^3}\partial_\mu\pi\partial_\nu\pi+\frac{1}{4\gamma^2}\tilde\kappa^{(2)}_2\chi\eta_{\mu\nu}-\frac{1}{\gamma^2}\frac{\tilde\kappa^{(2)}_3}{4\Lambda_3^3}\partial_\mu\chi\partial_\nu\chi\,,\\
&f_{\mu\nu}\to f_{\mu\nu}+\frac{1}{4}\tilde\kappa^{(3)}_2\chi'\eta_{\mu\nu}-\frac{\tilde\kappa^{(3)}_3}{4\Lambda_3^3}\partial_\mu\chi'\partial_\nu\chi'\,.
\end{split}
\ee
By using the relationships
\be
\mathcal E^{\alpha\beta}_{\mu\nu}\pi\eta_{\alpha\beta}=\frac{1}{2}(\varepsilon\varepsilon II\Pi)_{\mu\nu}\,,\qquad\mathcal E^{\alpha\beta}_{\mu\nu}\partial_\alpha\pi\partial_\beta\pi=-\frac{1}{2}(\varepsilon\varepsilon I\Pi\Pi)_{\mu\nu}\,,
\ee
we arrive at the following form of \eqref{DL_total}:
\be\label{DL_step1}
\begin{split}
g_*^2 S_{\text{DL}}=&-\gamma^2\int \d^4x\,h^{\mu\nu}\mathcal E^{\alpha\beta}_{\mu\nu}h_{\alpha\beta}-\int \d^4x\,f^{\mu\nu}\mathcal E^{\alpha\beta}_{\mu\nu}f_{\alpha\beta}\\
&+\frac{\gamma^2 x^4}{2}\int \d^4x\,\mathcal L_{\text{galileon}}\left(\pi,\tilde\kappa^{(1)}_n\right)+\frac{1}{2\gamma^2}\int \d^4x\,\mathcal L_{\text{galileon}}\left(\chi,\tilde\kappa^{(2)}_n\right)+\frac{1}{2}\int \d^4x\,\mathcal L_{\text{galileon}}\left(\chi',\tilde\kappa^{(3)}_n\right)\\
&+\frac{x^2}{\gamma^2}\int \d^4x\,\mathcal L_{\text{bigalileon}}\left(\pi,\tilde\kappa^{(1)}_n;\chi,\tilde\kappa^{(2)}_n\right)\,,
\end{split}
\ee
where $\mathcal L_{\text{galileon}}$ stands for the quartic Galileon Lagrangians \cite{Nicolis:2008in} defined as
\be
\mathcal L_{\text{galileon}}\left(\pi,\tilde\kappa^{(i)}\right)\equiv\pi\left[(\tilde\kappa_2^{(i)})^2\varepsilon\varepsilon I I I \Pi+2\tilde\kappa^{(i)}_2\tilde\kappa^{(i)}_3\frac{1}{\Lambda_3^3}\varepsilon\varepsilon I I \Pi\Pi+(\tilde\kappa^{(i)}_3)^2\frac{1}{\Lambda_3^6}\varepsilon\varepsilon I\Pi\Pi\Pi\right]\,,
\ee
and the mixed interactions $\mathcal L_{\text{bigalileon}}$ are given by
\be
\begin{split}
\mathcal L_{\text{bigalileon}}\left(\pi,\tilde\kappa^{(1)}_n;\chi,\tilde\kappa^{(2)}_n\right)\equiv\pi&\left[\tilde\kappa^{(1)}_2\tilde\kappa^{(2)}_2\varepsilon\varepsilon III\mathbb X+\tilde\kappa^{(1)}_2\tilde\kappa^{(2)}_3\frac{1}{\Lambda_3^3}\varepsilon\varepsilon I I \mathbb X\mathbb X+\tilde\kappa^{(1)}_3\tilde\kappa^{(2)}_2\frac{1}{\Lambda_3^3}\varepsilon\varepsilon I I \Pi\mathbb X\right.\\
&\left.+\tilde\kappa^{(1)}_3\tilde\kappa^{(3)}_3\frac{1}{\Lambda_3^6}\varepsilon\varepsilon I  \mathbb X\mathbb X\Pi\right]\,.
\end{split}
\ee
Note that the bi-Galileon Lagrangian is symmetric under exchanging $\mathcal L_{\text{bigalileon}}\left(\pi,\tilde\kappa^{(1)}_n;\chi,\tilde\kappa^{(2)}_n\right)=\mathcal L_{\text{bigalileon}}\left(\chi,\tilde\kappa^{(2)}_n;\pi,\tilde\kappa^{(1)}_n\right)$.

We next perform the Galileon duality transformation, inverse to the one we performed before in Eqns.~\eqref{defZ1} and \eqref{defZ2}, in the last term on the second line of \eqref{DL_step1}. We use the known relationships \cite{deRham:2013hsa} between the coefficients in the dual Galileon actions written in the form
\be
\int \d^4x\,\mathcal L_{\text{galileon}}\left(\chi',\tilde\kappa^{(3)}_n\right)=\Lambda_3^6\int \d^4x\,\sum_{n=2}^5 c_n\,\chi'\,\mathcal U_{n-1}\left(\frac{\mathbb X'}{\Lambda_3^3}\right)=\Lambda_3^6\int \d^4x\,\sum_{n=2}^5 p_n\,\chi\,\mathcal U_{n-1}\left(\frac{\mathbb X}{\Lambda_3^3}\right)\,
\ee
with
\be
c_2=(\tilde\kappa_2^{(3)})^2\,,\quad c_3=2\tilde\kappa^{(3)}_2\tilde\kappa^{(3)}_3\,,\quad c_4= (\tilde\kappa^{(3)}_3)^2\,,\quad c_5=0\,,
\ee
and the dual coefficients given by
\be
\begin{split}
p_2=c_2\,,\quad p_3=2c_2-c_3\,,\quad p_4=\frac{3}{2}c_2-\frac{3}{2}c_3+c_4\,,\quad p_5=\frac{1}{5}\left(2c_2-3c_3+4c_4-5c_5\right)\,.
\end{split}
\ee
Thus our final result for the decoupling limit action for the helicity-0/helicity-2 sector in a theory with line interactions becomes (see also \cite{Noller:2015eda}):
\be\label{DL_final}
\begin{split}
g_*^2 S_{\text{DL}}=&-\gamma^2\int \d^4x\,h^{\mu\nu}\mathcal E^{\alpha\beta}_{\mu\nu}h_{\alpha\beta}-\int \d^4x\,f^{\mu\nu}\mathcal E^{\alpha\beta}_{\mu\nu}f_{\alpha\beta}\\
&+\frac{\gamma^2 x^4}{2}\int \d^4x\,\mathcal L_{\text{galileon}}\left(\pi,\bar\kappa^{(1)}_n\right)+\frac{1}{2\gamma^2}\int \d^4x\,\mathcal L_{\text{galileon}}\left(\chi,\bar\kappa^{(2)}_n\right)\\
&+\frac{x^2}{\gamma^2}\int \d^4x\,\mathcal L_{\text{bigalileon}}\left(\pi,\chi\right)\,,
\end{split}
\ee
where the Galileon and bi-Galileon Lagrangians are defined respectively as
\be
\label{eq:Gal2}
\mathcal L_{\text{galileon}}\left(\pi,\bar\kappa^{(i)}\right)\equiv\pi\left[\bar\kappa^{(i)}_2\varepsilon\varepsilon I I I \Pi+\frac{\bar\kappa^{(i)}_3}{\Lambda_3^3}\varepsilon\varepsilon I I \Pi\Pi+\frac{\bar\kappa^{(i)}_4}{\Lambda_3^6}\varepsilon\varepsilon I\Pi\Pi\Pi+\frac{\bar\kappa^{(i)}_5}{\Lambda_3^9}\varepsilon\varepsilon \Pi\Pi\Pi\Pi\right]\,,
\ee
and
\be
\begin{split}
\label{eq:biGal2}
\mathcal L_{\text{bigalileon}}\left(\pi,\chi\right)\equiv\pi&\left[\bar\kappa_{11}\varepsilon\varepsilon III\mathbb X+\frac{\bar\kappa_{12}}{\Lambda_3^3}\varepsilon\varepsilon I I \mathbb X\mathbb X+\frac{\bar\kappa_{21}}{\Lambda_3^3}\varepsilon\varepsilon I I \Pi\mathbb X+\frac{\bar\kappa_{22}}{\Lambda_3^6}\varepsilon\varepsilon I  \mathbb X\mathbb X\Pi\right]\,,
\end{split}
\ee
where the various coefficients appearing in the above action are expressed in terms of the original coefficients $\tilde\beta^{(i)}_n$ in the line interactions \eqref{act_nonder_line2} in \eqref{eq:exp3} and \eqref{eq:exp4}.

\section{EFT for Multiple Interacting Spin-2 Fields}\label{sec:multi}

The vierbein interactions can easily be extended to arbitrary number of fields \cite{Hinterbichler:2012cn} and arise naturally in the dimensional deconstruction framework \cite{deRham:2013awa} (for earlier metric based work see \cite{ArkaniHamed:2003vb,Schwartz:2003vj,Deffayet:2003zk,Deffayet:2005yn}). A crucial observation for this is that in four spacetime dimensions there are only two additional interaction vertices of the double--epsilon structure that have to be added to \eqref{act_nonder} in the case when there are more fields present. These correspond to quartic interaction vertices mixing three and four of the dynamical vierbeins. If we label the set of $N$ dynamical vierbeins as $E^{(i)}$, $i=1,2,\dots, N$, the total non-derivative Lagrangian reads:
\be\label{act_full}
\begin{split}
g_*^2\mathcal L_{\text{non-der}}=&\sum _{i=1}^N\frac{m_i^2M_i^2}{4}\sum_{n=0}^4\beta^{(i)}_{n}\,\mathcal U_{n}(I,E^{(i)})\\
&+\sum_{\substack{i,j=1\\ i<j}}^{N}\frac{m^2M^2}{4}\sum_{n_i=1}^4\sum_{n_j=1}^{4-n_i}\beta^{(ij)}_{n_in_j}\,\mathcal U_{n_in_j}(I,E^{(i)},E^{(j)})\\
&+\sum_{\substack{i,j,k=1\\i<j<k}}^{N}\frac{m^2M^2}{4}\sum_{n_i=1}^4\sum_{n_j=1}^{4-n_i}\sum_{n_k=1}^{4-(n_i+n_j)}\beta^{(ijk)}_{n_in_jn_k}\,\mathcal U_{n_in_jn_k}(I,E^{(i)},E^{(j)},E^{(k)})\\
&+\sum_{\substack{i,j,k,l=1\\i<j<k<l}}^{N}\frac{m^2M^2}{4}\beta^{(ijkl)}_{1111}\,\mathcal U_{1111}(E^{(i)},E^{(j)},E^{(k)},E^{(l)})\,,
\end{split}
\ee
where on the last three lines $i,j,k=1,2,\dots, N$ and $m$ and $M$ are some mass scales that depend on the choice of normalization of the interaction terms. The first line are the individual mass terms for each of the dynamical metrics. The second line contains all the pairwise interactions between two of the dynamical vierbeins $E^{(i)}$, $E^{(j)}$, including both cubic and quartic interactions between the two. The third and the fourth lines show the couplings between three and four dynamical vierbeins respectively. To avoid the double counting between the terms contained in the different lines we start the sums on the last three lines from $n_i,n_j,n_k=1$. \\

Written in this form the action above is the most convenient for describing a line of interactions between the $N$ vierbeins and the reference vierbein $I$. If we choose to number the vierbeins as $E^{(1)},E^{(2)},\dots,E^{(N)}$ then the action for the line of interactions is given simply by
\be\label{act_multi_line}
\begin{split}
g_*^2\mathcal L_{\text{line}}=&\frac{m_{1}^2M_1^2}{4}\sum_{n=0}^4\beta^{(1)}_{n}\,\mathcal U_{n}(I,E^{(1)})+\sum _{i=2}^N\frac{m_i^2M_i^2}{4}\sum_{n=0}^4\beta^{(i)}_{n}\,\mathcal U_{n}(E^{(i-1)},E^{(i)})\,.
\end{split}
\ee
Following the decoupling limit analysis for the case of two interacting vierbeins it is clear that the main conclusions still hold for the case of $N$ vierbeins, provided we make the same assumption that the hierarchy between the masses $m_i$ and interaction scales $M_i$ is small in comparison to the hierarchy between masses. As long as this is true we can continue to define the same type of decoupling limit.
In particular, in this theory the constrained and unconstrained formalisms are equivalent, it is ghost-free, with the strong coupling scale $\Lambda_3$. In fact the full helicity-2/helicity-0 decoupling limit is worked out closely following the same methodology in \cite{Noller:2015eda}. The known absence of Boulware-Deser ghosts for line theories \cite{Hinterbichler:2012cn} implies that the decoupling limit analysis for the helicity-1/helicity-0 sector will be equally unproblematic. With this in mind, we may easily identify the typical EFT corrections in the form
\be \label{scaling5}
g_*^2 {S}_{\text{higher-der}} = \int \d^4 x  \Lambda_3^4 \, {\cal F}\[{E^{i}}^a_{\mu} ,\frac{\partial_{\mu}}{\Lambda_3},\frac{M R[E^i]_{abcd}}{\Lambda_3^3}\] \, ,
\ee
if the helicity-2 states also have $\Lambda_3$ interactions, or more typically
\be\label{scaling6}
g_*^2 {S}_{\text{higher-der}} = \int \d^4 x  \Lambda_3^4 \, {\cal F}\[{E^{i}}^a_{\mu} ,\frac{\partial_{\mu}}{\Lambda_3},\frac{R[E^i]_{abcd}}{\Lambda_3^2}\] \, ,
\ee
when the helicity-2 states are additionally $M$ suppressed.
 \\

For cycles of interactions it is more convenient to write the action \eqref{act_full} in terms of the vierbein `perturbations', \emph{i.e.} in terms of $(E^{(i)}-I)$ instead of $E^{(i)}$ as we did in \eqref{act_nonder2}. In the case of $N$ dynamical vierbeins this is given by
\be\label{act_multi_cycle}
\begin{split}
g_*^2\L_{\text{cycle}}&=\sum _{i=1}^N\frac{m_i^2M_i^2}{4}\,\sum_{n=0}^4\kappa^{(1)}_{n}\,\mathcal U_{n}(I,E^{(i)}-I)+\\
&+\sum_{\substack{i,j=1\\ i<j}}^{N}\frac{m^2M^2}{4}\,\sum_{n_i=1}^4\sum_{n_j=1}^{4-n_i}\kappa_{n_in_j}^{(ij)}\,\mathcal U_{n_in_j}(I,E^{(i)}-I,E^{(j)}-I)\\
&+\sum_{\substack{i,j,k=1\\i<j<k}}^{N}\frac{m^2M^2}{4}\sum_{n_i=1}^4\sum_{n_j=1}^{4-n_i}\sum_{n_k=1}^{4-(n_i+n_j)}\kappa^{(ijk)}_{n_in_jn_k}\,\mathcal U_{n_in_jn_k}(I,E^{(i)}-I,E^{(j)}-I,E^{(k)}-I)\\
&+\sum_{\substack{i,j,k,l=1\\i<j<k<l}}^{N}\frac{m^2M^2}{4}\kappa^{(ijkl)}_{1111}\,\mathcal U_{1111}(E^{(i)}-I,E^{(j)}-I,E^{(k)}-I,E^{(l)}-I)\,,
\end{split}
\ee
with $\kappa_{11}^{(ij)}=0$ to avoid linear mixing. Again, also in this case we still know that adding more fields do not change the main conclusions that we have learned in Sections~\ref{sec:cycle} and~\ref{sec:strong}. Most importantly, the perturbative unitarity in this theory with generic couplings $\kappa\sim\mathcal O(1)$ is broken by the terms on the second, third and fourth line of the above action at the scale $\Lambda_{7/2}$. That is because the dangerous helicity-1/helicity-0 interactions \eqref{bad_op} that arose before, occur whenever two distinct helicity-0 modes couple within the same interaction vertex to Lorentz \stu fields. In the case of multiple interacting spin-2 theories with any cycle interactions, we only increase the number of dangerous interactions of this form. Once again, these conclusions hold both in constrained and unconstrained formulation. As discussed in section \ref{sec:strong} we choose either to accept the lower cutoff scale and organize the EFT in a manner similar to \eqref{scaling3} and \eqref{scaling4}, or we may rescale the coefficients of the dangerous helicity-1/helicity-0 interactions as in \eqref{scaling} and define the cycle theory as a $\Lambda_3$ EFT. If this is done then the correct form of the EFT corrections is \eqref{scaling5} or \eqref{scaling6}. \\

We should note that in considering large $N$ spin-2 states, the actual scale at which perturbative unitarity is violated may additionally scale with some power of $N$. The precise scaling will depend on the details of the interactions between the spin-2 states and we do not consider this here (see for example \cite{Scargill:2015wxs,deRham:2013awa} for related discussions).

\section{Discussion}\label{sec:conclusions}
In this paper we have analyzed the EFT of interacting multiple massive spin-2 fields in Minkowski spacetime with the highest possible EFT cutoff, focussing in particular on the case of two spin-2 fields. Our principle focus is spin-2 states which acquire masses through the breaking of diffeomorphism symmetries. Unsurprisingly, the mixed interactions between the spin-2 fields must take the double--epsilon form characteristic to ghost-free massive gravity theories \cite{deRham:2010kj}. This result is inferred by performing a decoupling limit analysis in the helicity-zero/helicity-two mode sector. Even given this, there are two different classes of interacting theories: `Cycle theories', where the mass eigenstates are chosen to interact through non-derivative interactions, but whose kinetic interactions are diagonalized and standard, or `Line theories' where the mass eigenstates arise through mass mixing, hence leading to non-trivial kinetic mixing between mass eigenstates. Both classes of theories are acceptable starting point for EFT constructions of interactions between multiple spin-2 fields. \\

We perform the full decoupling limit analysis of both classes of theories in order to determine the most relevant interactions, and the cutoff of the two EFTs. For the case of cycle theories, a novel and previously unexpected result is that if these interactions are chosen to arise in the Lagrangian in a similar manner to the spin-2 self-interactions, then the cutoff of the EFT is lowered to the parametrically smaller energy scale $\Lambda_{7/2}$. This arises due to novel interactions in the helicity-one/helicity zero mode sector that have no analogue in the single spin-2 case. These interactions are higher derivative in nature, and are shown to contribute non-trivially to scattering amplitudes indicating that they are physical. This is consistent with the presented ADM Hamiltonian analysis in the metric (constrained vierbein) formalism and previous works on unconstrained vierbein formulations of the same theory which indicate the presence of a BD ghost. Our decoupling limit identifies the energy scale of this ghost, or more precisely the cutoff of the EFT beyond which the would-be BD ghost is banished. We show that by performing a technically natural tuning of the mixed interactions, it is possible to raise the cutoff of the combined theory to $\Lambda_3$, which is the maximum allowed scale for an interacting theory of two massive spin-2 particles in Minkowski spacetime. The line theories by contrast naturally give EFTs whose cutoff scale is $\Lambda_3$ (for all helicity states) without any further special tuning (beyond the double epsilon structure). This is consistent with previous decoupling limit results \cite{Fasiello:2013woa}.  In both cases we identify the generic form of EFT corrections that would come from either a weakly or strongly coupled UV completion.\\

The majority of these results extend straightforwardly to theories of multiple spin-2 particles provided that the hierarchy between the various physical masses $m_i$ and the various interaction scales $M_i$ is large in comparison to the hierarchy between the different physical masses. As long as this is true, it is still meaningful to talk about a $\Lambda_3$ (more generally $\Lambda_n$) decoupling limit. All of the tunings and scalings considered are technically natural, \emph{i.e.} stable under radiative corrections. Hence from an EFT point of view alone, there are no further constraints on the interaction coefficients (the various $\beta$'s and $\kappa$'s). This situation changes dramatically when we impose positivity bound requirements as we do in a forthcoming work \cite{AlberteTA}.

\bigskip
\noindent{\textbf{Acknowledgments:}}
AJT and CdR would like thank the Perimeter Institute for Theoretical Physics for its hospitality during part of this work and for support from the Simons Emmy Noether program.
LA is supported by the European Research Council under the European Union's Seventh Framework Programme (FP7/2007-2013),
ERC Grant agreement ADG 339140.
The work of AJT and CdR is supported by an STFC grant ST/P000762/1. JR is supported by an STFC studentship. CdR thanks the Royal Society for support at ICL through a Wolfson Research Merit Award. CdR is supported by the European Union's Horizon 2020 Research Council grant 724659 MassiveCosmo ERC-2016-COG and by a Simons Foundation award ID 555326 under the Simons Foundation's Origins of the Universe initiative, `\textit{Cosmology Beyond Einstein's Theory}'. AJT thanks the Royal Society for support at ICL through a Wolfson Research Merit Award.

\bigskip

\appendix

\section{Metric Formulation}\label{sec:metric}
\subsection{$\Lambda_3$ EFT}\label{drgt rev}
Here we review the highest cutoff EFT of a single massive spin-2 field. As was said in the main text, it is known that for generic interactions, the EFT of massive spin-2 field breaks (perturbative) unitarity  at the scale $\Lambda_5=(m^4 M)^{1/5}$ \cite{ArkaniHamed:2002sp,Creminelli:2005qk,Deffayet:2005ys}. There is however a unique set of fully nonlinear interactions for which the cutoff can be raised to much larger values, known as the $\Lambda_3$-EFT for a massive spin-2 field \cite{deRham:2010ik,deRham:2010kj}. That particular theory requires tunings that are not protected by any symmetry but remain stable under quantum corrections \cite{deRham:2012ew,deRham:2013qqa} and therefore leads to a meaningful quantum EFT.

In the context of the present paper the $\Lambda_3$ EFT is most conveniently written in the vierbein formalism \cite{Hinterbichler:2012cn}:
\be\label{act_vierbein}
\begin{split}
g_*^2S&=\frac{M^2}{2}\left[\int\frac{1}{4}\varepsilon_{abcd}\,E^a\wedge E^b\wedge R^{cd}\[E\]-m^2\int \d^4x\sum_{n=0}^4\frac{\beta_n}{n!(4-n)!}\,\mathcal U_{n}(I,E)\right]\,,
\end{split}
\ee
where $I^a=\delta^a_\mu \d x^\mu$ and $E^a=E^a_\mu \d x^\mu$ and the vierbein $E^a_\mu$ can be related to the `would-be' metric through $g_{\mu\nu}=E^a_\mu E^b_\nu\eta_{ab}$. The first term above then is the standard Einstein--Hilbert term associated to the `would-be' metric $g\mn$, and the non-derivative interactions are given by the double--epsilon potentials, conveniently written in the following notation:
\be\label{Unm}
\mathcal U_{n}(I,E)\equiv \varepsilon\varepsilon E^nI^{4-n}\equiv\varepsilon_{a_1\dots a_na_{n+1}\dots a_4}\varepsilon^{\mu_1\dots \mu_n \mu_{n+1}\dots\mu_4}E^{a_1}_{\mu_1}\dots E^{a_n}_{\mu_n}I^{a_{n+1}}_{\mu_{n+1}}\dots I^{a_4}_{\mu_4}\,,
\ee
where $\varepsilon_{abcd}$ is the flat space Levi-Civita symbol\footnote{Let us emphasize that here and henceforth we are using the anti-symmetric Euclidean Levi-Civita \emph{symbol}  with $\varepsilon_{i_1\dots i_ki_{k+1}\dots i_d}\varepsilon^{i_1\dots i_kj_{k+1}\dots j_d}=k!\delta_{i_{k+1}\dots i_d}^{j_{k+1}\dots j_d}$ where the generalized Kronecker delta is expressed as a determinant of a matrix built out of $\delta$'s. In a sense we are abusing the notation here, since it looks as if one could be lowering/raising the indices of the epsilons with a Minkowski metric. For this to be consistent we should add an overall minus sign everywhere because $\varepsilon^{\mu\nu\alpha\beta}_{\text{Euclidean}}=-\varepsilon^{\mu\nu\alpha\beta}_{\text{Lorentzian}}$. To avoid this, we shall never raise nor lower the indices directly on the Levi-Civita \emph{symbols}.} so that, \emph{e.g.},
\be
\int \d^4x\,\mathcal U_{3}(I,E)=\int \varepsilon_{abcd}\,\delta^a_\mu \d x^\mu\wedge E^b_\nu \d x^\nu\wedge E^c_\alpha \d x^\alpha\wedge E^d_\beta \d x^\beta=\int\varepsilon_{abcd}\,I^a\wedge E^b\wedge E^c\wedge E^d\,.
\ee
The action \eqref{act_vierbein} can thus be written explicitly as
\be\label{act_vierbein2}
\begin{split}
g_*^2S=\frac{M^2}{2}\varepsilon_{abcd}\int&\left[\frac{1}{4}E^a\wedge E^b\wedge R^{ab}\[E\]-m^2\left(\frac{\beta_0}{4!}I^a\wedge I^b\wedge I^c\wedge I^d\right.\right.\\
+&\left.\left.\frac{\beta_1}{3!}I^a\wedge I^b\wedge I^c\wedge E^d+\frac{\beta_2}{2!2!}I^a\wedge I^b\wedge E^c\wedge E^d\right.\right.\\
+&\left.\left.\frac{\beta_3}{3!}I^a\wedge E^b\wedge E^c\wedge E^d+\frac{\beta_4}{4!}E^a\wedge E^b\wedge E^c\wedge E^d\right)\right]\,.
\end{split}
\ee
The coefficients $\beta_n$ here are arbitrary constant coefficients; three of them are usually fixed by imposing the conditions for the absence of cosmological constant and a tadpole, and by the normalization of the mass of the spin-2 field. \\

It is straightforward to relate the vierbein to the symmetric spin-2 field perturbations $h_{\mu\nu}$ by defining\footnote{Throughout this appendix we choose not to canonically normalize the perturbation.}
\be
E^a_\mu\equiv\delta^a_\mu+h^a_\mu\,.
\ee
If the symmetric vierbein condition
\be\label{sym_vierb_gen}
\eta E =( \eta E )^T\,,
\ee
equivalent to $E_{a\mu}=E_{\mu a}$, is satisfied (or imposed) this implies that $h_{a\mu}=h_{\mu a}$ thus providing an equivalent mapping between the vierbein and metric formulations. In terms of the `would-be' metric $g\mn$ and of the tensor
\ba
\label{eq:K}
\mathcal K\mupn(g,\eta)=\delta^\mu_\nu-\left(\sqrt{g^{-1}\eta}\right)\mupn\;,
\ea
where $g^{\mu\nu}$ is the inverse of $g\mn$, and $\eta_{\mu\nu}$ is the Minkowski reference metric, the standard ghost-free massive gravity Lagrangian for a single massive spin-2 field is
\begin{equation}\label{act0}
g_*^2 \L_{\Lambda_3}[g]=\frac{M^{2}}{2}\sqrt{-g}\left[R[g]+\frac{m^2}{2}\,\sum_{n=0}^4\alpha_n\,\U_n\left[\mathcal K(g,\eta)\right]\right]\;,
\end{equation}
with $\alpha_0=\alpha_1=0$  (ensuring the absence of tadpole and cosmological constant)  and $\alpha_2$ = 1. The two remaining coefficients $\alpha_3\,,\alpha_4$ are the free parameters of the theory together with the graviton mass $m$. The relevant terms in the potential are defined as
\begin{align}
&\U_2(\mathcal K)=2\left([\K]^2-[\K^2]\right)\;,\\
&\U_3(\K)=[\K]^3-3[\K][\K^2]+2[\K^3]\;,\\
&\U_4(\K)=[\K]^4-6[\K^2][\K]^2+8[\K^3][\K]+3[\K^2]^2-6[\K^4]\;,
\end{align}
where the squared brackets denote the traces. The potential terms can be written in the double--epsilon form in terms of the flat space Levi-Civita symbol as
\begin{equation}\label{mass1}
\begin{split}
&\U_2(\K)=\varepsilon_{\mu\nu\alpha\beta}\varepsilon^{\mu\nu\alpha'\beta'}\K^{\alpha}_{\alpha'}\K^{\beta}_{\beta'}\,, \\
&\U_3(\K)=\varepsilon_{\mu\nu\alpha\beta}\varepsilon^{\mu\nu'\alpha'\beta'}\K^{\nu}_{\nu'}\K^{\alpha}_{\alpha'}\K^{\beta}_{\beta'}\,,\\
&\U_4(\K)=\varepsilon_{\mu\nu\alpha\beta}\varepsilon^{\mu'\nu'\alpha'\beta'}\K^{\mu}_{\mu'}\K^{\nu}_{\nu'}\K^{\alpha}_{\alpha'}\K^{\beta}_{\beta'}\;.
\end{split}
\end{equation}
We define the vierbein-inspired metric perturbations as \cite{deRham:2013qqa,deRham:2014zqa}
\be\label{def_vierb_pert}
g_{\mu\nu}=\left(\eta_{\mu\nu}+h_{\mu\nu}\right)^2\equiv\left(\eta_{\mu\alpha}+h_{\mu\alpha}\right)\eta^{\alpha\beta}\left(\eta_{\beta\nu}+h_{\beta\nu}\right)\,,
\ee
directly related to the constrained vierbein perturbations $E^a_\mu=\delta^a_\mu+h^a_\mu$ when the symmetric vierbein condition is imposed.
In terms of the metric perturbations defined in this way it is straightforward to take the square root of the following matrix
\be\label{Kmunu1}
\mathcal K^\mu_\nu(\eta,g )=\delta^\mu_\nu-\left(\sqrt{\eta^{-1}g}\right)^\mu_\nu=-\eta^{\mu\alpha}h_{\alpha\nu}\,,
\ee
so that the massive gravity action \eqref{act0} becomes
\be\label{drgt}
g_*^2 \L_{\Lambda_3}[g]=\frac{M^{2}}{2}\sqrt{-g}R+\frac{m^2M^{2}}{4}\sqrt{-\eta}\,\sum_{n=0}^4\kappa_n\,\U_n\left[\eta^{-1}h\right]\;
\ee
up to \emph{arbitrary} power in metric perturbations $h_{\mu\nu}$.
We stress that the 10 component vierbein defined as $\eta_{\mu\alpha}+h_{\mu\alpha}$ is a {\it constrained vierbein} since its symmetry is imposed from the outset. This should not be confused with the 16 component {\it unconstrained} vierbein that enters in first order formulations of GR and massive gravity as presented in the main text. The inequivalence of these two formulations in the case of two massive spin-2 particles is an important result \cite{Hinterbichler:2012cn,deRham:2015cha} which is why we must be clear from the outset which formalism we are working in.

The coefficients $\kappa_n$ are the same $\kappa$'s that appear in the vierbein action \eqref{act_nonder2} and are related to $\alpha_n$'s as $\kappa_n=\mathbb A_{nm} \alpha_m$ with $\mathbb A$ given by \cite{deRham:2014zqa}
\be\label{coeffsA}
\mathbb A=
\begin{pmatrix}
1&0&0&0&0\\
4&1&0&0&0\\
6&3&1&0&0\\
4&3&2&1&0\\
1&1&1&1&1
\end{pmatrix}\,.
\ee
For our choice of $\alpha_n$'s this means
\be
\kappa_0=\kappa_1=0\,,\quad\kappa_2=1\,,\quad\kappa_3 =2+\alpha_3 \,,\quad\kappa_4 = 1+\alpha_3+\alpha_4\,.
\ee
The coefficients $\alpha_n$'s in \eqref{act0} are related to the coefficients $\beta_n$ used in the vierbein action \eqref{act_vierbein2} as:
\be
\beta_n=\mathbb B_{nm}\alpha_m\,,\qquad\text{with}\qquad\mathbb B_{nm}=
\begin{pmatrix}
0&0&0&0&-12\\
0&0&0&3&12\\
0&0&-2&-6&-12\\
0&3&6&9&12\\
-12&-12&-12&-12&-12
\end{pmatrix}\,.
\ee
Finally, inserting $E^a_\mu=\delta^a_\mu+h^a_\mu$ in the action \eqref{act_vierbein2} we are lead back to \eqref{drgt} with the coefficients $\beta_n$ and $\kappa_n$ related as
\be\label{kappa_beta}
\kappa_n=\mathbb C_{nm}\beta_m\,,\qquad\text{with}\qquad\mathbb C_{nm}=-\frac{1}{12}
\begin{pmatrix}
1&4&6&4&1\\
0&4&12&12&4\\
0&0&6&12&6\\
0&0&0&4&4\\
0&0&0&0&1
\end{pmatrix}\,,
\ee
and satisfying $\mathbb C\mathbb B=\mathbb A$.

\subsection{Cycle of Interactions}\label{sec:metric_cycle}
An equivalent way of writing the cycle interactions given in \eqref{act_nonder_general} in the vierbein form is:
\be
g_*^2S_{\text{non-der}}=\frac{m^2M^2}{4}\int \d^4x\,\sum_{n=0}^4\sum_{m=0}^{4-n}\frac{\beta_{nm}}{n!(4-n)!m!(4-m)!}\,\mathcal U_{nm}(I,E,F)\,,\label{act_nonder}
\ee
where  we have extended the notations \eqref{Unm} as
\be\label{Unml}
\begin{split}
\mathcal U_{nm}(I,E,F)\equiv \varepsilon\varepsilon I^{4-(n+m)}E^nF^{m}\,,
\end{split}
\ee
that written explicitly in the index notation become
\be
\begin{split}
\varepsilon_{a_1\dots a_na_{n+1}\dots a_{n+m}a_{n+m+1}\dots a_4}\varepsilon^{\mu_1\dots \mu_n \mu_{n+1}\dots\mu_{n+m}\mu_{n+m+1}\dots\mu_4}E^{a_1}_{\mu_1}\dots E^{a_n}_{\mu_n}F^{a_{n+1}}_{\mu_{n+1}}\dots F^{a_{n+m}}_{\mu_{n+m}} I^{a_{n+m+1}}_{\mu_{n+m+1}}\dots I^{a_4}_{\mu_4}\,.\nn
\end{split}
\ee
The non-derivative terms written in \eqref{act_nonder} include both the individual mass terms as well as the interactions between the dynamical vierbeins $E^a$ and $F^a$. For instance, the $\beta_{n0}$ terms coincide with the potential terms in the Lagrangian of a single massive graviton given in \eqref{act_vierbein} and only involve pairwise interactions between the vierbein $E^a_\mu$ and the reference vierbein $I^a_\mu=\delta^a_\mu$ (similarly, the $\beta_{0n}$ terms are the mass terms for $F^a$).

The $\beta_{nm}$ terms with $n,m\neq 0$, however, also allow for mixed interactions between all \emph{three} vierbeins and also involve pairwise interactions directly between the two dynamical vierbeins $E$ and $F$:
\be
\begin{split}
\frac{m^2M^2}{4}&\int \varepsilon_{abcd}\left[\frac{\beta_{11}}{3!3!}I^a\wedge I^b\wedge E^c\wedge F^d+\frac{\beta_{12}}{3!2!2!}I^a\wedge E^b\wedge F^c\wedge F^d+\frac{\beta_{13}}{3!3!}E^a\wedge F^b\wedge F^c\wedge F^d\right.\\
&\left.+\frac{\beta_{21}}{3!2!2!}I^a\wedge E^b\wedge E^c\wedge F^c+\frac{\beta_{22}}{(2!)^4}E^a\wedge E^b\wedge F^c\wedge F^d+\frac{\beta_{31}}{3!3!}E^a\wedge E^b\wedge E^c\wedge F^d\right]\,.
\end{split}
\ee
Finally, let us remark that the coefficients $\kappa_{nm}$ used in \eqref{act_nonder_general} in the main text are related to $\beta_{nm}$ as
\be\label{kappa_beta_2}
\kappa_{nm}=\frac{1}{4}\mathbb C_{nk}\,\beta_{kl}\,(\mathbb C^T)_{lm}\,,
\ee
with $\mathbb C$ defined in \eqref{kappa_beta}.
with $\beta_{n0}=\beta_{0m}=0$. Note that in this action the first and the second line contains terms of similar structure because of our conventions when writing the action \eqref{act_nonder}.

The symmetric vierbein condition \eqref{sym_vierb_gen2} allows the identification \eqref{identify0} between the vierbeins and the symmetric spin-2 fields $h$ and $f$ equivalently expressed when defining the metric perturbations as
\begin{align}\label{def_perts}
g^{(1)}_{\mu\nu}\equiv(\eta_{\mu\nu}+h_{\mu\nu})^2\,,\qquad g^{(2)}_{\mu\nu}\equiv(\eta_{\mu\nu}+f_{\mu\nu})^2\,.
\end{align}
With this in mind the vierbein action for the cycle of interactions \eqref{act_nonder2} together with the corresponding kinetic terms becomes a nonlinear action describing the fields $h_{\mu\nu}$ and $f_{\mu\nu}$ as
\be\label{action}
\begin{split}
g_*^2 \L_{\text{cycle}}=&\frac{M_1^2}{2}\sqrt{-g^{(1)}}\,R[g^{(1)}]+\frac{m_1^2M_1^2}{4}\sqrt{-\eta}\,\sum_{n=0}^4\kappa^{(1)}_n\,\U_n\left[\eta^{-1}h\right]\\
+&\frac{M_2^2}{2}\sqrt{-g^{(2)}}\,R[g^{(2)}]+\frac{m_2^2M_2^2}{4}\sqrt{-\eta}\,\sum_{n=0}^4\kappa^{(2)}_n\,\U_n\left[\eta^{-1}f\right]+\frac{m^2M^2}{4}\L_{\text{int}}[h,f] +\L_{\text{h.d.}}\,,
\end{split}
\ee
with the interaction term
\be\label{def_int}
\L_{\text{int}}[h,f]=\kappa_{21}\L_{hhf}+\kappa_{12}\L_{hff}+\kappa_{22}\L_{hhff} + \kappa_{31} \L_{hhhf}+ \kappa_{13} \L_{hfff}\,,
\ee
and the various terms defined as
\begin{equation}\label{def_int_2}
\begin{split}
&\L_{hhf}=\varepsilon_{\mu\nu\alpha\beta}\varepsilon^{\mu\nu'\alpha'\beta'}h^\nu_{\nu'}h^{\alpha}_{\alpha'}f^{\beta}_{\beta'}\,, \\
&\L_{hff}=\varepsilon_{\mu\nu\alpha\beta}\varepsilon^{\mu\nu'\alpha'\beta'}h^\nu_{\nu'}f^{\alpha}_{\alpha'}f^{\beta}_{\beta'}\,, \\
&\L_{hhff}=\varepsilon_{\mu\nu\alpha\beta}\varepsilon^{\mu'\nu'\alpha'\beta'}h^\mu_{\mu'}h^\nu_{\nu'}f^{\alpha}_{\alpha'}f^{\beta}_{\beta'}\;,
\end{split}
\end{equation}
and similarly for $\L_{hhhf}$ and $\L_{hfff}$.
We also set $\kappa_0^{(i)}=\kappa_1^{(i)}=0$ as the no cosmological constant and no tadpole conditions and normalize the spin-2 masses as $\kappa_2^{(i)}=1$. Finally $\L_{\text{h.d.}}$ denotes higher derivative terms that arise in the effective theory.

We note that we could have written the action \eqref{action} without ever referring to the vierbeins. It is the action of two interacting massive spin-2 fields, such that in the absence of interaction terms $\mathcal L_{\text{int}}$ each is described by the standard ghost-free massive gravity Lagrangian. In particular, at quadratic level it reduces to two copies of the standard Fierz--Pauli Lagrangian for massive spin-2 fields, \eqref{eq:2FP}, while mixes the two metrics at nonlinear level, leading to a cycle of interactions (Fig.~\ref{Fig:cycle} [left]). Nevertheless, as we have shown above, the action \eqref{action} can be obtained starting from the vierbein formulation by imposing the symmetric vierbein condition \eqref{sym_vierb_gen2} on the otherwise unconstrained action \eqref{act_nonder2}.

\subsection{Line of Interactions}\label{sec:metric_line}
As presented in the main text the line of interactions are most conveniently described by the vierbein action \eqref{act_nonder_line} which can be easily related to the most general action \eqref{act_nonder} introduced in the previous subsection by the choice of coefficients $\beta_{0m}=0$, $\beta_{nm}=0$ for $n+m<4$ and the following mapping:
\be\label{rel_betas_1}
\tilde \beta^{(1)}_n=-\frac{1}{2}\frac{m^2M^2}{m_1^2M_1^2}\frac{\beta_{n0}}{4!}\,,
\ee
and
\be\label{rel_betas_2}
\tilde \beta^{(2)}_1=-\frac{1}{2}\frac{\beta_{31}}{3!}\,,\qquad\tilde \beta^{(2)}_2=-\frac{1}{2}\frac{\beta_{22}}{2!2!}\,,\qquad\tilde\beta^{(2)}_3=-\frac{1}{2}\frac{\beta_{13}}{3!}\,,\qquad\tilde\beta^{(2)}_4=-\frac{1}{2}\frac{\beta_{04}}{4!}\,.
\ee
In order to write the metric formulation of the vierbein action for line of interactions it is useful to further rotate the action \eqref{act_nonder_line2} in the form
\ba\label{act_nonder_line3}
g_*^2S_{\text{non-der}}=\frac{m_1^2M_1^2}{4}\int \d^4x\,&&\sum_{n=0}^4\tilde \kappa^{(1)}_{n}\,\mathcal U_{n}(I,E-I)\\
+\frac{m^2M^2}{4}\int  \d^4x \left(\det E\right)\,&&\sum_{n=0}^4\tilde \kappa^{(2)}_n\,\mathcal U_n\left(I, E^{-1}F-I\right)\,,\nn
\ea
with coefficients $\tilde\kappa_n^{(i)}$ related to $\tilde \beta^{(i)}_n$ as in \eqref{kappa_beta}.\\

As discussed in Section~\ref{sec:line_2} the symmetric vierbein conditions in this case read \eqref{sym_vierb_gen3} and imply that the symmetric metric perturbations should be related to the vierbeins as:
\be\label{sym_vierb_line4}
E^a_\mu=\delta^a_\mu+\tilde h^a_\mu\,,\qquad \delta^\mu_\nu+\left(g^{-1}_{(1)}\tilde f\right)^\mu_\nu=\left(E^{-1}F\right)^\mu_\nu\equiv E_a^\mu F^a_\nu\,.
\ee
To see why this is the case, we note that the above equation implies
\be
F^a_\nu=E^a_\nu+E^a_\mu g_{(1)}^{\mu\omega} \tilde f_{\omega \nu}
\ee
In component form, the symmetric vierbein condition \eqref{sym_vierb_gen3} is $F_{\mu}^a E_{\nu}^b \eta_{ab} - \mu \leftrightarrow \nu =0$. Now
\be
F_{\mu}^a E_{\nu}^b \eta_{ab} = (E^a_\mu+E^a_\alpha g_{(1)}^{\alpha\omega} \tilde f_{\omega \mu} ) E_{\nu}^b \eta_{ab}=g^{(1)}_{\mu \nu} + \tilde f_{\mu\nu}
\ee
and so the symmetric vierbein condition is satisfied for symmetric $ \tilde f_{\mu\nu}= \tilde f_{\nu\mu}$. \\

In metric language this corresponds to defining the metric perturbations as in \eqref{def_pert_new}:
\be
g_{\mu\nu}^{(1)}\equiv (\eta_{\mu\nu}+\tilde h_{\mu\nu})^2\,,\qquad g_{\mu\nu}^{(2)}\equiv (g_{\mu\alpha}^{(1)}+\tilde f_{\mu\alpha})g^{\alpha\beta}_{(1)}(g_{\beta\nu}^{(1)}+\tilde f_{\beta\nu})\,.
\ee

By using the equation \eqref{sym_vierb_line4}, it is now straightforward to rewrite the potential \eqref{act_nonder_line3} in metric form. Together with the Einstein--Hilbert kinetic terms this gives the full action of a line of perturbations:
\ba\label{act_metric_line}
g_*^2 \L_{\rm line}&=&\frac{M_1^2}{2}\sqrt{-g^{(1)}}\,R[g^{(1)}]
+\frac{m_1^2M_1^2}{4}\sqrt{-\eta}\,\sum_{n=0}^4\tilde\kappa_n^{(1)}\,\U_n\left[\eta^{-1}\tilde h\right]\\
&+&\frac{M_2^2}{2}\sqrt{-g^{(2)}}\,R[g^{(2)}]+\frac{m^2M^2}{4}\sqrt{-g^{(1)}}\,\sum_{n=0}^4\tilde \kappa^{(2)}_n\,\U_n\left[g^{-1}_{(1)}\tilde f\right]
+\L_{\text{h.d.}}\,.\nn
\ea
We also note that one could rewrite the above action in more traditional form used in the context of massive gravity by constructing the tensor $\mathcal K^\mu_\nu$ out of $g\mn^{(1,2)}$ as
\be\label{Kmunu}
\mathcal K(g^{(1)},g^{(2)})^\mu_\nu =\delta \mupn -\(\sqrt{g_{(1)}^{-1}g_{(2)}}\)\mupn=-g_{(1)}^{\mu\alpha}\tilde f_{\alpha\nu}=-\left(g^{-1}_{(1)}\tilde f\right)^\mu_\nu\,.
\ee
The symmetric vierbein condition then ensures that $\tilde \K\mn=g^{(1)}_{\mu\alpha}\K^\alpha_{\,\nu}$ is symmetric which is equivalent to $\tilde f_{\alpha\nu}=\tilde f_{\nu \alpha}$. Together with \eqref{Kmunu1} this makes the rewriting trivial. As expected, we see that the interactions between $g^{(1)}$ and $g^{(2)}$ that would have the highest possible cutoff are the double--epsilon polynomials of  $\K$.

\section{Integrating out the Lorentz St\"uckelberg Fields}\label{sec:Leom}

Here we consider the EFT of cycle of interactions \eqref{act_nonder2} in the unconstrained vierbein formalism. This theory involves genuine 16 component vierbeins $E^a{}_\mu$ (and $F^a{}_\mu$) and is not equivalent to the constrained vierbein theory that is obtained after imposing the symmetric vierbein conditions \eqref{sym_vierb_2}. In the unconstrained vierbein formulation of the theory, the Lorentz \stu fields are auxiliary fields and are determined by varying the action with respect to those fields. Unsurprisingly this leads to expressions for the Lorentz \stu fields that differ from the symmetric vierbein conditions \eqref{sym_vierb_2} leading to a theory which differs from its constrained version. While Section~\ref{sec:cycle} in the main text focused on the metric or constrained symmetric vierbein formulation of the cycle interactions, in what follows we shall explore briefly the unconstrained case and show the existence of interactions at the same scale.  \\

For the purpose of this section we use the non-derivative cycle interactions written in the form \eqref{act_nonder}:\\
\be\label{action_short}
g_*^2 S_{\text{non-der}}=-\frac{m^2\M^2}{2}\int \d^4x\,\sum_{n=0}^4\sum_{m=0}^n\frac{d_{nm}}{n!m!(4-n-m)!}\,\mathcal U_{nm}(I,E,F)\,,
\ee
with slightly different coefficients, related to the original $\beta_{nm}$'s as:
\be
d_{nm}=-\frac{1}{2}\,\beta_{nm}\,\frac{(4-n-m)!}{(4-n)!(4-m)!}\,.
\ee
We then make a further use of the relationship
\be
\frac{1}{n!m!(4-n-m)!}\mathcal U_{nm}(\mathbb I, \mathbb X,\mathbb Y)=\left.\frac{\partial^n}{\partial \mu^n}\frac{\partial^m}{\partial \nu^m}\det (\mathbb I+\mu \mathbb X+\nu\mathbb Y)\right|_{\mu=\nu=0}\,,
\ee
to write the action as
\be\label{action_short_2}
S_{\text{non-der}}=-\frac{m^2\M^2}{2}\int \d^4x\,\sum_{n=0}^4\sum_{m=0}^nd_{nm}\left.\frac{\partial^n}{\partial \mu^n}\frac{\partial^m}{\partial \nu^m}\det ( I+\mu   E+\nu F)\right|_{\mu=\nu=0}\,.
\ee

In order to derive the equations of motion for Lorentz St\"uckelberg fields, we introduce all the \stu fields in \eqref{action_short_2} (the diff and the Lorentz ones). Varying with respect to the Lorentz \stu $\Lambda$ then gives
\be
\begin{split}
\delta_\Lambda&\left(\det\left(I+ \mu \Lambda E\partial\phi+\nu \Gamma F\partial\psi\right)\right)=\\
&\det\left(I+\mu \Lambda E\partial\phi+\nu \Gamma F\partial\psi\right)\tr\left[\delta_\Lambda(\mu\Lambda E\partial\phi)\left(I+\mu\Lambda E\partial\phi+\nu\Gamma F\partial\psi\right)^{-1}\right]\,.
\end{split}
\ee
Similarly as in \cite{Ondo:2013wka} we further use the property
\be
\delta_\Lambda (\Lambda E\partial\phi)=(\delta_\Lambda\Lambda\Lambda^{-1}\eta)\eta(\Lambda E\partial\phi)
\ee
to write the equation of motion as
\be
\tr \left[(\delta_\Lambda\Lambda\Lambda^{-1}\eta)\eta(\Lambda E\partial\phi)\left(I+\mu\Lambda E\partial\phi+\nu\Gamma F\partial\psi\right)^{-1}\right]=0\,.
\ee
We then note that $(\delta\Lambda\Lambda^{-1}\eta)^T=-\delta\Lambda\Lambda^{-1}\eta$. Combined with the property of the trace that $\tr \mathbb X=\tr\mathbb X^T$ for any matrix $\mathbb X$ it then follows that for the equation of motion to be satisfied the following has to hold:
\be
\eta(\Lambda E\partial\phi)\left(I+\mu \Lambda E\partial\phi+\nu \Gamma F\partial\psi\right)^{-1}=\left[\eta(\Lambda E\partial\phi)\left(I+\mu \Lambda E\partial\phi+\nu\Gamma F\partial\psi\right)^{-1}\right]^T\,.
\ee
This can be rewritten as
\be\label{eqLambda}
(\Lambda E\partial\phi)^T\eta I-I^T\eta(\Lambda E\partial\phi)=\nu\left[(\Gamma F\partial\psi)^T\eta(\Lambda E\partial\phi)-(\Lambda E\partial\phi)^T\eta(\Gamma F\partial\psi)\right]\,.
\ee
We note that the above equation is independent on $\mu$ and the left hand side for $\nu=0$ gives the usual symmetric vierbein condition in massive gravity:
\be
(\Lambda E\partial\phi)^T\eta I=I^T\eta(\Lambda E\partial\phi)\,.
\ee
After performing the substitutions \eqref{stuck} in terms of the various helicities and taking the decoupling limit this takes the form
\be\label{sol_single}
\partial_a A_b-\partial_bA_a=2\omega_{ab}-\left(\omega_{da}\hat{\Pi}^d_b-\omega_{db}\hat{\Pi}^d_a\right)\,,
\ee
where we define $\hat{\Pi}^\mu_\nu\equiv\partial^\mu\partial_\nu\pi/\Lambda_{3}^3$\,. Similarly, varying the action \eqref{action_short_2} with respect to $\Gamma$ gives
\be\label{eqGamma}
(\Gamma F\partial\psi)^T\eta I-I^T\eta(\Gamma F\partial\psi)=-\mu\left[(\Gamma F\partial\psi)^T\eta(\Lambda E\partial\phi)-(\Lambda E\partial\phi)^T\eta(\Gamma F\partial\psi)\right]\,.
\ee
We note that the right hand sides of equations \eqref{eqLambda} and \eqref{eqGamma} coincide up to the coefficients $\mu,\nu$. In decoupling limit the equation \eqref{eqLambda} becomes
\be\label{EOM1}
\begin{split}
&\left(\delta^\mu_b\left(1+\frac{1}{\nu}\right)+\hat{\mathbb X}^\mu_b\right)\left(-\omega_{\mu a}+\partial_\mu A_a-\omega_{ca}\hat{\Pi}^c_\mu\right)\\
&-\left(\delta^\mu_a\left(1+\frac{1}{\nu}\right)+\hat{\mathbb X}^\mu_a\right)\left(\omega_{b\mu}+\partial_\mu A_b-\omega_{cb}\hat{\Pi}^c_\mu\right)\\
&=\left(\delta^\mu_b+\hat{\Pi}^\mu_b\right)\left(-\sigma_{\mu a}+\partial_\mu B_a-\sigma_{ca}\hat{\mathbb X}^c_\mu\right)-\left(\delta^\mu_a+\hat{\Pi}^\mu_a\right)\left(\sigma_{b\mu}+\partial_\mu B_b-\sigma_{cb}\hat{\mathbb X}^c_\mu\right)\,,
\end{split}
\ee
where we have defined $\hat{\mathbb X}^\mu_\nu\equiv\partial^\mu\partial_\nu\chi/\Lambda_{3}^3$. The equation \eqref{eqGamma} in turn becomes:
\be\label{EOM2}
\begin{split}
&\left(\delta^\mu_b+\hat{\mathbb X}^\mu_b\right)\left(-\omega_{\mu a}+\partial_\mu A_a-\omega_{ca}\hat{\Pi}^c_\mu\right)-\left(\delta^\mu_a+\hat{\mathbb X}^\mu_a\right)\left(\omega_{b\mu}+\partial_\mu A_b-\omega_{cb}\hat{\Pi}^c_\mu\right)\\
&=\left(\delta^\mu_b\left(1+\frac{1}{\mu}\right)+\hat{\Pi}^\mu_b\right)\left(-\sigma_{\mu a}+\partial_\mu B_a-\sigma_{ca}\hat{\mathbb X}^c_\mu\right)\\
&-\left(\delta^\mu_a\left(1+\frac{1}{\mu}\right)+\hat{\Pi}^\mu_a\right)\left(\sigma_{b\mu}+\partial_\mu B_b-\sigma_{cb}\hat{\mathbb X}^c_\mu\right)\,.
\end{split}
\ee
Importantly, the symmetric vierbein condition \eqref{sol_single} is not a solution when $\mu\,,\nu\neq 0$.

As already explained in the main text, in the analysis performed there we have a different situation in mind. Instead of treating the Lorentz St\"uckelberg fields as auxiliary fields with their own equations of motion we consider a \emph{constrained} version of the action \eqref{action_short_2}. In other words we impose the symmetric vierbein constraint \eqref{sol_single} on both Lorentz St\"uckelberg fields, $\omega$ and $\sigma$, and derive the decoupling limit action for the remaining fields. However, it is important to emphasize that had we used the actual equations of motion to integrate out $\omega$ and $\sigma$ in the decoupling limit action \eqref{DLvec3} for the cycle interactions it would not change the scale of the leading decoupling limit interactions. Indeed, at leading order in fields the equations of motion
\eqref{EOM1} and \eqref{EOM2} coincide with the symmetric vierbein conditions thus leading to the same conclusions about the strong coupling case as in the constrained theory.

\section{Bi-Galileon Coupling Constants}
\label{App:biGalileondetails}

The bi-Galileon coupling constants $\tilde \kappa^{(i)}_n$ introduced in \eqref{eq:BiGalX_1} are given in terms of the $\tilde \beta^{(i)}_n$'s as follows
\ba
\label{eq:exp1}
&\tilde\kappa_2^{(1)}=-\tilde\beta_1^{(1)}-2\tilde\beta_2^{(1)}-\tilde\beta^{(1)}_3\,,\quad\tilde\kappa_3^{(1)}=-\tilde\beta_1^{(1)}-\tilde\beta_2^{(1)}\,,\quad\tilde\kappa_4^{(1)}=-\frac{1}{3}\tilde\beta_1^{(1)}\,,\\
&\tilde\kappa_2^{(2)}=-\tilde\beta_1^{(2)}-2\tilde\beta_2^{(2)}-\tilde\beta^{(2)}_3\,,\quad\tilde\kappa_3^{(2)}=-\tilde\beta_2^{(2)}-\tilde\beta_3^{(2)}\,,\quad\tilde\kappa_4^{(2)}=-\frac{1}{3}\tilde\beta_3^{(2)}\,,\\
&\tilde\kappa_2^{(3)}=-\tilde\beta_1^{(2)}-2\tilde\beta_2^{(2)}-\tilde\beta^{(2)}_3\,,\quad\tilde\kappa_3^{(3)}=-\tilde\beta_1^{(2)}-\tilde\beta_2^{(2)}\,,\quad\tilde\kappa_4^{(3)}=-\frac{1}{3}\tilde\beta_1^{(2)}\,.
\ea
Note that while the first two lines of coefficients were already introduced in Section~\ref{sec:hel02_line}, the third one is new, but very similar in structure. In particular, if we relate
\be
\label{eq:exp2}
\tilde\kappa^{(3)}_n=\mathbb D_{nm}\tilde \beta^{(2)}_m \,\qquad\text{with}\qquad\mathbb D_{nm}=
\begin{pmatrix}
0&0&0&0&0\\
0&-1&-2&-1&0\\
0&-1&-1&0&0\\
0&-\frac{1}{3}&0&0&0\\
0&0&0&0&0
\end{pmatrix}\,,
\ee
then also $\tilde\kappa^{(1)}_n=\mathbb D_{nm}\tilde\beta^{(1)}_m$ and $\tilde\kappa^{(2)}_n=\mathbb D_{nl}\mathbb J_{lm}\tilde\beta^{(2)}_m$ where $\mathbb J$ is the exchange matrix (\emph{i.e.} the backward identity matrix).

In Eqns.~(\ref{eq:Gal2}, \ref{eq:biGal2}),
the coefficients  $\bar \kappa^{(i)}$'s  are expressed in terms of the original coefficients $\tilde\beta^{(i)}_n$ in the line interactions \eqref{act_nonder_line2} as follows,
\be
\label{eq:exp3}
\begin{split}
&\bar\kappa^{(1)}_2=(\tilde\beta^{(1)}_1+2\tilde\beta^{(1)}_2+\tilde\beta^{(1)}_3)^2\,,\\
&\bar\kappa^{(1)}_3=2(\tilde\beta^{(1)}_1+\tilde\beta^{(1)}_2)(\tilde\beta^{(1)}_1+2\tilde\beta^{(1)}_2+\tilde\beta^{(1)}_3)\,\\
&\bar\kappa^{(1)}_4=(\tilde\beta^{(1)}_1+\tilde\beta^{(1)}_2)^2\,,\quad\bar\kappa^{(1)}_5=0\,,\\
&\bar\kappa^{(2)}_2=\left(\frac{1}{\gamma^2}+1\right)(\tilde\beta^{(2)}_1+2\tilde\beta^{(2)}_2+\tilde\beta^{(2)}_3)^2\,,\\
&\bar\kappa^{(2)}_3=2\left(\frac{1}{\gamma^2}+1\right)(\tilde\beta^{(2)}_2+\tilde\beta^{(2)}_3)(\tilde\beta^{(2)}_1+2\tilde\beta^{(2)}_2+\tilde\beta^{(2)}_3)\,,\\
&\bar\kappa^{(2)}_4=-\frac{1}{2}(\tilde\beta^{(2)}_1+\tilde\beta^{(2)}_2)^2+\left(\frac{3}{2}+\frac{1}{\gamma^2}\right)(\tilde\beta^{(2)}_2+\tilde\beta^{(2)}_3)^2\,,\\
&\bar\kappa^{(2)}_5=-\frac{2}{5}(\tilde\beta^{(2)}_1-\tilde\beta^{(2)}_2)(\tilde\beta^{(2)}_2+\tilde\beta^{(2)}_3)\,,
\end{split}
\ee
and
\be
\label{eq:exp4}
\begin{split}
&\bar\kappa_{11}=(\tilde\beta^{(1)}_1+2\tilde\beta^{(1)}_2+\tilde\beta^{(1)}_3)(\tilde\beta^{(2)}_1+2\tilde\beta^{(2)}_2+\tilde\beta^{(2)}_3)\,,\\
&\bar\kappa_{12}=(\tilde\beta^{(1)}_1+2\tilde\beta^{(1)}_2+\tilde\beta^{(1)}_3)(\tilde\beta^{(2)}_2+\tilde\beta^{(2)}_3)\,,\\
&\bar\kappa_{21}=(\tilde\beta^{(1)}_1+\tilde\beta^{(1)}_2)(\tilde\beta^{(2)}_1+2\tilde\beta^{(2)}_2+\tilde\beta^{(2)}_3)\,,\\
&\bar\kappa_{22}=(\tilde\beta^{(1)}_1+\tilde\beta^{(1)}_2)(\tilde\beta^{(2)}_2+\tilde\beta^{(2)}_3)\,.
\end{split}
\ee

\bibliographystyle{JHEP}
\bibliography{references}

\end{document}